\documentclass[english,11pt]{article}
\pdfoutput=1
\usepackage{jcappub}
\usepackage{lmodern}

\usepackage{graphicx}
\usepackage{esint}
\usepackage[toc,page]{appendix}
\usepackage{array}
\usepackage{longtable}
\usepackage{booktabs}
\usepackage{titlesec}
\usepackage{slashed}

\usepackage{tocloft}   


\begin{document}

    \title{\huge{ Cosmic Microwave Background Anisotropy numerical solution (CMBAns) I:  An introduction to $C_l$ calculation}}
\author[a,b]{Santanu Das}
\author[a]{Anh Phan}
\affiliation[a]{Department of Physics, University of Wisconsin, Madison, 1150 University Avenue
Madison, WI 53706, USA}
\affiliation[b]{Fermi National Accelerator Laboratory, PO Box 500, Batavia, IL 60510, USA}
\emailAdd{sdas33@wisc.edu, anh@wisc.edu}


\abstract{
Cosmological Boltzmann codes 
are often used by researchers for calculating the CMB angular power spectra from different theoretical models, for cosmological parameter estimation, etc. Therefore, the accuracy of a Boltzmann code is of utmost importance. 
Different Markov Chain Monte Carlo based parameter estimation algorithms typically require $10^3 \-- 10^4$ iterations of Boltzmann code. This makes the time complexity of such codes another critical factor.
In the last two decades, several Boltzmann packages, such as \texttt{CMBFAST}, \texttt{CAMB}, \texttt{CMBEasy}, \texttt{CLASS} etc., have been developed. In this paper, we present a new cosmological Boltzmann code, \texttt{CMBAns}, that can be used for accurate calculation of the CMB power spectrum and BipoSH coefficients. At present, \texttt{CMBAns} is developed for a flat background matrix.  It is mostly written in the \texttt{C} language. However, we borrowed the  concept of class from \texttt{C++}. This gives researchers the flexibility to develop their own independent package based on \texttt{CMBAns}, without an in-depth understanding of the source code. We also develop multiple stand-alone facilities which can be directly compiled and run on a given parameter set. In this paper, we discuss all the mathematical formulation, approximation schemes, integration methods etc., that are used in \texttt{CMBAns}. The package will be made available through \texttt{github} for public use in the near future.   
}

\maketitle

\section{Introduction}

Since the discovery of the Cosmic Microwave Background (CMB) by Penzias and Wilson, the CMB has become an invaluable probe for understanding the physics in the early universe. Several cosmological theories, proposed in the past, failed to explain the origin of CMB. Hence, they were rejected as feasible cosmological theories. Others like Big Bang cosmology, with some assumptions, provide a more complete explanation of the origin of CMB radiation. These models later, after several theoretical modifications, were accepted as the standard cosmological models. Hence, the discovery of CMB marked the path for the birth of standard cosmology.

The precision of the CMB observation has improved over the years. In the past decade, several ground-based and space-based experiments like WMAP, Planck, BICEP, ACT etc. have measured the CMB temperature to an exquisite precision.  Future experiments like SPT-3G and Simons Array will provide even better measurements of CMB temperature and polarization~\citep{abazajian2016cmb}. To analyze this influx of data and to test different cosmological models, we also need more accurate Boltzmann codes. 

The theory of cosmological perturbations for standard model cosmology was first developed by Lifshitz~\citep{Lifshitz1946} and later was reviewed by many others~\citep{Lifshitz1963}. The subsequent research works are summarized in review articles~\citep{Kodama1984,Mukhanov1992}, in books and in theses~\citep{Weinberg2008cosmology,Tassev2011}.
Lifshitz used the synchronous gauge for formulating the linear perturbation
theory. Later, Bardeen and others developed the perturbation theory
in the conformal gauge due to some complications with the synchronous
gauge, such as the appearance of the coordinate singularity \citep{Kodama1984,Bardeen1980} etc.
The conformal gauge is more frequently used for analytical calculations
of the cosmological perturbation equations. However, the synchronous gauges are preferred for the numerical calculations due to the stability issues~\citep{Montani2011,Hu2004}. 

The Boltzmann codes have been in use in cosmology for a long time to calculate
the CMB angular power spectrum. The first of such code provided in the public domain was  \texttt{COSMICS} \citep{Ma1995}, written
by Ma and Bertschinger. 
Later, Seljac and Zaldarriaga developed \texttt{CMBFAST} \citep{Seljak1996,Zaldarriaga2000}, in which the line-of-sight integration method was used to make the power spectrum
calculation faster. Since then, several packages utilizing Boltzmann codes, such as \texttt{CAMB}~\citep{CAMB-1}, \texttt{CMBEasy} \citep{Doran2005}, \texttt{CLASS}~\citep{lesgourgues2011cosmic, blas2011cosmic, lesgourgues2011cosmic1, lesgourgues2011cosmic2}, \texttt{PyCosmo}~\citep{refregier2018pycosmo} etc., have come into existence. In this paper, we describe a new Boltzmann code, called \texttt{CMBAns} (Cosmic Microwave Background Anisotropy numerical solution). The package is based on \texttt{CMBFAST} and was initially developed in 2010 for a variety of CMB studies~\citep{Das2010,Das2013a,Das:2013sca}. 

There are three principal motivations behind developing \texttt{CMBAns}. First of all, in future CMB missions, the precision of the CMB measurements will improve drastically. Hence, the Boltzmann packages should be able to calculate the CMB power spectrum very accurately up to high multipoles. Secondly, different Markov Chain Monte Carlo (MCMC) packages,
such as \texttt{CosmoMC} \citep{Lewis2002,Lewis2013}, \texttt{SCoPE} \citep{scope},
\texttt{AnalyzeThis} \citep{Doran2004}, etc, which are often used to estimate the cosmological
parameters, typically require $10^{3}$ - $10^{4}$ evaluations of Boltzmann codes. Therefore, 
the Boltzmann code should be able to calculate the CMB power spectrum fast and efficiently. 
Thirdly, most of the present Boltzmann codes follow a monolithic architecture design and are not modular. Therefore, it is difficult to add any new feature in the package and the functions cannot be used independently. Users cannot write their own packages and use existing functions without an extensive knowledge of the entire source code.  To overcome this limitation, \texttt{CLASS} code introduced a modular architecture. \texttt{CMBAns} is also written in a modular format. It consists of several stand-alone codes, as well as some user-defined functions that users can use to write their codes. 

\texttt{CMBAns} solves the linear Boltzmann equations for different constituents
of the universe and thereafter uses the line-of-sight integration approach
to calculate the source terms and the brightness fluctuations. These are then convolved with the primordial power spectrum to get the CMB angular
power spectrum. 
\texttt{CMBAns} can calculate the cosmological power spectrum for different dark energy
models (both perturbed and unperturbed), two-field inflation model, etc. 
\texttt{CMBAns} also comes with a MATLAB GUI, where the Hubble parameter of the universe can be visually modified as a function of redshift. \texttt{CMBAns} translates the modified Hubble parameter into the dark energy equation of state (EOS) and then computes the CMB power spectrum for that particular model \citep{Das:2013sca}. Apart from these, \texttt{CMBAns} can also calculate the isotropy violation signals by calculating the BipoSH coefficients along with the angular power spectrum in presence of anisotropic inflation~\citep{Das2014a}. 

This is the first paper, in a series of papers about to come on CMBAns. In this paper, we describe all the mathematical equations and the discretization techniques that are used in \texttt{CMBAns}.  We also briefly describe the cosmological perturbation
theory in the synchronous gauge and  discuss the equations and approximation schemes
used in developing \texttt{CMBAns}. The paper is divided into eight sections. In the second section, we describe the conformal time calculations between any two eras in the universe. In the third section, we describe different recombination processes, calculating the baryon temperature, sound speed etc. The fourth section discusses CMB perturbation calculations and different approximation schemes  used in \texttt{CMBAns}, and how they affect the power spectrum calculations. Different scalar and tensor initial conditions are discussed in the fifth section. The sixth and seventh sections are for line-of-sight integration and calculating the power spectra and the BipoSH coefficients for anisotropic inflation. The numerical techniques, time and wave number space grid etc. are discussed in the eighth section. The final section is for discussion and conclusion. 



\section{Conformal time calculation}
In cosmology, the redshift $z$ is often used for measuring time. However, for numerical calculation of the perturbation equations, line-of-sight integration, etc. the conformal time plays an important role. It is straight-forward to calculate the conformal time under the assumption of a matter-dominated or dark energy-dominated universe. However, in the presence of all the components of the universe, the calculations can be complicated and an analytical solution may not exist. 

In this paper, we denote the conformal time as $\tau$. The Hubble parameter $H(\tau)$ is defined as

\begin{equation}
H^{2}(\tau)  =  \left(\frac{1}{a^{2}}\frac{\mathrm{d}a}{\mathrm{d}\tau}\right)^{2}\,,
\end{equation}

\noindent where $a$ is the scale factor. From the FLRW equation, we can write the Hubble parameter as 

\begin{equation}
\frac{H(\tau)^{2}}{H_{0}^{2}}  =  \Omega_{0,m}a^{-3}+\Omega_{0,\gamma}a^{-4}+\Omega_{0,\nu}a^{-4}+\Omega_{\nu_{m}}+\Omega_{d}\,.
\end{equation}

\noindent The above two equations give

\begin{equation}
\frac{\mathrm{d}a}{\mathrm{d}\tau}=\sqrt{a^{4}H_{0}^{2}\Bigg(\Omega_{0,m}a^{-3}+\Omega_{0,\gamma}a^{-4}+\Omega_{0,\nu}a^{-4}+\Omega_{\nu_{m}}+\Omega_{d}\Bigg)}\,,\label{eq:dadtau}
\end{equation}

\noindent where $\Omega_{0,m}$,  $\Omega_{0,\gamma}$, and $\Omega_{0\nu}$  are the density parameters for present-day matter 
(which include both cold dark matter and baryonic mater), photon, and massless neutrinos respectively. The density parameter of 
massive neutrinos and dark energy at a scale factor $a$ are $\Omega_{\nu_{m}}$ and $\Omega_{d}$, respectively. 
The density parameters are defined as the ratios of the respective densities over the critical density:
\[
\Omega_{0,m}=\frac{\rho_{0,m}}{\rho_{cr}},\;\;\;\;\;\Omega_{0,\gamma}=\frac{\rho_{0,\gamma}}{\rho_{cr}},\;\;\;\;\;\Omega_{0,\nu}=\frac{\rho_{0,\nu}}{\rho_{cr}},\;\;\;\;\;\Omega_{\nu_{m}}=\frac{\rho_{\nu_{m}}}{\rho_{cr}}\;\;\;\;\;\Omega_{d}=\frac{\rho_{d}}{\rho_{cr}}
\]
\noindent where, the critical density $\rho_{cr}$ is given by ${\displaystyle \rho_{cr}=\frac{3H_{0}^{2}}{8\pi G}}$. The densities of matter and radiation at any era are scaled as $a^{-3}$, $a^{-4}$, respectively with their densities at the present era. For calculating the density of the massive neutrinos, we need to use the Fermi-Dirac statistics. For $\Lambda$ dark energy model, the density of the dark energy will be constant. However, for any other dark energy model, we need to calculate the density variation from its equation of state (eos).   

\subsection{Matter density}

The first term in Eq.~\ref{eq:dadtau} can be calculated by evaluating 
$aH_{0}^{2}\Omega_{0,m}$. As the CDM and baryon density parameters, $\Omega_{0,c}$ 
and $\Omega_{0,b}$ are the input parameters, we can calculate $\Omega_{0,m}=\Omega_{0,c}+\Omega_{0,b}$.
$H_{0}$ is also an input parameter, but its unit is $\texttt{km/sec/Mpc}$. In \texttt{CMBAns},  we use \texttt{Mpc} as the unit for both the spatial 
and temporal dimensions. In order to convert the Hubble parameter in $\rm{\texttt{Mpc}^{-1}}$, we  multiply $H_0$ with $1/c^{2}=1.11265\times10^{-11}$$\rm{(\texttt{km}/\texttt{sec})^{-2}}$.


\subsection{Radiation density for photons}

The radiation density consists of two components: 
photon density $\rho_{\gamma}$ and the massless (relativistic) neutrino
density $\rho_{\nu}$.  The photon number density as a function of frequency
can be derived from the Planck radiation law:

\begin{equation}
n_\gamma(\nu)\,\mathrm{d}\nu=\frac{8\pi\nu^{2}\,\mathrm{d}\nu}{e^{h\nu/k_{B}T_{0}}-1}\,,
\end{equation}

\noindent where $k_{B}$ is the Boltzmann constant, $h$ is the Planck constant,
and $T_{0}$ is the current CMB temperature. The photon energy density
can be calculated as 
\begin{equation}
\rho_{0,\gamma}c^{2}=\int_{0}^{\infty}h\nu n_\gamma(\nu)\,\mathrm{d}\nu=a_{B}T_{0}^{4}\,,
\end{equation}

\noindent where
$a_{B}=\frac{8\pi^{5}k_{B}^{4}}{15h^{3}c^{3}}=7.56577\times10^{-16}\;\; \texttt{J}\texttt{m}^{-3}\texttt{K}^{-4}$ is the radiation constant. 
 We also know that
\begin{equation}
\rho_{cr}=\frac{3H_{0}^{2}}{8\pi G}=1.87847\times10^{-30} \,H_{0}^{2}\;\;\texttt{kg}\,\texttt{m}^{-3}(\texttt{km}/\texttt{sec}/\texttt{Mpc})^{-2}.
\end{equation}

\vspace{.7em}
\noindent Therefore, the second term in Eq.~\ref{eq:dadtau} can be calculated by evaluating 
$H_{0}^{2}\Omega_{0,\gamma}$ as follows

\begin{equation}
\Omega_{0,\gamma}H_{0}^{2}=\frac{\rho_{0,\gamma}}{\rho_{cr}}H_{0}^{2}=\frac{a_{B}}{c^{2}\rho_{cr}}T_{0}^{4}=4.98613\times10^{-14}\times T_{0}^{4}\;\texttt{Mpc}^{-2}\;.
\end{equation}

\vspace{.3em}
\subsection{Radiation density for massless neutrinos}

Massless neutrinos follow Fermi-Dirac statistics with neutrino temperature
$T_{\nu}$. The distribution function is given by 

\begin{equation}
n_\nu(\nu)\,\mathrm{d}\nu=\frac{8\pi\nu^{2}\,\mathrm{d}\nu}{e^{h\nu/k_{B}T_{\nu}}+1}\,.
\end{equation}

\noindent We can calculate the radiation density of the massless neutrinos as 
\begin{equation}
\rho_{0,\nu}c^{2}=\int_{0}^{\infty}h\nu n_\nu(\nu)\,\mathrm{d}\nu=\left(\frac{7}{8}\right)a_{B}T_{\nu}^{4}\;.
\end{equation}

For relating the temperatures between photon and neutrinos, consider the era before neutrino and photon decoupling. In that ultra high energy regime, 
as photon and neutrino were coupled, the medium in which they existed had a fixed temperature.
Other species in the medium were electrons (2 spin states), positrons (2 spin states), neutrinos (1 spin state for each of the three generations), and antineutrinos (1 spin states for each of the three generations). 
Shortly after the photon and neutrino decoupling, the temperature drops below the electron mass, and the forward reaction $e^+ + e^- \longleftrightarrow \gamma + \gamma$ (annihilation) becomes strongly favored. This heats up the photons. 
We can assume that this entropy transfer did not affect the
neutrinos because they were already completely decoupled. Using entropy conservation of the electromagnetic plasma, we can calculate the change in the photon temperature before and after $e^\pm$ annihilation. This gives~\citep{dodelson2003modern} 






\[
\frac{T_{\nu}}{T_{0}}=\left(\frac{4}{11}\right)^{1/3}\,.
\]

\noindent The neutrino density is related to the photon density by 
\[
\rho_{0,\nu}=N_{\mathrm{eff}}\left(\frac{7}{8}\right)\left(\frac{4}{11}\right)^{4/3}\rho_{0,\gamma}\;,
\]

\noindent where $N_{\mathrm{eff}}$ is the effective number of neutrinos. 
Theoretically, there are 3 neutrino families. However, due to non-instantaneous decoupling and QED effects etc. the effective neutrino density will be slightly higher then this value. This can be accounted for by considering $N_{\text{eff}}>3$. Considering a general framework for neutrino decoupling, it can be shown that for
non instantaneous neutrino decoupling, $N_{\text{eff}} \approx 3.034$. In addition, the QED effects contribute about $\Delta N_{\text{eff}} \approx 0.011$. Assuming these two effects can be added linearly,
the final value of $N_{\text{eff}} \approx 3.045$~\citep{Dolgov1997,Dolgov1998,Mangano2001,Grohs2015,Escudero2018,Salas2016}.

Therefore, the third term in Eq.~\ref{eq:dadtau} can be calculated as

\begin{equation}
\Omega_{0,\nu}H_{0}^{2}=\frac{\rho_{0,\nu}}{\rho_{cr}}H_{0}^{2}=N_{\text{eff}}\,\frac{7}{8}\left(\frac{4}{11}\right)^{4/3}\frac{a_{B}}{c^{2}\rho_{cr}}T_{0}^{4}=1.1324\times N_{\mathrm{eff}}\times10^{-14}\times T_{0}^{4}\;\texttt{Mp\ensuremath{c^{-2}}}
\end{equation}


\subsection{Radiation density for massive neutrinos}
In standard model of particle physics, neutrinos are massless. However, different experiments point toward a small nonzero mass for the neutrinos. For massive neutrinos, the Fermi-Dirac distribution function contains mass
term, and it is not analytically integrable. 
Therefore, to get the density $\rho_{\nu_{m}}$ at any given redshift, the distribution function must be integrated numerically. 




Assuming that all the neutrino species have equal mass, the mass of the neutrinos is given by

\begin{equation}
m_{\nu_{m}}  =  \frac{\rho_{0,\nu_{m}}}{N_{\mathrm{\text{eff}}}n_{0,\nu_{m}}}
  =  \frac{\Omega_{0, \nu_{m}}}{N_{\text{eff}}}\frac{\rho_{cr}}{n_{0,\nu_{m}}}\,,
\end{equation}

\noindent where $N_{\text{eff}}$ is the effective number of neutrinos. $\rho_{0,\nu_{m}}$ and $n_{0,\nu_{m}}$
are the massive neutrino density and number density at present time respectively. $\rho_{cr}$ is the critical density.

The neutrino number density can be calculated by integrating the
Fermi-Dirac distribution function:

\begin{equation}
n_{\nu_{m}}=\frac{8\pi }{h^{3}}\int_{0}^{\infty}\frac{p^{2}\mathrm{d}p}{\exp(\sqrt{p^{2}c^{2}+m^{2}c^{4}}/k_{b}T_{\nu_{m}})+1)}\,.
\end{equation}

\noindent For neutrinos $pc\gg mc^{2}$, and we can ignore the term $mc^{2}$ in the above equation. This  simplifies to

\begin{equation}
n_{\nu_{m}}=\frac{8\pi}{h^{3}}\int_{0}^{\infty}\frac{p^{2}\mathrm{d}p}{\exp(pc/k_{b}T_{\nu_{m}})+1}=\frac{8\pi}{h^{3} c^{3}}k_{b}^{3}T_{\nu_{m}}^{3}\int_{0}^{\infty}\frac{\xi^{2}\mathrm{d}\xi}{e^{\xi}+1}=\frac{8\pi c^{3}}{h^{3}}k_{b}^{3}T_{\nu_{m}}^{3}\zeta(3)\Gamma(3)\,,
\end{equation}

\noindent where $\zeta(3)$ is the Riemann Zeta function and $\Gamma(3)$ is
the Gamma function. $\Gamma(3)=2!=2$. 


\noindent The density and pressure of massive neutrinos at any given redshift can be written as 
\begin{eqnarray}
\rho & = & \frac{8\pi}{h^{3}c^{3}}k_B^4 T_{\nu_{m}}^4\int_0^\infty q{}^{2}f(q)\epsilon(q)\,\mathrm{d}q\label{eq:neutrino_density}\,,\\
P & = & \frac{8\pi}{h^{3}c^{3}}k_B^4 T_{\nu_{m}}^4\int_0^\infty q^{2}f(q)\frac{q^{2}}{3\epsilon}\,\mathrm{d}q\,,\label{eq:neutrino_pressure}
\end{eqnarray}
where $q = apc$ and,
\begin{equation}
\epsilon=\frac{a}{k_BT_{\nu_{m}}}\sqrt{m^2_{\nu_{m}}c^{4} + (pc)^{2}}\,.
\end{equation}
\noindent Here, in Eq.~\ref{eq:neutrino_pressure} the factor of 3 comes because we consider 3 spatial dimensions\footnote{For an ideal gas, the pressure can be found by $n m v^2 /3$. $n$ is the number density of the gas molecule, $v$ is the velocity, and $m$ is the mass of each gas molecules. The factor of 3 arises because we have considered 3 special dimensions and we consider that the velocity distribution of the gas is isotropic,  i.e. $v_x^2 =v_y^2=v_z^2=v^2/3$. Eq.~\ref{eq:neutrino_pressure} can also be derived in a similar way, where $q$ corresponds to the momentum.}.
Simple re-arrangements of the above equations give us the massive neutrino density and pressure in terms of massless neutrino density, as

\begin{eqnarray}
\rho & = & \left(\frac{7}{8}\right)a_{B}T_{\nu}^{4}\rho_{\mathrm{dl}}=\left(\frac{7}{8}\right)a_{B}T_{0,\nu}^{4}a^{-4}\rho_{\mathrm{dl}}=\frac{7}{8}\left(\frac{4}{11}\right)^{4/3}a_{B}T_{0}^{4}a^{-4}\rho_{\mathrm{dl}}\,,\\
P & = & \left(\frac{7}{8}\right)a_{B}T_{\nu}^{4}p_{\mathrm{dl}}=\left(\frac{7}{8}\right)a_{B}T_{0,\nu}^{4}a^{-4}P_{\mathrm{dl}}=\frac{7}{8}\left(\frac{4}{11}\right)^{4/3}a_{B}T_{0}^{4}a^{-4}P_{\mathrm{dl}}\,.
\end{eqnarray}

\noindent Here $\rho_{\mathrm{dl}}$ and $p_{\mathrm{dl}}$ are dimensionless density and pressure and are expressed as

\begin{eqnarray}
\rho_{\mathrm{dl}} & = &\frac{1}{\Upsilon} \int_0^\infty q{}^{2}f(q)\epsilon(q)\,\mathrm{d}q\,,\label{rhodl}\\
P_{\mathrm{dl}} & = & \frac{1}{\Upsilon} \int_0^\infty q^{2}f(q)\frac{q^{2}}{3\epsilon}\,\mathrm{d}q\,.\label{pdl}
\end{eqnarray}

\begin{figure}
\includegraphics[width=0.49\textwidth,trim = 320 320 320 320, clip]{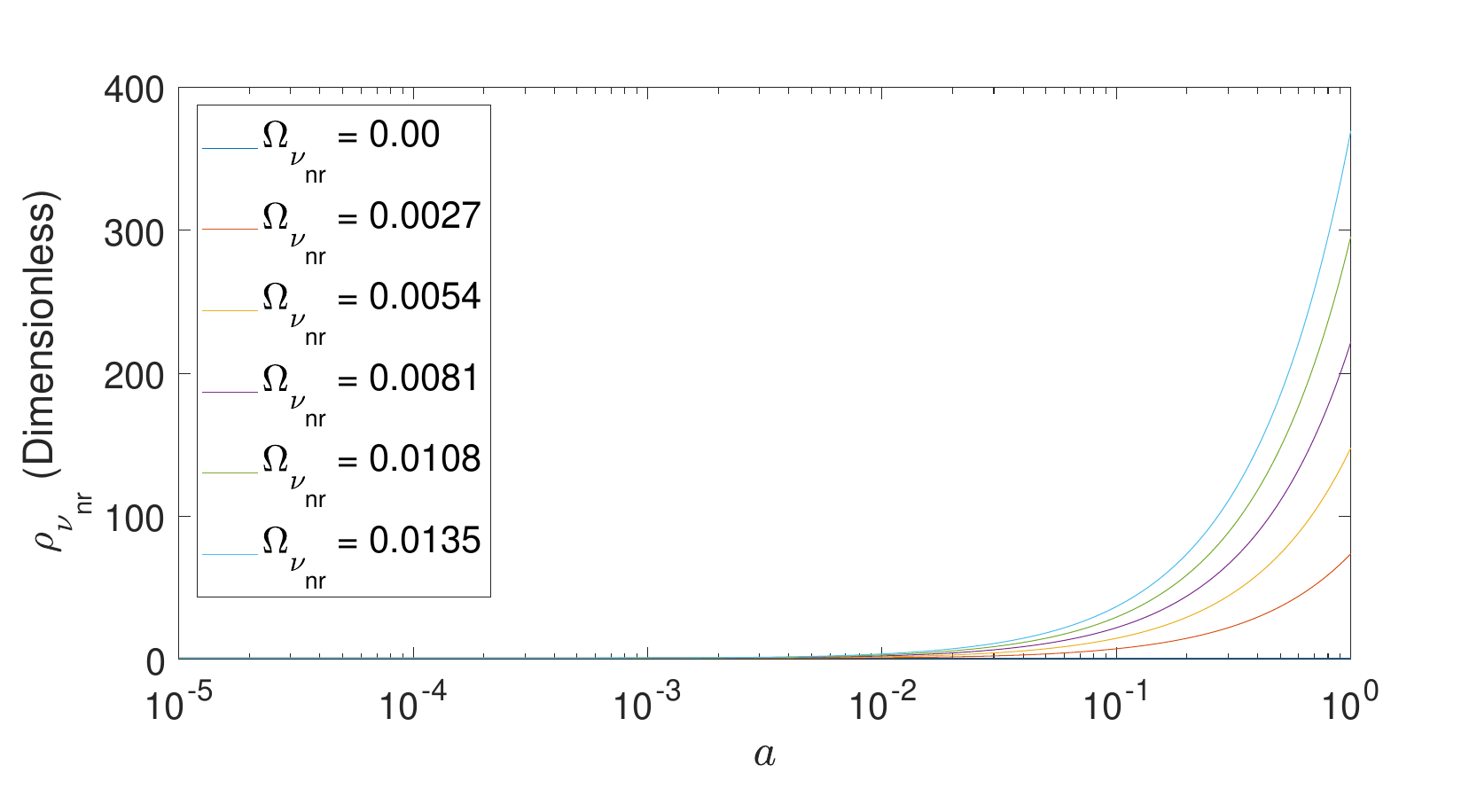}
\includegraphics[width=0.49\textwidth,trim = 320 320 320 320, clip]{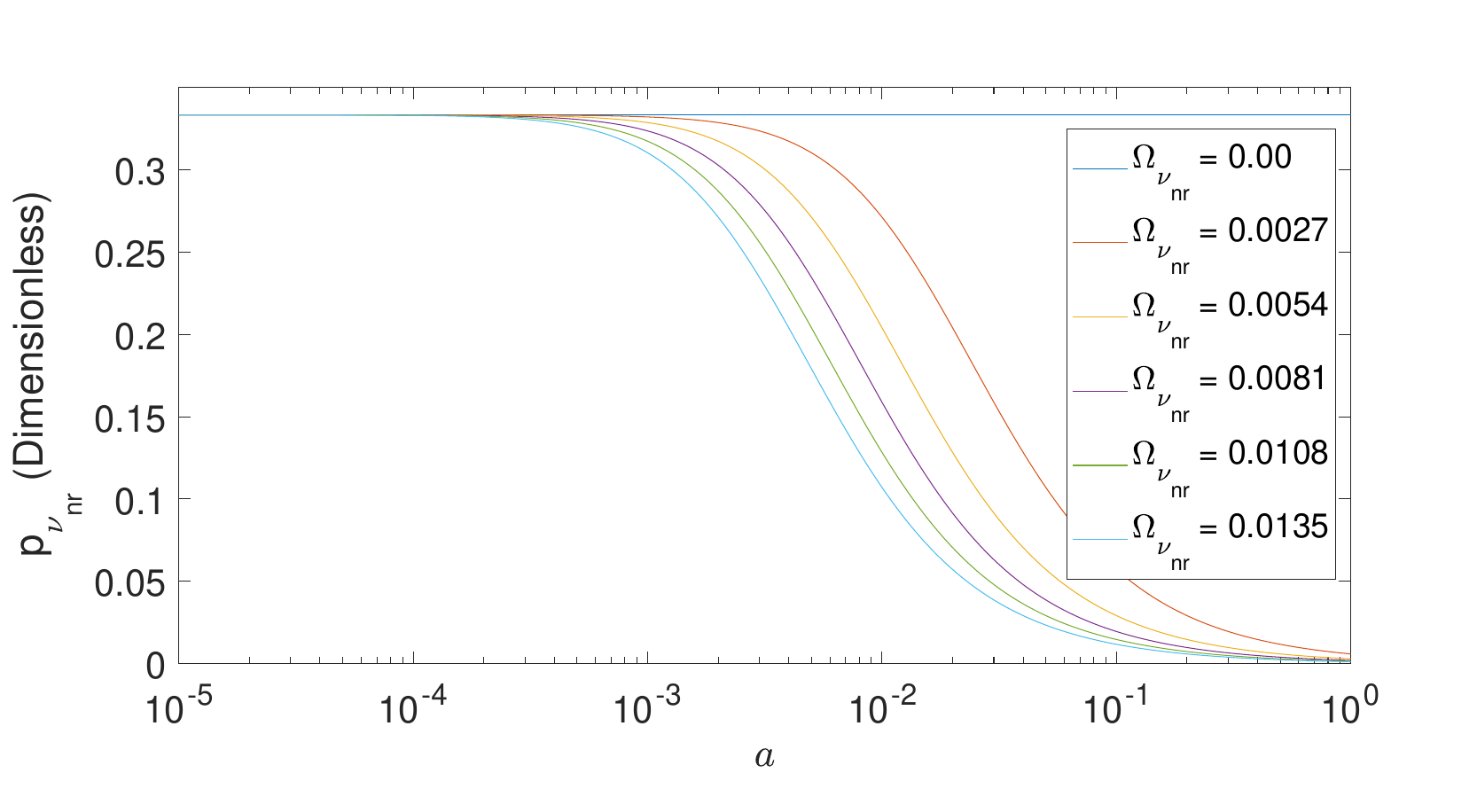}
\caption{Dimensionless neutrino density $\rho_{\mathrm{DL}}$ and $p_{\mathrm{DL}}$, given by Eq.~\ref{rhodl} and Eq.~\ref{pdl}\label{dimensionlessdensity}, for different massive neutrino density parameters. The massless neutrinos are shown in dark blue curve for a reference.}
\end{figure}

\begin{figure}\centering
\includegraphics[width=0.49\textwidth,trim = 0 200 30 210, clip]{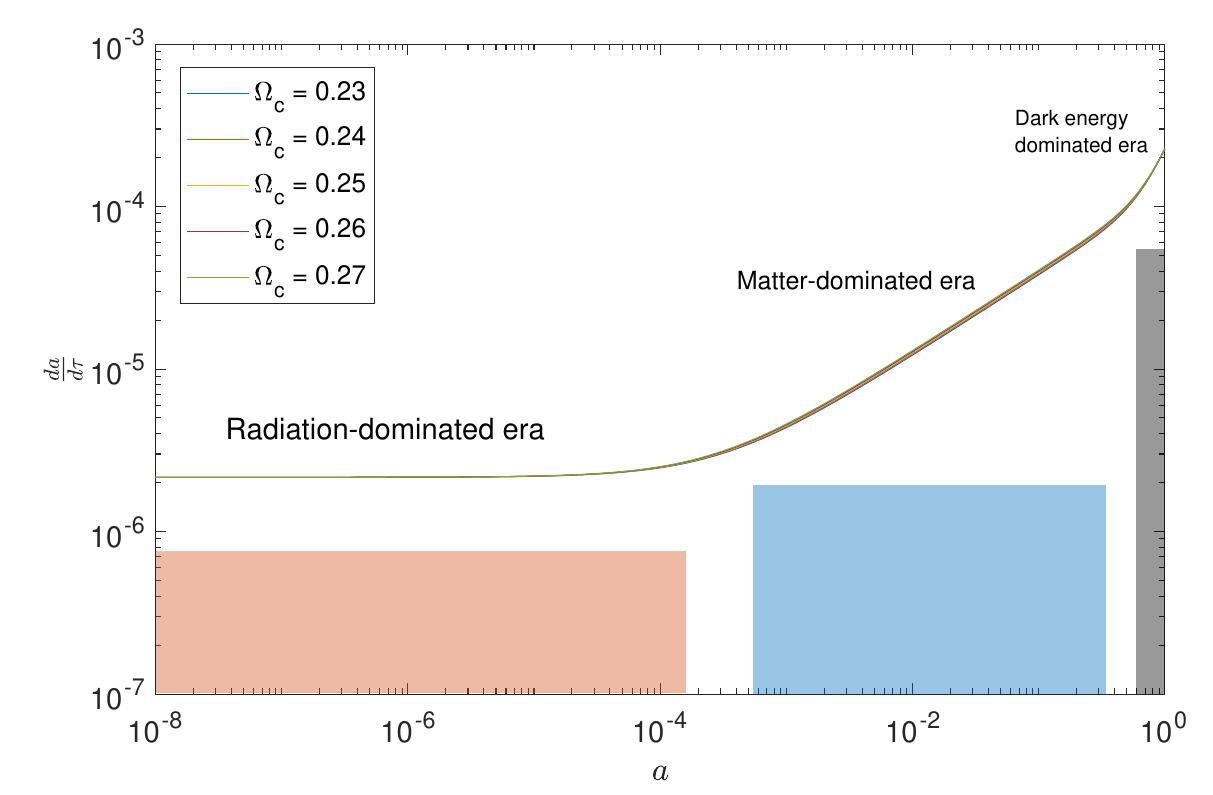}
\includegraphics[width=0.49\textwidth,trim = 0 200 30 210, clip]{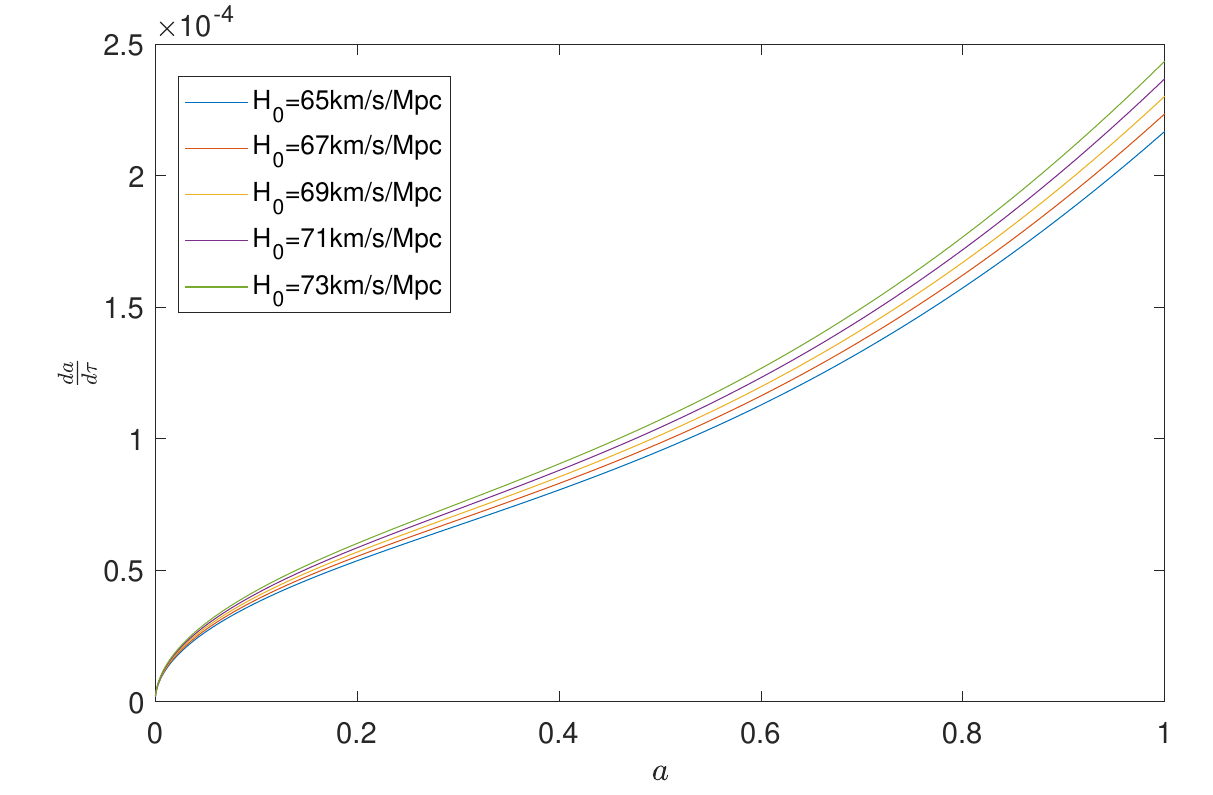}
\caption{Plot of $\frac{\mathrm{d}a}{\mathrm{d}\tau}$ for different $\Omega_c$ (left) and $H_0$ (right). The radiation dominated era, matter dominated era and the dark energy dominated era are clearly shown in the left plot. As shown in Eq.~\ref{eq:dadtau}, $\frac{\mathrm{d}a}{\mathrm{d}\tau}$ is constant in the radiation dominated era (orange), varies as $a^{\frac{1}{2}}$ in the matter dominated era (blue) and varies as $a^2$ in the dark energy dominated era (gray).\label{omegacdadt}}
\end{figure}

\noindent where $\Upsilon = \frac{7}{8}\frac{\pi^4}{15}$ \footnote{Note that for calculating $\Upsilon$ we need the Bose-Einstein integration formula, $\int_{0}^{\infty}\frac{\xi^{3}\mathrm{d}\xi}{e^{\xi}+1}=\frac{\pi^4}{15}$.}. In Fig.~\ref{dimensionlessdensity}, we plot the dimensionless density and pressure for massive neutrinos for different density parameters, $\Omega_{\nu_{m}}$ (note that $\sum m_{\nu_{m}}/93.14 \texttt{eV} = \Omega_{\nu_{m}}h^2$, where $h$ is the the Hubble parameter in units of $100  \texttt{ km/s/Mpc}$)~\citep{mangano2005relic}. In the early universe, where the temperature is high, $pc \ll mc^2$, the neutrinos behave like massless particles and $\rho_\mathrm{dl}\rightarrow 1$ and  $P_\mathrm{dl}\rightarrow \frac{1}{3}$. However, later, where $mc^2$ dominates, the massive neutrinos start behaving like matter particles and $P_\mathrm{dl}\rightarrow 0$ and $\rho_\mathrm{dl}\propto a$, i.e. the actual density of the massive neutrinos goes as $a^{-3}$. 

\begin{figure}
\centering
    \includegraphics[width=0.6\textwidth,trim = 0 80 30 80, clip]{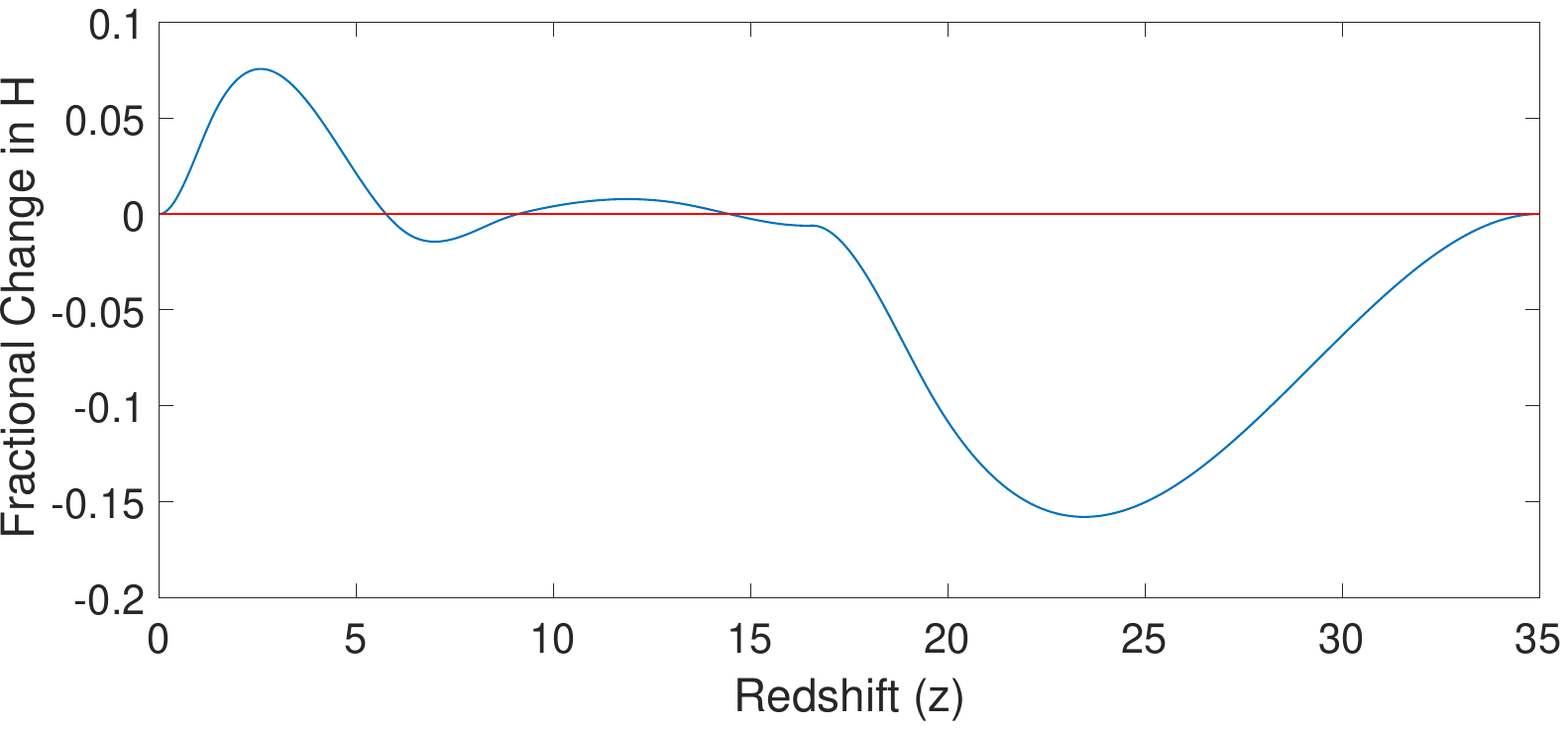} 
    \caption{Fractional change in the Hubble parameter ($f(z) = \Delta H(z) / H_\Lambda (z)$) is shown here. We try to keep the distance to the last scattering surface to be constant. At low redshift we use a Bump kind of features and at high redshift we use a dip feature. }
    \label{fig:HzVariation}
\end{figure}
\begin{figure}
\centering
    \includegraphics[width=0.59\textwidth,trim = 10 80 30 80, clip]{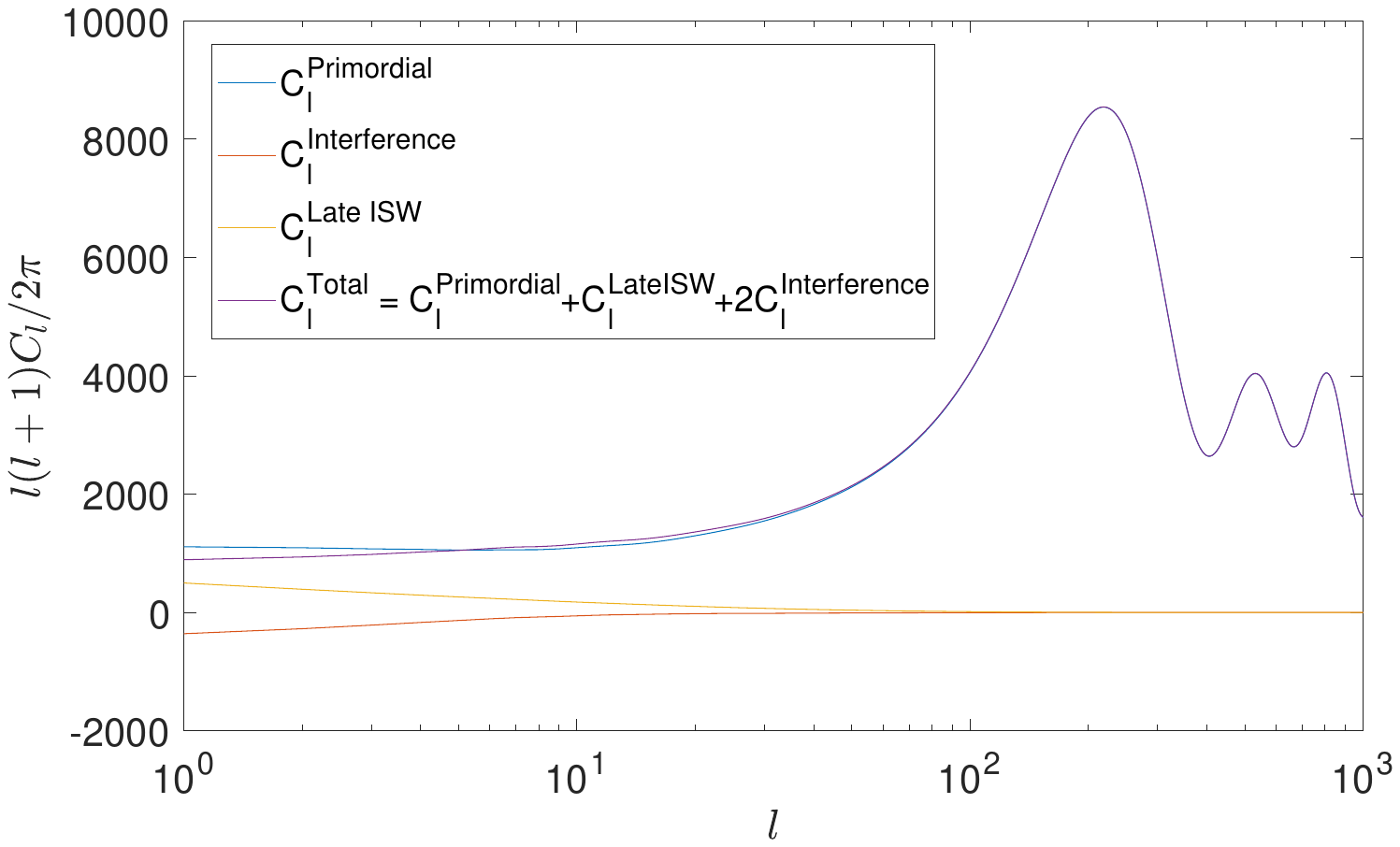}
    \caption{We have shown the Primordial (Sachs Wolf, velocity term and the early ISW term) and the ISW term separately in this plot. The interference term is the cross term between the primordial and the ISW brightness fluctuation functions (check Sec.~\ref{Section-6}). 
    However, this particular $f(z)$ is interesting because this particular shape gives a negative ISW contribution to the CMB temperature power spectrum. }
    \label{fig:Cl_HzVariation}
\end{figure}

\subsection{\label{darkenergy}Contribution from dark energy}

The last term in Eq.~\ref{eq:dadtau} is the contribution from the dark energy. We can use the approximation $\Omega_{0,d}\approx1-\Omega_{0,m}$
(since $\Omega_{0,\gamma}$, $\Omega_{0,\nu}$, $\Omega_{\nu_{m}}$
are of the order of $10^{-5}$). 
For a $\Lambda$CDM model, the equation of state for dark energy is $w_d = -1$. 
However, several dark energy models have been proposed over years based on a single scalar field, a mixture of multiple scalar fields, e.g. quintessence~\citep{Ratra:1987rm}, K-essence~\citep{Armendariz2000,Chiba2000,Armendariz2001}, tachyon~\citep{Padmanabhan2002,Bagla2003}, dilatonic models~\citep{Kamenshchik:2001cp};  massive vector field~\citep{Koivisto:2008xf,Boehmer:2007qa} etc. 
For different dark energy models, the equation of state for dark energy may vary as a function of scale factor, i.e. $w_d(a)$. In such cases we can write the generalized form of $\Omega_d$ as 
\begin{eqnarray}
\Omega_d = \Omega_{0,d}\exp{\Bigg(-3\int_1^a \frac{\mathrm{d}a}{a}\left[1-w(a)\right]\Bigg)}\,.
\end{eqnarray}

\noindent \texttt{CMBAns} is capable of handling both the constant $w_{d}$ or a varying equation of state, $w_{d}(a)$, models of dark energy. Presently there is tension between the Hubble parameter measured using the Planck and using supernova data. Astronomers are trying to model the Hubble parameter as a function of redshift and modify the dark energy accordingly. In \texttt{CMBAns} we add modules which allow users to provide Hubble parameter as a function of redshift using Matlab-GUI input. \texttt{CMBAns} translate the Hubble parameters as equation of state of dark energy~\citep{Das:2013sca} and calculate the CMB power spectrum. \texttt{CMBAns} also provides the primordial, late time ISW and interference term between the primordial and late time ISW along with the full power spectrum to show the effect of late time expansion history of the universe on the CMB power spectrum. 

Fig.~\ref{omegacdadt} shows the variation of $\frac{\mathrm{d}a}{\mathrm{d}\tau}$ as a function of scale factor for different values of $\Omega_c$ and $H_0$  for standard $\Lambda$CDM model. The conformal time between two given redshifts can be calculated by numerically integrating Eq.~\ref{eq:dadtau}. 
For various dark energy models the shape of the CMB power spectra changes. In Fig.~\ref{fig:HzVariation} and Fig.~\ref{fig:Cl_HzVariation} we demonstrate the special feature of \texttt{CMBAns} where we can select the deviation in Hubble parameter from the standard $\Lambda$CDM Hubble parameter ($f(z) = H(z)/ H_\Lambda (z)$) as a function of redshift using a Matlab GUI input. The deviation that we have selected for this illustration is shown in Fig.~\ref{fig:HzVariation}. We try to keep the distance to the last scattering surface to be constant. The temperature power spectrum that we get for this particular deviation deviation of Hubble parameter is shown in Fig.~\ref{fig:Cl_HzVariation}.  An interesting fact for such kind of deviation is that the ISW part provides a very small contribution, sometimes even negative contribution to the $C_l^{TT}$. For this particular illustration, we have not consider any re-ionization. This is because  purpose of the illustration is to show the particular form of the late time ISW effect which is not an well known phenomenon. As the polarization part don't have any ISW contribution, there will be no such effect on the polarization power spectra~\citep{Das:2013sca}. Without a GUI input, exploring such model would have been immensely difficult.


\section{Recombination and Reionization}
For calculating the baryon sound speed, optical depth, and visibility function, we need to calculate the recombination and the reionization process very accurately. \texttt{CMBAns} provides functions for calculating the recombination using the Saha equation, Peebles equation, $\texttt{recfast}$ or $\texttt{CosmoRec}$ method.  

\begin{figure}
\centering
\includegraphics[width=0.70\textwidth,trim = 10 200 30 220, clip]{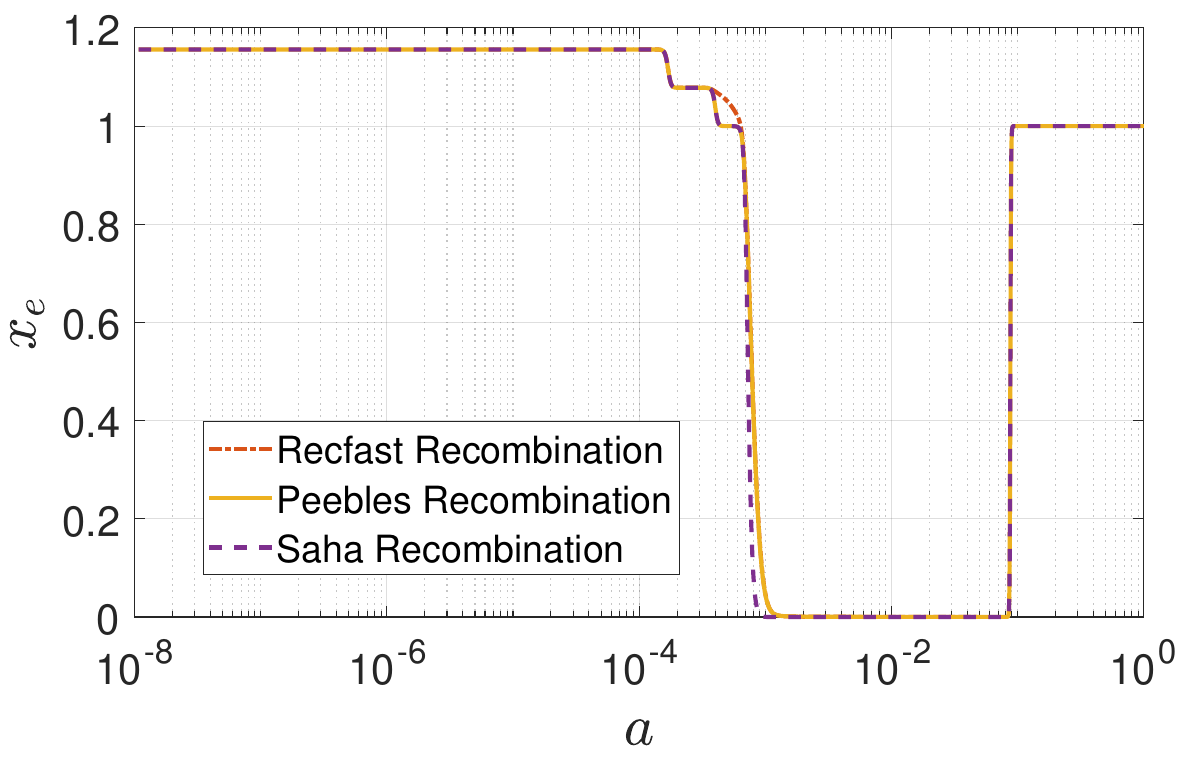}
\caption{\label{recombination} Ionization fractions for Saha, Peebles and $\texttt{recfast}$ recombination processes are shown as a function of the scale factor. The reionization is considered as a step function. The first step in the left is for $\text{He}^{++} \rightarrow \text{He}^+$. The  second step is for $\text{He}^{+} \rightarrow \text{He}$ recombination. In $\texttt{recfast}$ recombination, the second step is absent as it considers an extra fuse function.}
\end{figure}

\subsection{Saha Equation}
Saha equation provides a very rough estimate of the recombination epoch. It assumes the recombination reaction $p+e^{-}\longleftrightarrow H+\gamma$ is fast enough that it proceeds near thermal equilibrium, i.e. it ignores the expansion of the universe. According to the Saha equation,

\begin{equation}
n_{\text{H}}\frac{x_{e}^{2}}{1-x_{e}}=\left(\frac{ k_{B}m_{e}T_b}{2\pi\hbar^{2}}\right)^{3/2}e^{-B_{1}/k_{B}T_b}\,,
\end{equation}

\noindent where $x_e$ is the hydrogen ionization fraction.  $n_{\text{H}}$ is the number density of the hydrogen atoms, i.e. $n_H = n_{1s} + n_p$, where $n_{1s}$ and $n_p$ are number density of neutral hydrogen and ionized hydrogen, respectively. $B_{1}=m_{e}e^{4}/(2\hbar^{2})=13.6\,\texttt{eV}$ is the ionization potential of the hydrogen atom. $T_b$ is the baryon temperature. 

The hydrogen number density can be calculated as 

\begin{equation}
n_{\text{H}}=n_{b}\left(1-Y_{\text{He}}\right)=\frac{\rho_{b}}{m_{\text{H}}}\left(1-Y_{\text{He}}\right)=\frac{3}{8\pi G}\Omega_{0,b}a^{-3}H_{0}^{2}\frac{\left(1-Y_{\text{He}}\right)}{m_{H}}
\end{equation}

\noindent where $Y_{\text{He}}$ is the helium fraction after the Big Bang nucleosynthesis.
In Fig.~\ref{recombination} we show the recombination result using the Saha equation. The plot shows that the recombination of the Hydrogen is almost instantaneous.  For the helium recombination, we separately use the Saha equation given by  Eq.~\ref{helium_recombination}.


\subsection{Peebles' Recombination}
Peebles' equation provides a very accurate estimate of the 
recombination history of hydrogen. The calculations are done using effective three-level atom calculations. Peebles' formalism is based on the assumptions that 
\begin{itemize}
    \item Direct recombinations to the ground state of hydrogen are very inefficient: each such event leads to a photon with energy greater than $13.6\, \texttt{eV}$, which almost immediately re-ionizes a neighboring hydrogen atom. Electrons therefore only efficiently recombine to the excited states of hydrogen, from which they cascade very quickly down to the first excited state, with principal quantum number $n = 2$.
    \item From the first excited state, electrons can reach the ground state $n = 1$ through two pathways:
\begin{enumerate}
    \item  Decay from the 2p state by emitting a Lyman-$\alpha$ photon. This photon will almost always be reabsorbed by another hydrogen atom in its ground state. However, cosmological redshifting systematically decreases the photon frequency, and hence there is a small chance that it escapes reabsorption if it gets redshifted far enough from the Lyman-$\alpha$ line resonant frequency before encountering another hydrogen atom.
    \item Decay from the $2s$ to $1s$ state, which is only possible using an electron double transition. The rate of this transition is very slow, $8.22\,\texttt{s}^{-1}$. It is however competitive with the slow rate of Lyman-$\alpha$ escape in producing ground-state hydrogen.
    \end{enumerate}
    \item Atoms in the first excited state may also be re-ionized by the ambient CMB photons before they reach the ground state, as if the recombination to the excited state did not happen in the first place. To account for this possibility, Peebles defines the factor $C$ as the probability that an atom in the first excited state reaches the ground state through either of the two pathways described above before being photo-ionized.
\end{itemize}

Accounting for these processes, the recombination history is then described by the differential effect \citep{peebles1968recombination}

\begin{equation}
\frac {\mathrm{d}x_{\text{e}}}{\mathrm{d}t}=-aC\left(\alpha^{(2)} (T_{b})n_{\text{p}}x_{e}-4(1-x_{\text{e}})\beta(T_{b})e^{-E_{21}/T}\right)
\end{equation}

\noindent where

\begin{equation}
\beta(T_{b})=\left(\frac{m_{e}k_{{\rm B}}T_{b}}{2\pi\hbar^{2}}\right)^{3/2}e^{-B_{1}/k_{{\rm B}}T_{b}}\,\alpha^{(2)}(T_{b})\label{eq:beta}
\end{equation}

\noindent The recombination rate to excited states \citep{Ma1995} is taken as

\begin{equation}
\alpha^{(2)}(T_{b})=\frac{64\pi}{(27\pi)^{1/2}}\frac{e^{4}}{m_{e}^{2}c^{3}}\left(\frac{k_{{\rm B}}T_{b}}{B_{1}}\right)^{-1/2}\,\phi_{2}(T_{b})\ ,\quad\quad\phi_{2}(T_{b})\approx0.448\,\ln\left(\frac{B_{1}}{k_{{\rm B}}T_{b}}\right)\ .\label{eq:recombrate}
\end{equation}

\noindent This expression for $\phi_{2}(T_{b})$ provides a good approximation at low temperature. 
At high temperature this expression
underestimates $\phi_{2}$, but the amount is negligible. For $T_{b}>B_{1}/k_{{\rm B}}=1.58\times10^{5}\,\texttt{K}$,
we set $\phi_{2}=0$.
\begin{equation}
C = \frac{\Lambda_\alpha + \Lambda_{2s\rightarrow1s}}{\Lambda_\alpha + \Lambda_{2s\rightarrow1s} +\beta^{(2)}(T_b)} 
\end{equation}

\noindent where 

\begin{equation}
\beta^{(2)}(T_b) = \beta(T_b)e^{+hc/\lambda_\alpha k_BT_b}\;,   \;\;\;\;\;\;\;\;\;\;\;\;\;\; \Lambda_\alpha = \frac{8\pi\dot{a}}{a^2\lambda^3_\alpha n_{1s}}\,.
\end{equation}

\noindent $\lambda_\alpha = \frac{8\pi\hbar c}{3B_1} =  1.21567 \times 10^{-7} \texttt{m}$, is the wavelength for Lyman-$\alpha$    
emission.  Over-dot represents the derivative with respect to the conformal time. $\Lambda_{2s\rightarrow 1s}$ is the rate of hydrogen double transition from $2s$ to $1s$. $\Lambda_{2s\rightarrow 1s} = 8.227 \texttt{s}^{-1} = 8.4678 \times 10^{14} \texttt{Mpc}^{-1}$.

\begin{eqnarray}
\Lambda_{2s\rightarrow1s}/\Lambda_{\alpha} &=&\frac{\Lambda_{2s\rightarrow1s}\lambda_{\alpha}^{3}a^{2}n_{1s}}{8\pi\dot{a}}=\frac{\Lambda_{2s\rightarrow1s}\lambda_{\alpha}^{3}(1-x_{e})a^{3}n_{H}}{8\pi\dot{a}a} 
=(1-x_e)(1-Y_{He})\frac{\Lambda_{2s\rightarrow 1s}\lambda_\alpha^{3}}{8\pi\dot{a}a}
\frac{a^{3}\rho_{m}}{m_{H}} \nonumber \\
&=&\Lambda_{2s\rightarrow1s}\left(\frac{\lambda_{\alpha}^{3}}{8\pi}\frac{3}{8\pi G}\frac{1}{m_{H}}\right)\frac{(1-x_{e})}{\dot{a}a}(1-Y_{He})\Omega_{m0}H_{0}^{2} \nonumber \\
&=&\left(8.4678 \times 10^{14}\right) \times \left(8.0230194 \times 10^{-26}\right)\frac{(1-x_{e})}{\dot{a}a}(1-Y_{He})\Omega_{m0}H_{0}^{2}
\end{eqnarray}
\noindent Similarly, $\beta^{(2)}(T_b)/\Lambda_{\alpha}$  can be calculated using 

\begin{eqnarray}
\frac{\beta^{(2)}(T_b)}{\Lambda_{\alpha}} &=& T_b\phi_2(T_b) \mathcal{K} e^{ - 0.25T_{ion}/T_b}\left( 8.0230194 \times 10^{-26} \right)\frac{(1-x_{e})}{\dot{a}a}(1-Y_{He})\Omega_{m0}H_{0}^{2} \nonumber
\end{eqnarray}

\noindent where  $\mathcal{K} = \left(\frac{64\pi}{\left(27\pi\right)^{1/2}}\frac{e^{4}}{m_{e}^{2}c^{3}}\left(\frac{k_{{\rm B}}}{B_{1}}\right)^{-1/2}\left(\frac{m_{e}k_{{\rm B}}}{2\pi\hbar^{2}}\right)^{3/2}\right) = 5.13 \times 10^{18}$.
Here, $H_0$ is in $\texttt{km}/\texttt{sec}/\texttt{MPc}$ unit, and $\dot{a}$ has unites of $\texttt{MPc}^{-1}$ unit. The numerical values are converted to match these units.





\subsubsection{Helium Recombination}
For calculating the He recombination, we use the Saha Equation \citep{Ma1995}. 
\begin{equation}
\frac{n_{e}x_{n+1}}{x_{n}}=\frac{2g_{n+1}}{g_{n}}\left(\frac{m_{e}k_{B}T_{b}}{2\pi\hbar^{2}}\right)^{3/2}e^{-\chi_{n}/k_{B}T_{b}}\,,
\label{helium_recombination}
\end{equation}

\noindent where $n\in (0, 1)$, and $x_0 = 1 - x_1 - x_2$. The helium ionization fractions $x_{1}=n\left(\text{He}^{+}\right)/n\left(\text{He}\right)$
and $x_{2}=n\left(\text{He}^{++}\right)/n\left(\text{He}\right)$,
where $n\left(\text{He}\right)$ is the total number density of helium
nuclei. $n_e$ is the free electron number density. $g_0 = g_1 = 1$ and $g_2 = 2$. $\frac{\chi_1}{k_B} = T^{ion}_{1}= 2.855 \times 10^5\,\texttt{K}$ and $\frac{\chi_2}{k_B} = T^{ion}_{2} = 6.313 \times 10^5\,\texttt{K}$ are the first and second ionization temperature of He. 
\begin{figure}
\centering
\includegraphics[width=0.48\textwidth,trim = 0 190 10 200, clip]{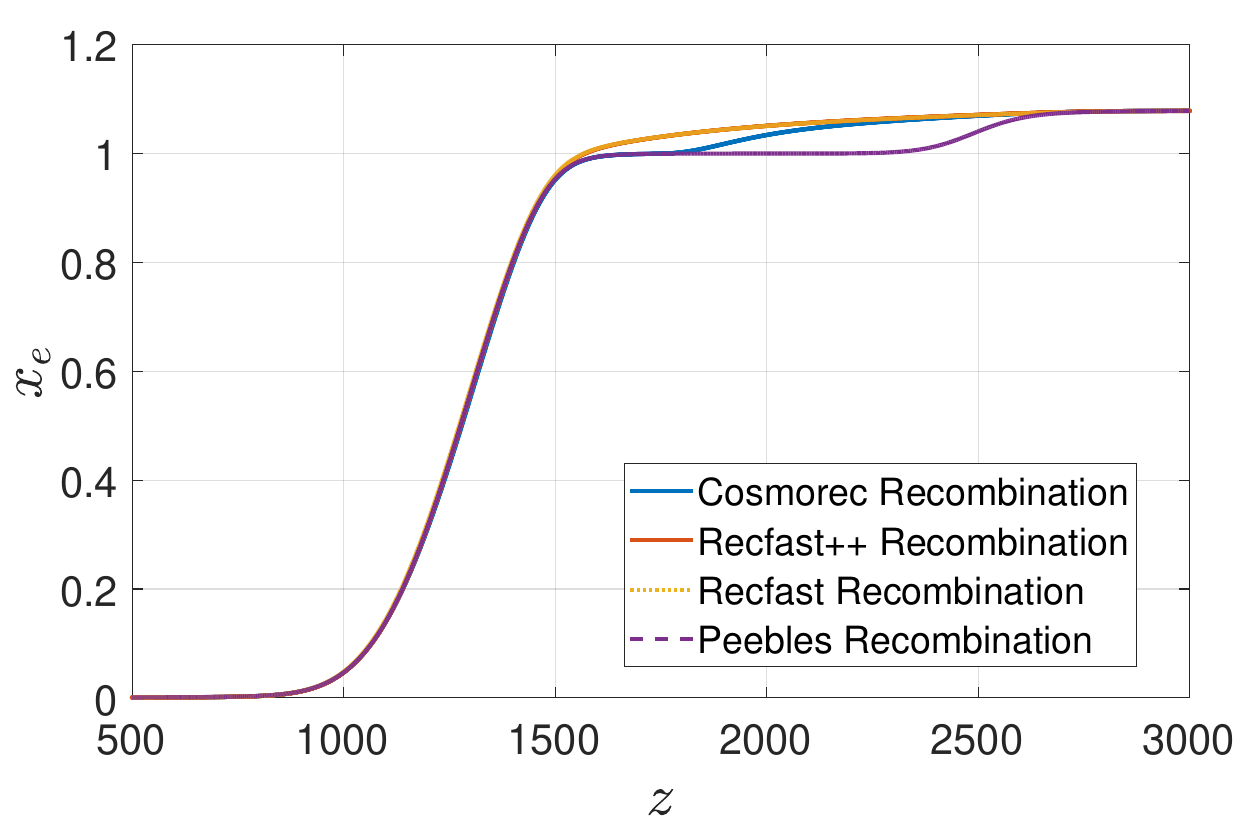}
\includegraphics[width=0.48\textwidth,trim = 0 190 10 200, clip]{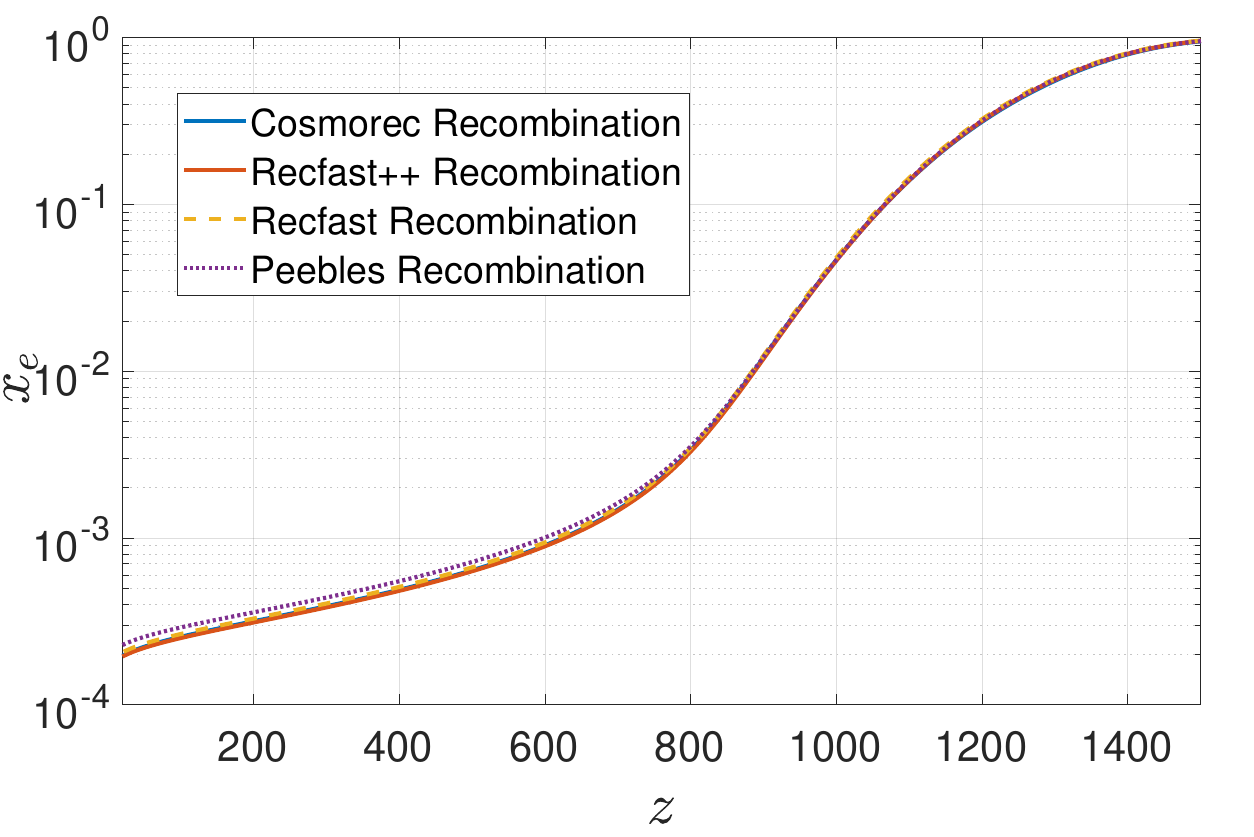}
\caption{\label{recombinationCosmorec} Comparison between the ionization fractions from different modern recombination routine $\texttt{recfast}$, $\texttt{recfast++}$ and $\texttt{CosmoRec}$. For $\texttt{CosmoRec}$, we choose the  dark matter annihilation efficiency to be $10^{-24}\,\texttt{eV/sec}$ and all the other parameters are set to default settings. Left: Ionization fraction is plotted with a linear scale to show the $\text{He}^{+}$ recombination. Right: Ionization fraction is plotted with a log scale to amplify the effect at low redshift after the $\text{H}^+$ recombination.}
\end{figure}
\subsection{Recfast, CosmoRec}
Peebles' three-level atom model accounts for the most important physical processes. However, these approximations may lead to errors on the predicted recombination history at a level as high as $10\%$. This can also alter the temperature and polarization power spectra up to $3-5\%$ at high multipoles. Several research groups have revisited the details and proposed different models like $\texttt{recfast}$\footnote{\url{https://www.cfa.harvard.edu/~sasselov/rec/}}\citep{Seager1999, Seager1999a}, $\texttt{CosmoRec}$\footnote{\url{http://www.jb.man.ac.uk/~jchluba/Science/CosmoRec/Welcome.html}}\citep{Chluba2010,Ali-Haimoud2010,Chluba2010a,Switzer2008,Grin2010,Martin2010}, \texttt{HyRec}\footnote{\url{https://cosmo.nyu.edu/yacine/hyrec/hyrec.html}}\citep{Ali-Haimoud2010a} etc. These packages can calculate the recombination history up to $0.1\%$ accuracy. We use the available $\texttt{CosmoRec}$ code in \texttt{CMBAns} as a default case. However, users can choose to use Saha, Peebles or $\texttt{recfast}$  routine which are also available in $\texttt{CMBAns}$. The other packages can also be easily added in the \texttt{CMBAns} or run separately. In the later case  the ionization fraction, and baryon temperature can be stored in a file as a function of scale factor and pass it to \texttt{CMBAns}. 

In Fig.~\ref{recombination}, we show the ionization fraction from different recombination methods. We use a smooth reionization, where we join an ionization fraction before and after the reionization using a $\tanh(...)$ function.  In Fig.~\ref{recombinationCosmorec}, we show the differences between $\texttt{recfast}$, $\texttt{recfast++}$ and $\texttt{CosmoRec}$ recombination. This small change in the ionization fraction can change the $C_l$ at high multipoles.

\subsection{Calculating baryon temperature}
For calculating ionization fraction during the recombination, we need the baryon temperature at each scale factor. The rate of change of the baryon temperature can be calculated as (check Appendix~\ref{baryontempappendix})

\begin{equation}
\dot{T}_{b}=-2\left(\frac{\dot{a}}{a}\right)T_{b}+\frac{8\pi^{2}}{45}\frac{k_B^4}{c^4 \hbar^3}\frac{\sigma_{T}T_{\gamma}^{4}}{ m_{e}}f_e\left(T_{\gamma}-T_{b}\right)\,,
\label{fulltemperature}
\end{equation}

\noindent where $\sigma_{T}$ is the Thomson scattering cross section. $f_e$ is given by 

\begin{equation}
    f_e = \frac{\left( 1-Y_{He}\right)x_e^{tot}}{1 -\frac{3}{4} Y_{He} + (1-Y_{He})x_e^{tot}}\,.
\end{equation}

\noindent $x_e^{tot}$ is the total ionization fraction and is given by 
\begin{equation}
    x_e^{tot} = x_e+\frac{1}{4}Y_{He}\frac{(x_1+2x_2)}{(1.0-Y_{He})}\;.
\end{equation}

The constant term in Eq.~\ref{fulltemperature} is given by  $\frac{8\pi^{2}}{45}\frac{k_B^4}{c^4 \hbar^3}\frac{\sigma_{T}}{ m_{e}} = 5.0515\times10^{-8}\texttt{K}^{-4}\texttt{Mpc}^{-1}$. We can see that the baryon temperature depends on the ionization fraction of the electrons. Therefore, we need to jointly evaluate the baryon temperature and ionization fraction. The temperature of the photons at any era is $T_\gamma = a^{-1}T_{0 \gamma}$. We can consider  $T_b = T_\gamma$ before recombination (in the tight coupling era), and we can use it as the initial condition for solving Eq.~\ref{fulltemperature}. 

\subsection{Baryon sound speed, optical depth and visibility}
Calculating the baryon acoustic oscillations require the speed of sound in the plasma, $c_{s}$. If we consider the plasma as a single fluid, then the pressure, density and the temperature of the fluid will be related as $P_b=\frac{k_B}{m}\rho_b T_b$. We can calculate the sound speed in the plasma as 

\begin{eqnarray}
c_{s}^{2}=\frac{\mathrm{d}P_{b}}{\mathrm{d}\rho_{b}} \Bigg|_{adiabatic}&=&\frac{k_{B}T_{b}}{m}\left(1-\frac{1}{3}\frac{\mathrm{d}(\ln T_{b})}{\mathrm{d}(\ln a)}\right) \nonumber\\
&=& \frac{k_BT_{b}}{m_p}\Bigg[1.0-\frac{3}{4}Y_{He}+(1.0-Y_{He})x_e^{tot} \Bigg]\left(1-\frac{1}{3}\frac{\mathrm{d}(\ln T_{b})}{\mathrm{d}(\ln a)}\right)\;.\label{baryonsoundspeed}
\end{eqnarray}

\noindent Here $m$ is the mean molecular weight of the fluid, and $m_p$ is the mass of a proton \footnote{For all our calculations, we consider the mass of H and H$^+=m_p$ , and mass of He, He$^+$, He$^{++}=4m_p$, i.e. we consider that the mass of electron is negligible and the mass of proton and neutron are the same.}. The mean molecular weight is calculated assuming the fluid contains free electrons, H, H$^+$, He, He$^+$, He$^{++}$. Here one should note that a more accurate formulation of the sound speed was proposed by~\cite{lewis2007linear}, and are used in $\texttt{CAMB}$ and $\texttt{CLASS}$. We are in process of implementing it in $\texttt{CMBAns}$. 

The optical depth from the present time ($\tau_0$) to any conformal time $\tau$ is given by\
\begin{eqnarray}
\kappa = \int_\tau^{\tau_0}an_e\sigma_T \mathrm{d}\tau = \int_\tau^{\tau_0}\Bigg(\frac{H_0^2c^2}{8\pi G }\Bigg)\Bigg(\frac{\Omega_b}{m_Ha^2}\Bigg)\sigma_T (1-Y_{He}) \mathrm{d}\tau\;.
\end{eqnarray}

\begin{figure}
\centering
\includegraphics[width=0.70\textwidth,trim = 340 340 300 350, clip]{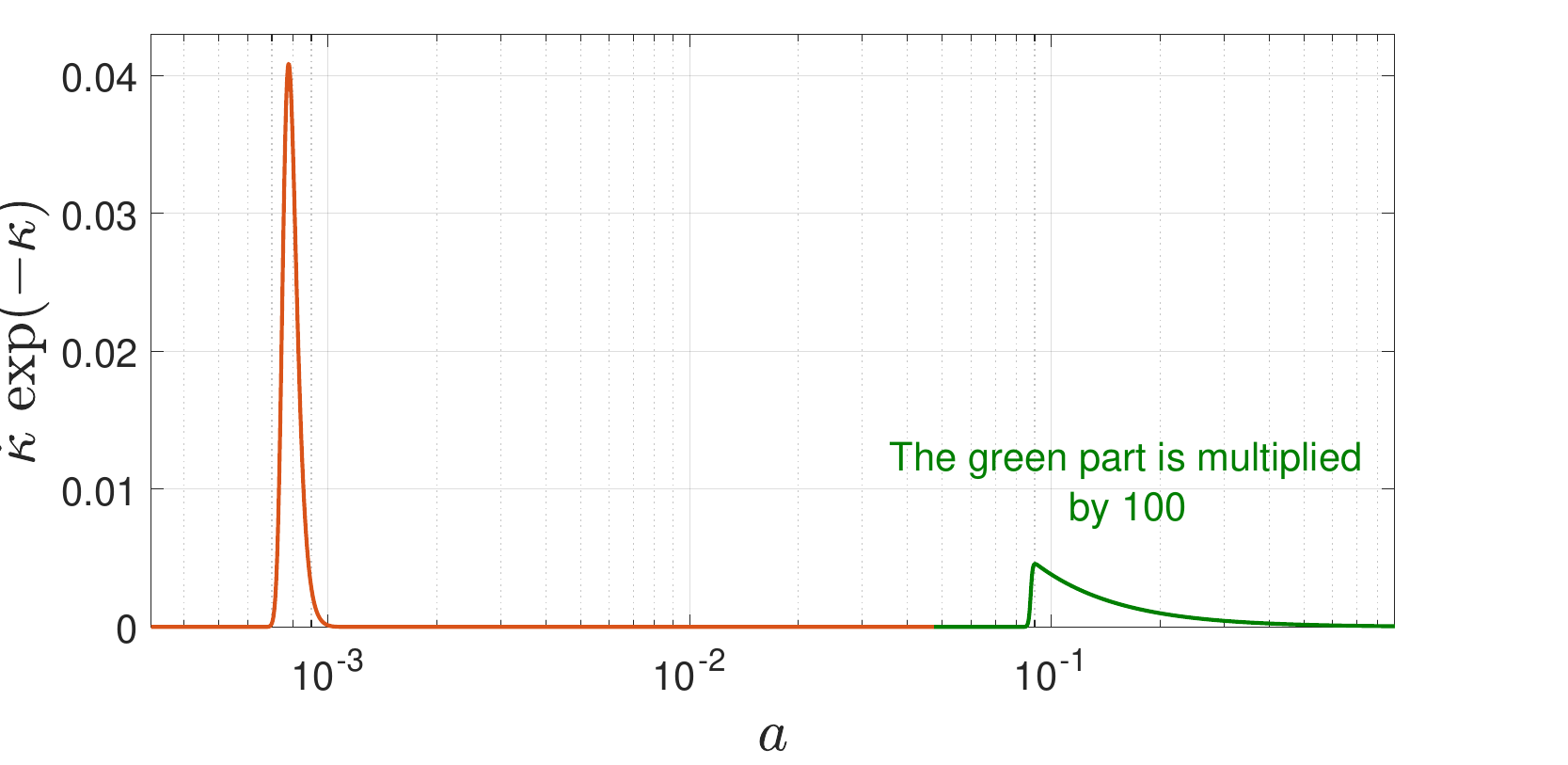}
\caption{\label{visibility} Visibility function ($g=\dot{\kappa}\exp(-\kappa)$) as a function of red-shift. The green section of the plot is multiplied by $100$ for displaying it on the same plot.}
\end{figure}

The visibility function at any conformal time $\tau$ can be calculated as $g = \dot{\kappa}\exp(-\kappa)$. In Fig.~\ref{visibility} we show the visibility function vs the scale factor. The visibility function is nonzero only during the recombination and reionization process. 
The change in the visibility function is significantly smaller during reionization, than recombination. To show both on the same plot, we multiply the reionization part by $100$. 

\section{A brief overview of the cosmological perturbations}

\texttt{CMBAns} is developed only for the flat background metric, i.e. $\Omega_{k}=0$. We can represent a completely isotropic and homogeneous expanding universe using FLRW metric, $\mathrm{d}s^{2}=a^{2}g_{\mu\nu}\mathrm{d}x^\mu \mathrm{d}x^\nu$, where $g_{\mu\nu} = a(\tau)^2\eta_{\mu\nu}$, and $\eta_{\mu\nu}$ is the Minkowski metric. Since our universe is not completely homogeneous and isotropic, we need to add some 
perturbation to the metric. The line element in the perturbed metric is given by
\begin{equation}
    \mathrm{d}s^2 = a^2(\tau)\left(\eta_{\mu\nu} + h_{\mu\nu}\right)\mathrm{d}x^\mu \mathrm{d}x^\nu\,.
\end{equation}
Here $h_{\mu\nu}$ is the perturbation in the metric. For simplicity of calculation, the metric perturbation is expanded in the spatial and temporal parts. This can be done by breaking $h_{\mu\nu}$ in $(1 + 3)$ dimensional format as~\citep{Weinberg2008cosmology}
\begin{eqnarray}
h_{00} &=& E \nonumber\\
h_{i0} &=& \frac{\partial F}{\partial x^i} + G_i \nonumber\\
h_{ij} &=& A\delta_{ij} + \frac{\partial^2 B}{\partial x^i\partial x^j} + \frac{\partial C_i}{\partial x^j}  + \frac{\partial C_j}{\partial x^{i}} + D_{ij} \label{metricmerturbation}
\end{eqnarray}
\noindent where $A$, $B$, $E$, $F$, $C_i$, $G_i$ are $D_{ij}$ are the perturbation variables and 
\begin{equation}
    \frac{\partial C_i}{\partial x^i} = \frac{\partial G_i}{\partial x^i} = 0, \;\;\;\;\;\;\;\;  \frac{\partial D_{ij}}{\partial x^i} = 0, \;\;\;\;\;\;\;\; D_{ii} = 0\,.
\end{equation}


From Einstein's equation, we get $G_{\mu\nu} = \frac{8\pi G}{c^2}T_{\mu\nu}$, where $G_{\mu\nu}$ is the Einstein tensor, and $T_{\mu\nu}$ is the stress-energy tensor. Perturbing the equations up to the first order, we get
$G_{\mu\nu} + \delta G_{\mu\nu} = \frac{8\pi G}{c^2}\left(T_{\mu\nu} + \delta T_{\mu\nu}\right)$.
We can use the above perturbation variables to calculate the perturbations in the Christoffel symbols. The perturbation to the Einstein's tensor, $\delta G_{\mu\nu}$ can then be computed from the Christoffel symbols.
For calculating the perturbation in $\delta T_{\mu\nu}$, we need to know the perturbation in the pressure and density of the different components in the Universe, i.e. baryons, photons, neutrinos, dark matter, and dark energy, etc. For a perfect fluid, the stress-energy tensor is given by 

\begin{equation}
    T_{\mu\nu} = pg_{\mu\nu} + (p + \rho)u^\mu u^\nu\,.
\end{equation}

Similar to the metric tensor, the perturbation in the stress-energy tensor can also be expanded into spatial and temporal parts
\begin{eqnarray}
\delta T_{00} &=& -\rho + h_{00}\delta \rho   \nonumber \\
\delta T_{0i} &=& ph_{i0} -\left(\rho + p\right)\left(\frac{\partial\delta u}{\partial x^i} + \delta u_i^V\right) \nonumber \\
\delta T_{ij} &=& ph_{ij} + \left[\delta_{ij}\delta p + \frac{\partial^2 \pi^S}{\partial x^i \partial x^j}  + \frac{\partial \pi_i^V}{\partial x^j}  + \frac{\partial \pi_j^V}{\partial x^i} + \pi^T_{ij}\right] \label{stressperturbation}
\end{eqnarray}
in which
\begin{equation}
    \frac{\partial \pi_i^V}{\partial x^i} = \frac{\partial \delta u_i^V}{\partial x^i} = 0, \;\;\;\;\;\;\;\;  \frac{\partial \pi_{ij}^T}{\partial x^i} = 0, \;\;\;\;\;\;\;\; \pi_{ii}^T = 0\,.
\end{equation}
   
\noindent We can match both the sides in  $\delta G_{\mu\nu} = \frac{8\pi G}{c^2}\left(\delta T_{\mu\nu}\right)$ and separate out:
\begin{itemize}
    \item The terms containing $A$, $B$, $E$, $F$, $\delta \rho$, $\delta p$, $\pi^S$ and $\delta u$. These involve all the scalar quantities and are called the scalar perturbations. 
    \item The terms containing $C_i$, $G_i$, $\pi^V_i$ and $\delta u^V_i$. These involve all the vector quantities in the spatial dimension and are called the vector perturbations. These vector modes decay and hence have a small contribution to the CMB power spectrum~\citep{Weinberg2008cosmology}. 
    \item The terms involving $D_{ij}$ and $\pi^T_{ij}$. These terms behave as tensor quantities in 3-spatial dimensions and are called the tensor perturbations. 
\end{itemize}

\subsection{\label{PerturbationEquationfluid}Theory of scalar perturbations}
We can obtain the scalar perturbation equations by separating out the terms involving the scalar perturbation variables, i.e. $A$, $B$, $E$, $F$, $\delta \rho$, $\delta p$, $\pi^S$ and $\delta u$. However,
 these terms are not all independent ~\citep{Weinberg2008cosmology}. Also, there can be unphysical modes due to the choice of the coordinate system. 
These problems can be resolved by fixing a proper coordinate system, and adopting suitable conditions on the full perturbed metric and energy-momentum tensor. This process is called  gauge fixing~\citep{Weinberg2008cosmology, Ma1995}. In \texttt{CMBAns}, we do all the calculations involving the scalar perturbation in synchronous gauge~\citep{Lifshitz1946}.

In synchronous gauge, the scalar component of the perturbed metric can be written as  $\mathrm{d}s^{2}=a^{2}(\tau)\{-\mathrm{d}\tau^{2}+(\delta_{ij}+h_{ij})\mathrm{d}x^{i}\mathrm{d}x^{j}\}$,
where $\delta_{ij}$ is the Kronecker delta, and $h_{ij}$ is the scalar part of the metric perturbation in synchronous gauge. $h_{ij}$ can be represented using only two scalar
fields $h(\vec{k},\tau)$ and $\eta(\vec{k},\tau)$, in ($\vec{k}$,$\tau$)
space, where 

\begin{equation}
h_{ij}(\vec{x},\tau)=\int \mathrm{d}^{3}ke^{i\vec{k}\cdot\vec{x}}\left\{ \hat{k}_{i}\hat{k}_{j}h(\vec{k},\tau)+(\hat{k}_{i}\hat{k}_{j}-\frac{1}{3}\delta_{ij})\,6\eta(\vec{k},\tau)\right\}\,.
\end{equation}

\noindent $\vec{k}=k\hat{k}$, $\hat{k}$ is the unit vector along direction
of vector $\vec{k}$ and $k$ is its amplitude. We can relate $h(\vec{k},\tau)$ and $\eta(\vec{k},\tau)$ with the real space perturbation variables $A$, $B$, $E$, $F$. This gives $E=0$, $F=0$, because in synchronous gauge we take constant time hyperspace, $A(\vec{x}, \tau) =  \int \left(2\eta(\vec{k},\tau)\right) \mathrm{d}^{3}ke^{i\vec{k}\cdot\vec{x}} $ and $B(\vec{x}, \tau) =  \int \left(h(\vec{k},\tau) + 6\eta(\vec{k},\tau)\right) \mathrm{d}^{3}ke^{i\vec{k}\cdot\vec{x}} $. The growth of these
perturbation variables can be computed using the perturbed Einstein equations $\delta G_{\mu\nu}=\frac{8\pi G}{c^2}\delta T_{\mu\nu}$.
The perturbation in $G_{\mu\nu}$ can be calculated using the metric
perturbation variables~\citep{Ma1995} as
\begin{eqnarray}
    k^2\eta - {1\over 2}{\dot{a}\over a} \dot{h}
        &=& \frac{4\pi G}{c^2} a^2 \delta T^0{}_{\!0}
    \,,\label{ein-syna}\\
    k^2 \dot{\eta} &=& \frac{4\pi G}{c^2}a^2 (\bar{\rho}+\bar{P})
    \theta     \,,\label{ein-synb}\\
    \ddot{h} + 2{\dot{a}\over a} \dot{h} - 2k^2
    \eta &=& -8\frac{\pi G}{c^2} a^2 \delta T^i{}_{\!i}
    \,,\label{ein-sync}\\
    \ddot{h}+6\ddot{\eta} + 2{\dot{a}\over a}\left(\dot{h}+6\dot{\eta}
    \right) - 2k^2\eta &=& -\frac{24\pi G}{c^2} a^2 (\bar{\rho}+\bar{P})
    \sigma    \,.\label{ein-synd}
\end{eqnarray}

\noindent Here $\theta$ is defined as $\theta(\vec{k},\tau)  =  \frac{ik^j\delta T^0_j}{\bar{\rho} + \bar{P}}$.  For a fluid, this is simply the divergence of its velocity field, i.e. $\theta=ik^j v_j$. 
The $\sigma$ in Eq.~\ref{ein-synd} can be defined as $\sigma(\vec{k},\tau) = \frac{2\Pi \bar{P}}{3(\bar{\rho}+\bar{P})}$, where $\Pi$ is the anisotropic stress term, corresponding to its real space quantity $\pi^S$ shown in Eq.~\ref{stressperturbation}.

We also need to conserve the perturbation in the stress-energy tensor, i.e. $\delta T^{\mu\nu}_{;\mu} = 0$. 
For non-relativistic perfect fluid, these conservation equations  give us

\begin{eqnarray}
\label{fluid}
    \dot{\delta} &=& - (1+w) \left(\theta+{\dot{h}\over 2}\right)
      - 3{\dot{a}\over a} \left({\delta P \over \delta\rho} - w
      \right)\delta  \,,\nonumber\\
    \dot{\theta} &=& - {\dot{a}\over a} (1-3w)\theta - {\dot{w}\over
         1+w}\theta + {\delta P/\delta\rho \over 1+w}\,k^2\delta
         - k^2 \sigma\,, \label{matter_prtb}
\end{eqnarray}

\noindent where $\delta = \frac{\delta \rho}{\rho}$ is the perturbation in the density of the fluid. $w$ is the equation of state of the fluid.  Non-relativistic fluids like the CDM and baryon should obey these equations. 

To calculate the stress-energy tensor for the relativistic particles, we need to use the distribution function. Photons follow the Bose-Einstein distribution, whereas the neutrinos follow the Fermi-Dirac distribution. Considering $f$ as the phase space distribution function, the stress-energy tensor is given by 
\begin{equation}
    T_{\mu\nu} = \int \mathrm{d}P_1\mathrm{d}P_2\mathrm{d}P_3(-g)^{-\frac{1}{2}}\frac{P_\mu P_\nu}{P^0} f(x^i,P_j,\tau)\,.
\end{equation}

\noindent $P$ is the four-momentum. $P_0=-\epsilon = a(p^2+m^2)^{\frac{1}{2}} = (q^2+a^2m^2)^{\frac{1}{2}}$. $P_j = aq_j = qn_j$. $q$, $n_i$ are the magnitude and the direction of the momentum and $n^in_i = 1$. 
It is convenient to write the distribution function as 
\begin{equation}
    f(x^i,P_j,\tau) = f_0(q)\Bigg[1+\Psi(x^i,q,n_j,\tau) \Bigg]\;.
\end{equation}

\noindent Here $\Psi(x^i,q,n_j,\tau)$ in the first order perturbation in the distribution function.  In synchronous gauge $(-g)^\frac{1}{2}=a^{-4}(1-\frac{1}{2}h)$, and $\mathrm{d}P_1\mathrm{d}P_2\mathrm{d}P_3 = (1+\frac{1}{2}h)q^2\mathrm{d}q\mathrm{d}\Omega$. 
The phase-space distribution evolves as the Boltzmann equation. If we convert everything to  Fourier space ($\vec{k}$ space), then the first-order perturbations of the Boltzmann equation in the synchronous gauge can be written as

\begin{equation}
    \frac{\partial \Psi}{\partial\tau} + \frac{q}{\epsilon}(\vec{k}.\hat{n})\Psi + \frac{\mathrm{d}(\ln f_0)}{\mathrm{d}(\ln q)}\Bigg[\dot{\eta}-\frac{\dot{h}+6\dot{\eta}}{2}(\vec{k}.\hat{n})^2\Bigg] = \frac{1}{f_0}\Bigg(\frac{\partial f}{\partial \tau}\Bigg)_C  \label{radiationPerturbation}
\end{equation}

\noindent Photons and neutrinos will follow this equation. The term on the right-hand side is the collision term. Neutrinos are collisionless in our domain as they decouple long back in the radiation dominated era. However, before decoupling, the photon and baryon collisions will provide some contribution to this term. 

$\delta$ and $\theta$ for photons and neutrinos can be calculated in terms of $\Psi$, the detail of which is discussed in Sec.~\ref{secneutrino}.

\subsubsection{Scalar perturbations in the metric}
In the previous section, we describe the equations for the metric perturbation in terms of $T_{\mu\nu}$. However, there are the different components of the universe, i.e. CDM, baryon, photon, neutrino, DE, etc. and we need the metric perturbations in terms of perturbations of these individual components. We use the subscript $c,b,\gamma,\nu$, $\nu_m$ and $d$ for representing cold dark matter, baryonic
matter, photon, massless neutrinos, massive neutrinos, and dark energy respectively. Following Eq.~\ref{ein-syna} and Eq.~\ref{ein-synb}, the equations for the metric perturbation can be written in terms of density ($\delta$) and velocity perturbations ($\theta$) of the individual components as 


\begin{eqnarray}
\frac{\dot{a}}{a}\dot{h} & = & 2k^{2}\eta+\frac{8\pi G}{c^{2}}\left[\rho_{cr}\left(\Omega_{c}\frac{\delta_{c}}{a}+\Omega_{b}\frac{\delta_{b}}{a}+\Omega_{d}\delta_{d}a^2\right)\right.\nonumber\\
 &  &+\left(\frac{4\sigma_{B}T^{4}}{c^{3}}\right) \left(\frac{\delta_{\gamma}}{a^{2}}\right.+N_{\text{eff}}\left(\frac{7}{8}\right)\left(  \left(\frac{4}{11}\right)^{4/3}\frac{\delta_{\nu}}{a^{2}}\right.
 \left.\left.\left.+\left(\frac{43}{11}\frac{1}{g_{A}}\right)^{4/3}\frac{\delta_{\nu_m}}{a^{2}}\right)\right)\right]\,, \\
2k^{2}\dot{\eta}&=&\frac{8\pi G}{c^{2}}\left[\rho_{cr}\left(\Omega_{b}\frac{\theta_{b}}{a}+\Omega_{d}\delta_{d}a^2\right)
+\left(\frac{4\sigma_{B}T^{4}}{c^{3}}\right)\left(\frac{\theta_{\gamma}}{a^{2}}+N_{\text{eff}}\left(\frac{7}{8}\right)\left(\left(\frac{4}{11}\right)^{4/3}\frac{\theta_{\nu}}{a^{2}} \right.\right.\right.\nonumber\\ 
& &+\left(\frac{43}{11}\frac{1}{g_{A}}\right)^{4/3}  \left.\left.\left.\times\left(\frac{\theta_{\nu_m}}{a^{2}}\right)\right)\right)\right]\,,
\end{eqnarray}

\noindent where, over-dot ($\dot{\ldots}$) represents the derivative with respect
to the conformal time and $g_{A}$ is the effective number of spin
states before neutrino decoupling. For $g_{A} = 10.75$,  we get $\left(\frac{43}{11}\frac{1}{g_{A}}\right) = \frac{4}{11}$. However, in presence of any other particle in the early universe, the above expression will correct for that. 

Instead of the last two equations (Eq.~\ref{ein-sync} and Eq.~\ref{ein-synd}), we take the conservation equations ($\delta T^{\mu\nu}_{;\mu} = 0$) for different components of the universe, which evolve independently except before decoupling, when the baryon and photon evolve together as a single fluid. The conservation equations for
different components in terms of their density and velocity perturbations ($\delta$ and $\theta$) are given below.

\begin{figure}[t!]
\centering
\includegraphics[width=0.47\textwidth,trim = 80 250 80 280, clip]{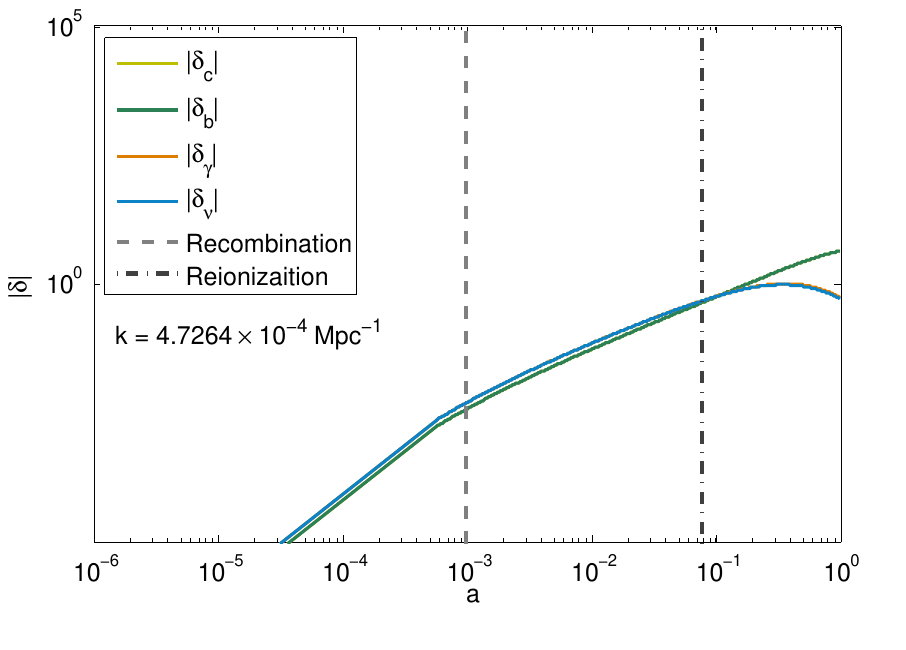}
\includegraphics[width=0.47\textwidth,trim = 80 250 80 280, clip]{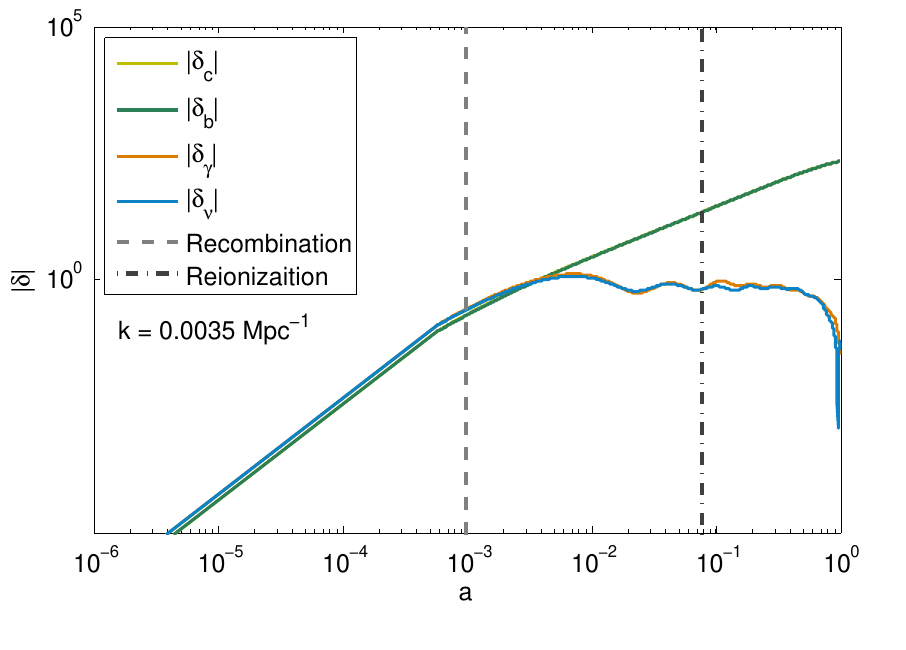}
\includegraphics[width=0.47\textwidth,trim = 80 250 80 280, clip]{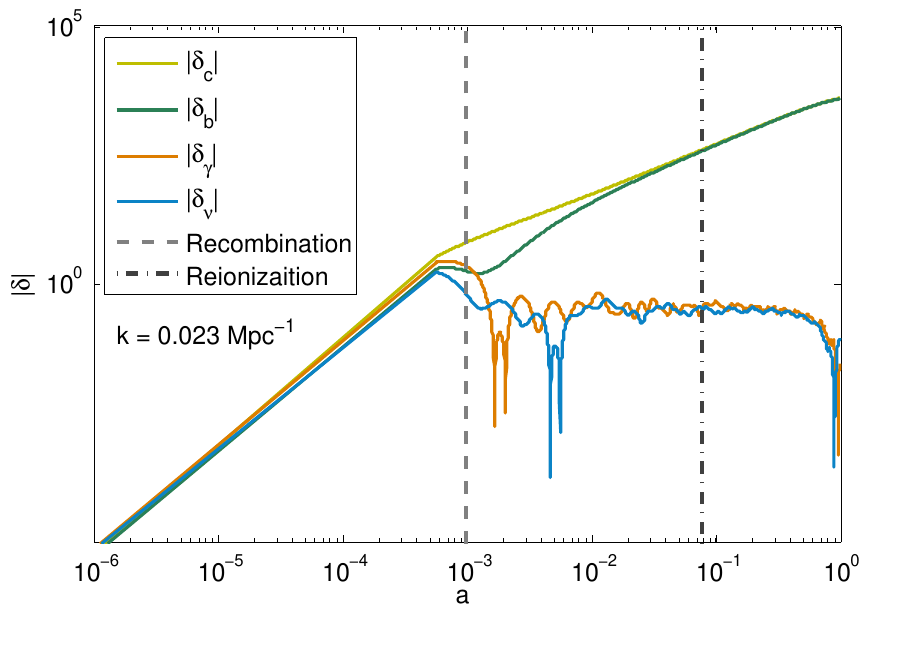}
\includegraphics[width=0.47\textwidth,trim = 80 250 80 280, clip]{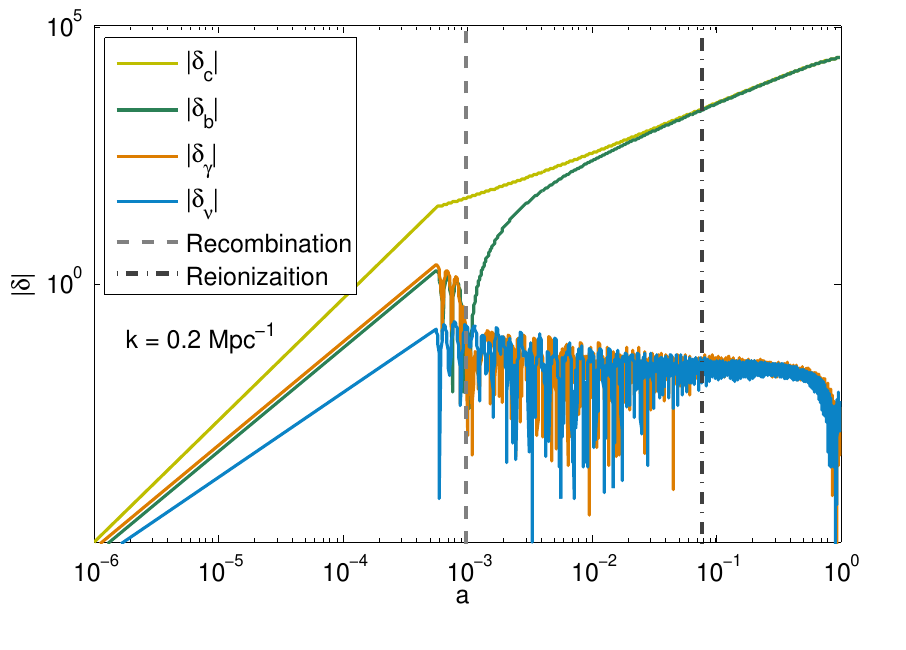}

\caption{\label{fig:delta}The evolution of the density fields for CDM, baryon,
photon and massless neutrinos in the synchronous gauge are shown for
four different modes $k=4.7\times10^{4}$$\mbox{Mpc}^{-1}$, $k=0.0035$$\mbox{Mpc}^{-1}$,
$k=0.023$$\mbox{Mpc}^{-1}$, $k=0.2$$\mbox{Mpc}^{-1}$of the density
perturbation. The initial amplitudes of the $\delta$'s
are related by the adiabatic initial condition $\delta_{\gamma}=\delta_{\nu}=\frac{4}{3}\delta_{c}=\frac{4}{3}\delta_{b}$.
When the modes are outside the horizon they grow as $\delta\propto\tau^{2}$.
In the radiation dominated universe as $a(\tau)\propto\tau$, $\delta$
grows in proportion to $a(\tau)^{2}$. After the matter radiation
equality ($a_{eq}\sim4\times10^{-4}$), as in the matter dominated universe $a(\tau)\propto\tau^{2}$, we have $\delta\propto a(\tau)$. The perturbation wavelength comes within causal contact after crossing the horizon.
For modes that enter the horizon before the recombination ($a_{rec}\sim10^{-3}$,
$k=\frac{2\pi}{\tau_{rec}}\sim0.23$), the baryon and photon
coupled together by Thomson scattering. The photons `dragging' against
baryons leads to Silk damping, which is slightly visible in $k=0.2$ $\text{Mpc}^{-1}$ plot.
After recombination, the baryon perturbations start rapidly growing
as they fall into the potential wells formed by CDM. 
For this plot we use $\Omega_{b}h^{2}=0.022068$, $\Omega_{m}h^{2}=0.14236$,
$h=0.6711$, optical depth $\kappa=0.0925$ and $n_{s}=0.9624$
}
\end{figure}

\subsubsection{Conservation equation for Cold Dark Matter (CDM)}

CDM can be treated as a pressure-less perfect fluid and it interacts
with other particles only through gravity. Therefore, for CDM $P = w = 0$ and the conservation equations, i.e Eq.~\ref{matter_prtb}, for CDM gives 
\begin{eqnarray}
\dot{\delta_{c}}  =  -\theta_{c}-\frac{1}{2}\,\dot{h}\label{eq:cdmdelta}\,, \quad\quad\enskip\quad\quad\enskip \dot{\theta}_{c} & = & -\frac{\dot{a}}{a}\,\theta_{c}\label{eq:cdmtheta}\,.
\end{eqnarray}

\noindent As CDM does not interact with other particles, if $\theta_{c}=0$
is fixed as the initial condition then the values of $\theta_{c}$
will remain $0$ through the entire era. Setting $\theta_{c}=0$ also solves another purpose by removing the extra gauge mode in the synchronous gauge (Sec.~\ref{sub:initialcond_scalar}). Thus the CDM equation becomes
as $\dot{\delta_{c}}=-\frac{1}{2}\,\dot{h}$. In Fig.~\ref{fig:delta}
we show the evolution of the CDM density perturbation for different
$k$ modes.

\subsubsection{Conservation equations for massless neutrinos\label{secneutrino}}

According to the standard model of particle physics the neutrinos
are massless. However, neutrino oscillation provides evidence for non-zero mass, but only $\mathrm{d}m^2$ values.
Therefore, cosmologists are interested
in checking the consequences of both the massive and massless neutrinos
in the cosmic fluid. The perturbation equation for the massive and the
massless neutrinos differ significantly. 

For massless neutrinos, the density and pressure are related as $\rho_{\nu}=a^{-4}\int q^{2}f_\text{FD}(x^i,q,n_j,\tau)\mathrm{d}q\mathrm{d}\Omega$ $=3P_{\nu}$ ,
where, $f_\text{FD}(x^i,q,n_j,\tau)$ is the Fermi-Dirac distribution function and $q$
is the momentum in the co-moving frame. Unlike CDM, neutrinos have
pressure. Therefore, solving the growth equations for neutrinos is
difficult as there are several direction-dependent variables. However, for solving 
Eq.~\ref{ein-syna}~--~Eq.~\ref{ein-synd}, we only need $\delta_\nu$, $\theta_\nu$ and $\sigma_\nu$ and these can
be simplified by defining a new variable 

\begin{equation}
F_{\nu}(\vec{k},\hat{n},\tau)=\frac{\int q^{2}f_{0}(q)q\Psi \mathrm{d}q}{\int q^{2}f_{0}(q)q\mathrm{d}q},    \label{Fterm}
\end{equation}

\noindent where $\Psi$ is the perturbation in the Fermi-Dirac distribution
function i.e. $f_\text{FD}=f_{0}(1+\Psi)$, $f_{0}$ being the $0$th order
term of the Fermi-Dirac distribution function. 

According to Eq.~\ref{radiationPerturbation}, $\vec{k}$ and $\hat{n}$ always appear together as $(\vec{k}.\hat{n})$ , showing that only the magnitude and the angle between them are important, not the individual values of $\vec{k}$ and $\hat{n}$. We can expand $F_{\nu}(\vec{k},\hat{n},\tau)$
in terms of Legendre Polynomials as 
\begin{equation}
F_{\nu}(\vec{k},\hat{n},\tau)=\sum_{l^{\nu}=0}^{\infty}(-i)^{l^{\nu}}(2l^{\nu}+1)F_{\nu l^{\nu}}(k,\tau)P_{l^{\nu}}(\hat{k}.\hat{n})    
\end{equation}

where $P_{l^{\nu}}(\hat{k}.\hat{n})$ are the Legendre polynomials and $F_{\nu l^{\nu}}(k,\tau)$ are their coefficients. 

\noindent The term $(-i)^{l^{\nu}}(2l^{\nu}+1)$ is taken out to simplify the perturbation equations. For calculating the perturbation equations in terms of these variables, we can use Eq.~\ref{radiationPerturbation} and Eq.~\ref{Fterm} and expand them in the Legendre polynomials. The perturbation equations for massless neutrinos
take the form \citep{Ma1995}

\begin{eqnarray}
\dot{F}_{\nu\,0} &=& -\frac{4}{3}\theta_{\nu}-\frac{2}{3}\dot{h}  =\dot{\delta}_{\nu}\,,\\
\dot{F}_{\nu\,1} &=& k\left(\frac{1}{3}\delta_{\nu}-\frac{4}{3}\sigma_{\nu}\right) =\frac{4}{3k}\dot{\theta}_{\nu}\,,\\
\dot{F}_{\nu\,2} &=&\frac{8}{15}\theta_{\nu}-\frac{3}{5}kF_{\nu\,3}+\frac{4}{15}\dot{h}+\frac{8}{5}\dot{\eta} =  2\dot{\sigma}_{\nu}\,,\\
\dot{F}_{\nu\, l^{\nu}} &=&  \frac{k}{2l^{\nu}+1}\left[l^{\nu}F_{\nu\,(l^{\nu}-1)}-(l^{\nu}+1)F_{\nu\,(l^{\nu}+1)}\right]\,,\quad l^{\nu}\geq3\,.\label{eq:Fnu8}
\end{eqnarray}

\noindent Here $l^{\nu}$ goes up to $\infty$. The truncation condition can
be taken as~\citep{Ma1995}
\begin{equation}
F_{\nu(l_{max}^{\nu}+1)}\thickapprox\frac{2l_{max}^{\nu}+1}{k\tau}F_{\nu l_{max}^{\nu}}-F_{\nu(l_{max}^{\nu}-1)}\label{eq:Fnu9} \,.
\end{equation}

\begin{figure}
\includegraphics[width=0.49\textwidth,trim = 0 80 0 100, clip]{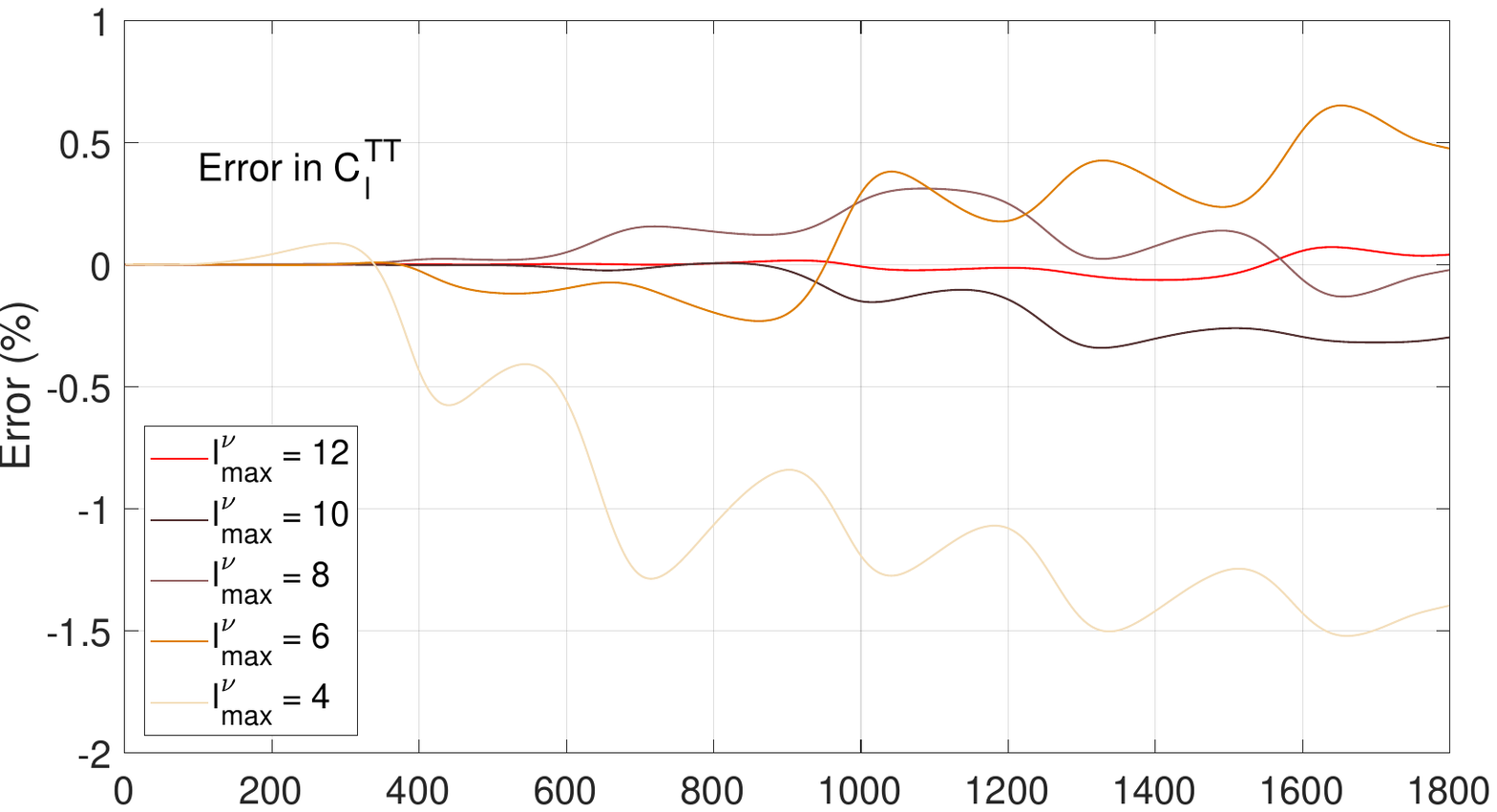}
\includegraphics[width=0.49\textwidth,trim = 0 80 0 100, clip]{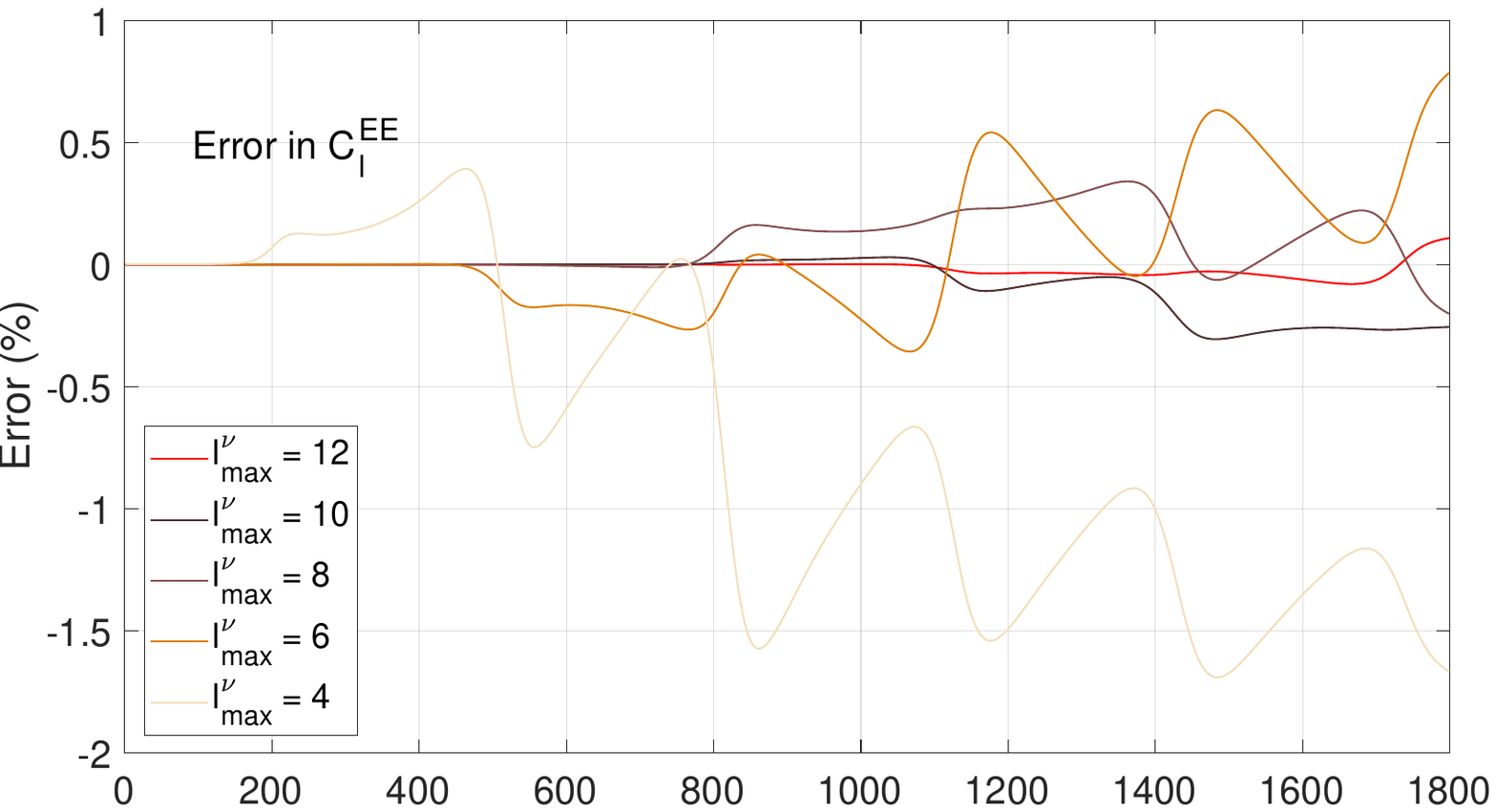}

\caption{\label{fig:neutrinoerror}The error $\left(\frac{(C_{l}^{l_{max}^{\nu}}-C_{l}^{l_{max}^{\nu}=14})}{C_{l}^{l_{max}^{\nu}=14}}\times100\%\right)$
involved in the power spectrum calculation for truncating the neutrino
multipole equations at different $l_{max}^{\nu}$.
The left and right plot show the error involved in $C_{l}^{TT}$
and $C_{l}^{EE}$ calculations respectively. We use $\Omega_{b}h^{2}=0.022068$, $\Omega_{m}h^{2}=0.14236$,
$h=0.6711$, optical depth $\kappa=0.0925$, $n_{s}=0.9624$}
\end{figure}

\noindent 
However, such a choice of truncation condition can lead to a large error in the final power spectrum calculation, due to the propagation error if the Boltzmann hierarchy is truncated after some low $l_{max}^{\nu}$.
For truncating the set of equations, in \texttt{CMBAns} we use $l_{max}^{\nu}=12$ 
(which the user can change to any higher value).
In Fig.~\ref{fig:neutrinoerror} we show the percentage errors $\left(\frac{(C_{l}^{l_{max}^{\nu}}-C_{l}^{l_{max}^{\nu}=14})}{C_{l}^{l_{max}^{\nu}=14}}\times100\%\right)$ involved 
in the CMB power spectrum calculation for truncating the equations at different $l_{max}^{\nu}$, considering
$l_{max}^{\nu}=14$ to be standard. The plots show that the error
introduced due to the truncation of the neutrino equations slowly decreases
with the increase in $l_{max}^{\nu}$. 

In Fig.~\ref{fig:delta}, we show the growth of the density fluctuations
in neutrino. As they don't interact with any other species, their
evolution is completely independent of others. The modes $\delta_{\nu}$
evolve as $\propto\tau$ before horizon crossing. After horizon
crossing, when they come in causal contact they start oscillating.

\subsubsection{Conservation equations for massive neutrinos}

For the massive neutrinos $\epsilon=(q^{2}+a^{2}m_{\nu_m}^{2})^{\frac{1}{2}}$, where
$q$ is the momentum in the co-moving frame and $m_{\nu_m}$ is the
mass of the neutrino. Hence, unlike massless neutrinos, the expressions
$\int q^{2}f_{0}(q)\frac{q^{2}}{\epsilon}\Psi \mathrm{d}q$ cannot be
integrated analytically. Thus, we expand the perturbation $\Psi$
directly in terms of Legendre polynomials as \citep{Ma1995}:

\begin{equation}
\Psi(\vec{k},\hat{n},q,\tau)=\sum_{l^{\nu_m}=0}^{\infty}(-i)^{l^{\nu_m}}(2l^{\nu_m}+1)\Psi_{l^{\nu_m}}(\vec{k},q,\tau)P_{l^{\nu_m}}(\hat{k}\cdot\hat{n})\,\label{psisubl}\,.
\end{equation}

\noindent Substituting this in Eq.~\ref{radiationPerturbation}, the Boltzmann
equations for the massive neutrinos become 

\begin{eqnarray}
\dot{\Psi}_{0} & = & -\frac{qk}{\epsilon}\Psi_{1}+\frac{1}{6}\dot{h}\frac{\mathrm{d}(\ln f_{0})}{\mathrm{d}(\ln q)}\,,\nonumber \\
\dot{\Psi}_{1} & = & \frac{qk}{3\epsilon}\left(\Psi_{0}-2\Psi_{2}\right)\,,\nonumber  \\
\dot{\Psi}_{2} & = & \frac{qk}{5\epsilon}\left(2\Psi_{1}-3\Psi_{3}\right)-\left(\frac{1}{15}\dot{h}+\frac{2}{5}\dot{\eta}\right)\frac{\mathrm{d}(\ln f_{0})}{\mathrm{d}(\ln q)}\,,\\
\dot{\Psi}_{l^{\nu_m}} & = & \frac{qk}{(2l^{\nu_m}+1)\epsilon}\left[l^{\nu_m}\Psi_{l^{\nu_m}-1}-(l^{\nu_m}+1)\Psi_{l^{\nu_m}+1}\right]\,,\quad l^{\nu_m}\geq3\,\nonumber 
\end{eqnarray}

\noindent For truncating the series we use the condition 

\begin{equation}
\Psi_{(l_{{\rm max}}^{\nu_m}+1)}\approx\frac{(2l_{{\rm max}}^{\nu_m}+1)\epsilon}{qk\tau}\,\Psi_{l_{{\rm max}}^{\nu_m}}-\Psi_{(l_{{\rm max}}^{\nu_m}-1)}\ \label{truncmnu}\,.
\end{equation}

\noindent Here, $\Psi_l$ are the functions of $k,\tau$ and $q$. The integration
over $q$ is performed numerically after calculating different massive
neutrino multipoles for different $q$. For the massive neutrinos
the modes are truncated after $l_{max}^{\nu_m}=5$.

\subsubsection{Photons \label{photonbrightnessfunction}}

Photons 
evolve differently before recombination, when they were tightly coupled
with baryons and after recombination, when they were free streaming.
The evolution of the photons can be treated in the same way as the massless
neutrinos except the collision terms will be present. Thompson scattering by a density perturbation can introduce polarization in an unpolarized photon field and in a density perturbation. Therefore, along with the total intensity, we also have to consider the polarization component for the photons. 
The details of the polarization is discussed are Appendix~\ref{AppendixA}. 

Defining the intensity perturbations as $\Delta_{T}$
and the polarization as $\Delta_{P}$, the Boltzmann equations for photons
in the synchronous gauge can be written as \citep{Ma1995}

\begin{eqnarray}
\dot{\Delta}_{T\,0} & = & \dot{\delta}_{\gamma} = -\frac{4}{3}\theta_{\gamma}-\frac{2}{3}\dot{h}\,,\nonumber \\
\frac{4k}{3}\dot{\Delta}_{T\,1} & = & \dot{\theta}_{\gamma} = k^{2}\left(\frac{1}{4}\delta_{\gamma}-\sigma_{\gamma}\right)+an_{e}\sigma_{T}(\theta_{b}-\theta_{\gamma})\,,\nonumber \\
\dot{\Delta}_{T\,2} & = & 2\dot{\sigma}_{\gamma}=\frac{8}{15}\theta_{\gamma}-\frac{3}{5}k\Delta_{T\,3}+\frac{4}{15}\dot{h}+\frac{8}{5}\dot{\eta}-\frac{9}{5}an_{e}\sigma_{T}\sigma_{\gamma}+\frac{1}{10}an_{e}\sigma_{T}\left(\Delta_{P\,0}+\Delta_{P\,2}\right)\,,\nonumber\\
\dot{\Delta}_{T\, l^{\gamma}} & = & \frac{k}{2l^{\gamma}+1}\left[l^{\gamma}\Delta_{T\,(l^{\gamma}-1)}-(l^{\gamma}+1)\Delta_{T\,(l^{\gamma}+1)}\right]-an_{e}\sigma_{T}\Delta_{T\, l^{\gamma}}\,,\quad l^{\gamma}\geq3\,,\nonumber \\
\dot{\Delta}_{P\, l^{\gamma}} & = & \frac{k}{2l^{\gamma}+1}\left[l^{\gamma}\Delta_{P\,(l^{\gamma}-1)}-(l^{\gamma}+1)\Delta_{P\,(l^{\gamma}+1)}\right]+an_{e}\sigma_{T}\left[-\Delta_{P\, l^{\gamma}}\right. \label{eq:k}\\
&&+\left.\frac{1}{2}\left(\Delta_{T\,2}+\Delta_{P\,0}+\Delta_{P\,2}\right)\left(\delta_{l^{\gamma}0}+\frac{\delta_{l^{\gamma}2}}{5}\right)\right]\,,\nonumber 
\label{photonmomentum}
\end{eqnarray}

\noindent where $\sigma_{T}$ is the Thomson scattering cross-section, and
$n_{e}$ is the free electron number density. The truncation of the Boltzmann
equations is done in the same way as that of the massless neutrinos. For $l^{\gamma}=l_{max}^{\gamma}$
we can replace the $\Delta_{T\, l^{\gamma}}$ and the $\Delta_{P\, l^{\gamma}}$
by the following equations 

\begin{eqnarray}
\dot{\Delta}_{T\, l_{max}^{\gamma}} & = & k\Delta_{T\,(l_{max}^{\gamma}-1)}-\frac{l_{max}^{\gamma}+1}{\tau}\Delta_{T\, l_{max}^{\gamma}}-an_{e}\sigma_{T}\Delta_{T\, l_{max}^{\gamma}}\nonumber\,, \\
\dot{\Delta}_{P\, l_{max}^{\gamma}} & = & k\Delta_{P\,(l_{max}^{\gamma}-1)}-\frac{l_{max}^{\gamma}+1}{\tau}\Delta_{P\, l_{max}^{\gamma}}-an_{e}\sigma_{T}\Delta_{P\, l_{max}^{\gamma}}\,.\label{eq:Photontrancation}
\end{eqnarray}

\noindent The evolution of the photons is shown in Fig.~\ref{fig:delta}.
Before recombination, the photons are tightly coupled with the
baryons, and they evolve together as a single fluid. However, after decoupling, the photons follow a similar pattern
as that of the neutrinos. 

\begin{figure}
\includegraphics[width=0.49\textwidth,trim = 70 270 70 280, clip]{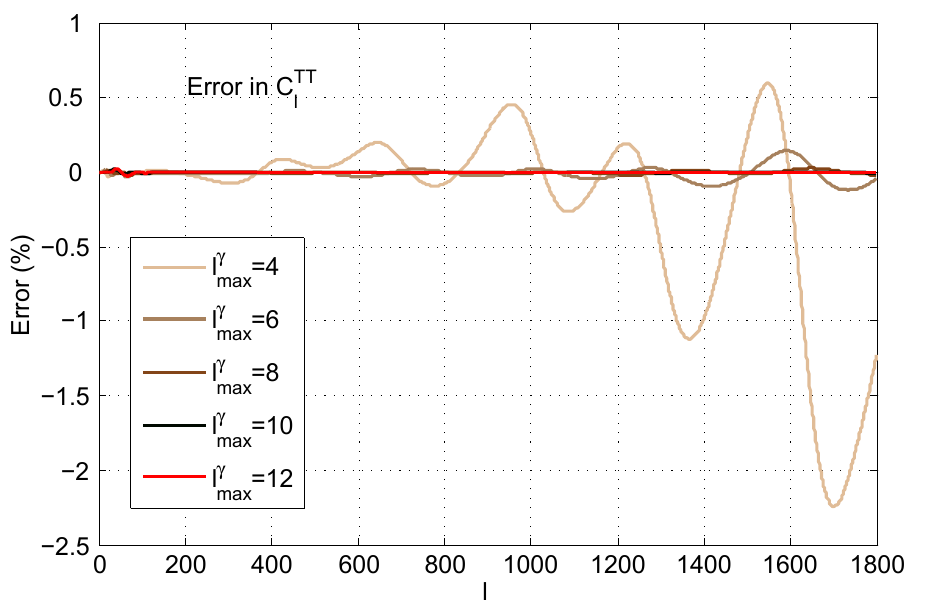}
\includegraphics[width=0.49\textwidth,trim = 70 270 70 280, clip]{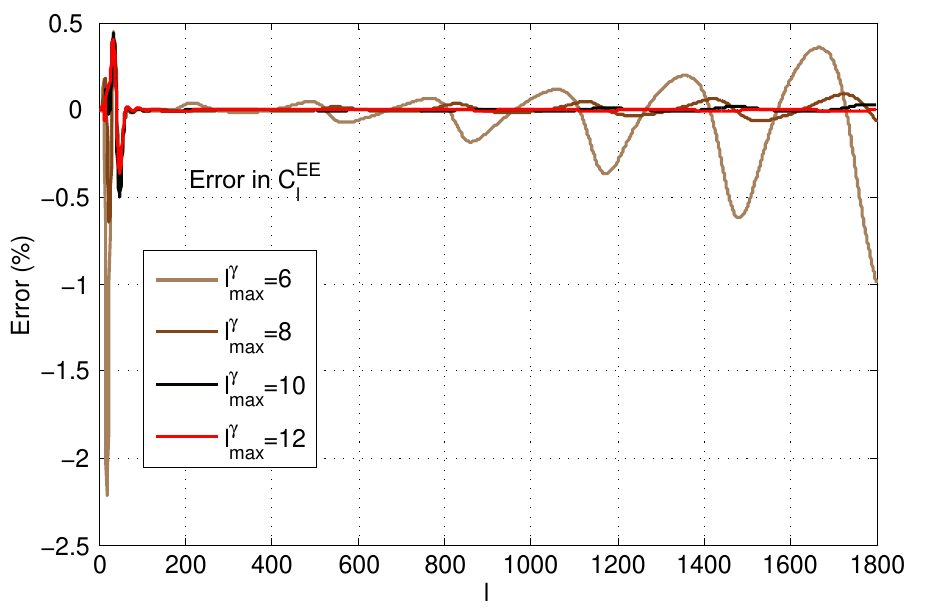}

\caption{\label{fig:photonerror}The error $\left(\frac{(C_{l}^{l_{max}^{\gamma}}-C_{l}^{l_{max}^{\gamma}=13})}{C_{l}^{l_{max}^{\gamma}=13}}\times100\%\right)$
involved in the power spectrum calculation for truncating the photon
multipole equations at different $l_{max}^{\gamma}$ are shown here.
The left and right plot shows the error involved in $C_{l}^{TT}$
and $C_{l}^{EE}$ calculations. Here, $\Omega_{b}h^{2}=0.022068$, $\Omega_{m}h^{2}=0.14236$,
$h=0.6711$, optical depth $(\kappa)=0.0925$, $n_{s}=0.9624$}
\end{figure}

For photons we choose the truncation at $l_{max}^{\gamma}=12$. The error
involved due to the truncation of the photon multipole equations,
i.e. $\frac{(C_{l}^{l_{max}^{\gamma}}-C_{l}^{l_{max}^{\gamma}=13})}{C_{l}^{l_{max}^{\gamma}=13}}\times100\%$
is shown in Fig.~\ref{fig:photonerror}. The errors are calculated
considering $l_{max}^{\gamma}=13$ to be standard. We can see that
if we truncate the evaluation in low $l_{max}^{\gamma}$, such as
$l_{max}^{\gamma}=4$ or $6$, then the error involved can be as high as $0.5\%$. However,
if we increase $l_{max}^{\gamma}$, the error decreases drastically. At $l_{max}^{\gamma} = 12$, the error in the final $C_l$ is less than $0.01\%$. 

\subsubsection{Baryons}

Before recombination, the baryons and photons were tightly coupled and
they evolved together as a single fluid. However, after recombination, the baryons decouple from the photons and evolve separately. The baryon
density and the velocity perturbations are governed by the following
equations \citep{Ma1995}

\begin{eqnarray}
\dot{\delta}_{b} & = & -\theta_{b}-\frac{1}{2}\dot{h}\,,\nonumber \\
\dot{\theta}_{b} & = & -\frac{\dot{a}}{a}\theta_{b}+c_{s}^{2}k^{2}\delta_{b}+\frac{4\rho_{\gamma0}}{3\rho_{b0}}an_{e}\sigma_{T}(\theta_{\gamma}-\theta_{b})\,,
\end{eqnarray}

\noindent where, $\rho_{\gamma0}$ and $\rho_{b0}$ are the background density
for photons and baryons respectively. $c_{s}$ is the baryon sound speed shown in Eq~\ref{baryonsoundspeed}.

In the tight coupling era, the photon and baryon equations
cannot be solved independently using the Runge-Kutta methods, as this may lead to large numerical errors. Therefore, a separate set of equations is used for solving the
baryon perturbation in the tight coupling era. 

\begin{figure}
\includegraphics[width=0.49\textwidth,trim = 70 240 70 270, clip]{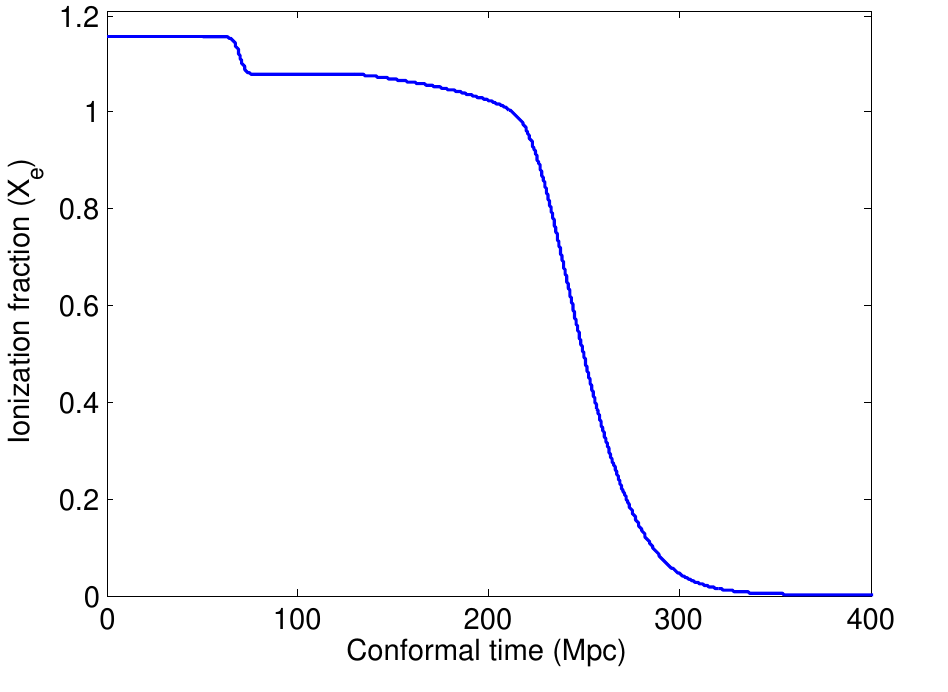}
\includegraphics[width=0.49\textwidth,trim = 70 240 70 270, clip]{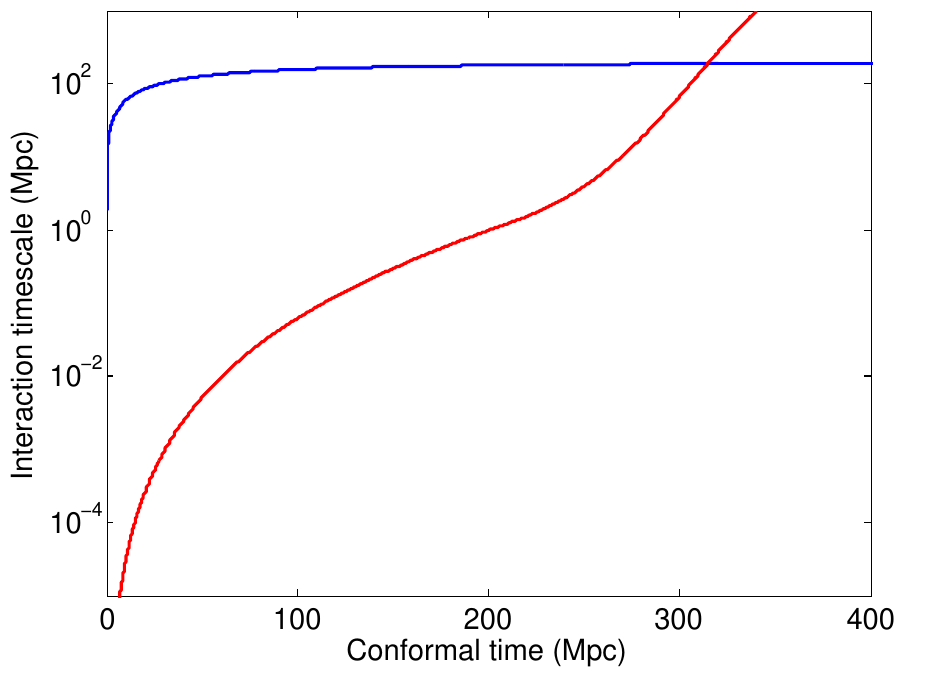}

\caption{\label{fig:ionizationfrac}Left : The ionization fraction $x_{e}=n_{e}/n_{H}$
from the $\texttt{recfast}$ recombination routine. Right : Change of the photon
baryon interaction time scale $\left(\tau_{c}=\left(an_{e}\sigma_{T}\right)^{-1}\right)$
over time is plotted in red and the time scale at which the modes
in the super-Hubble scale evolve ($\tau_{H}=a/\dot{a}$) is plotted
in blue. In the region where $\tau_{c}\ll\tau_{H}$, the photons
and baryons are tightly coupled to each other. We choose conformal time to be zero at redshift $10^8$. }
\end{figure}

\subsubsection{Recombination and the Tight Coupling Approximation }

Before recombination the opacity $\dot{\mu}=an_{e}\sigma_{T}$
was very large. Hence, the photons and baryons evolve together during
the tight coupling approximation. The conformal time scale for the
photon baryon interaction is $\tau_{c}=\left(an_{e}\sigma_{T}\right)^{-1}$.
In the tight coupling era, this interaction time scale is much smaller than the timescale on
which the modes in the super-Hubble scale evolve, i.e. $\tau_{H}=a/\dot{a}$
and the modes for the sub-Hubble scale i.e. $\tau_{k}=1/k$. The standard
numerical integration is not efficient to integrate the baryon and
photon perturbation equations independently in this era. Therefore,
Peebles and Yu, \citep{Ma1995, Peebles1970} developed a new set of equations
for the baryon perturbations which is valid in the tight coupling
regime i.e. where $\tau_{c}/\tau_{H}$ or $\tau_{c}/\tau_{k}\ll 1$. Instead of using the standard baryon and photon equations, the idea is to integrate a coupled form of differential
equation, given by 

\begin{equation}
(1+R)\dot{\theta}_{b}+\frac{\dot{a}}{a}\theta_{b}-c_{s}^{2}k^{2}\delta_{b}-k^{2}R\left(\frac{1}{4}\delta_{\gamma}-\sigma_{\gamma}\right)+R(\dot{\theta}_{\gamma}-\dot{\theta}_{b})=0\,,\label{momentum}
\end{equation}

\noindent where $R=(4/3)\rho_{\gamma0}/\rho_{b0}$, and $\left(\dot{\theta}_{\gamma}-\dot{\theta}_{b}\right)$
is given by the following equation

\begin{eqnarray}
\dot{\theta}_{b}-\dot{\theta}_{\gamma} &=& \frac{2R}{1+R}\frac{\dot{a}}{a}(\theta_{b}-\theta_{\gamma})+\frac{\tau_{c}}{1+R}\left[-\frac{\ddot{a}}{a}\theta_{b}-\frac{\dot{a}}{a}k^{2}\left(\frac{1}{2}\delta_{\gamma}\right)\right. \nonumber\\
& &+\left. k^{2}\left(c_{s}^{2}\dot{\delta}_{b}-\frac{1}{4}\dot{\delta}_{\gamma}\right)\right]+O(\tau_{c}^{2}).\label{sliprate}
\end{eqnarray}

\noindent For obtaining $\dot{\theta}_\gamma$, we can use the following equation~\citep{Ma1995}

\begin{equation}
    \dot{\theta}_\gamma = - R^{-1}\Bigg(\dot{\theta}_b + \frac{\dot{a}}{a}\theta_b - c_s^2k^2\delta_b\Bigg)+k^2\Bigg(\frac{1}{4}\delta_\gamma-\sigma_\gamma\Bigg)
\end{equation}

\noindent In the tight coupling limit the higher multipoles of the photon distribution, i.e. $F_{\gamma 3}$, $F_{\gamma 4}$, ... and $G_{\gamma 0}$, $G_{\gamma 1}$, $G_{\gamma 2}$, ... can be taken as $0$. We also consider $\dot{\sigma}_\gamma = 0$, which from Eq.~\ref{photonmomentum} gives

\begin{equation}
    \sigma_\gamma = \frac{\tau_c
}{9}\Bigg(\frac{8}{3}\theta_\gamma +\frac{4}{3}\dot{h} + 8\dot{\eta}\Bigg) \end{equation}

\begin{figure}
\centering
\includegraphics[width=0.7\textwidth,trim = 50 200 50 200, clip]{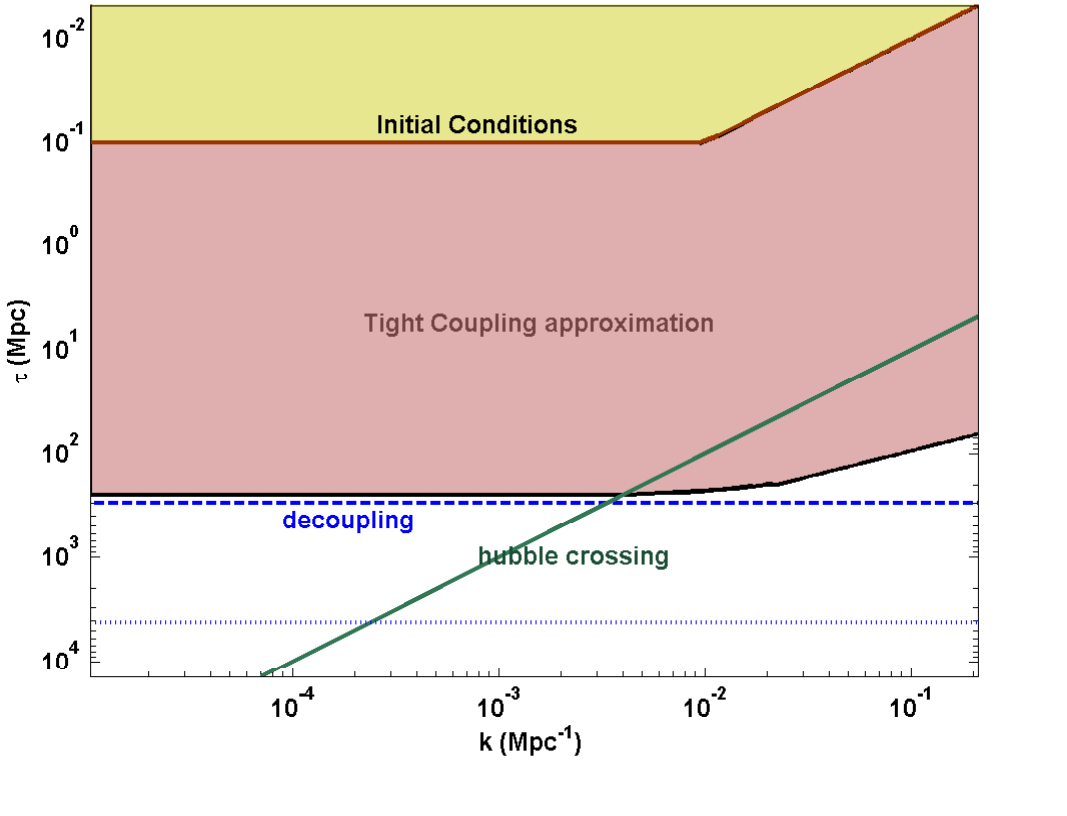}
\caption{\label{fig:initialcond}The plot shows different epoch of the universe.
The $x$ axis shows the wave-numbers and the $y$ axis shows
conformal time. The blue dotted line shows the beginning of the reionization epoch
and the dashed line shows the beginning of the recombination epoch in the universe.
The green line depicts the Hubble crossing of the modes. The light red
region shows the era where the tight coupling approximation equations
are used. The initial conditions in \texttt{CMBAns} are specified along the
brown line at the interface of the light red and yellow region. A similar analysis for $\texttt{CLASS}$ can be found in \cite{blas2011cosmic}.}
\end{figure}

\noindent In Fig.~\ref{fig:initialcond}, we show in light red color the region where we use the tight coupling approximation using light red color.

\subsubsection{DE perturbation}

Under fluid assumption, a general closed-form solution \citep{Bean2004,Weller2003,Hannestad2005}
for the density and the velocity perturbation 
of dark energy is given by 

\begin{eqnarray}
\dot{\delta}_d &=& -(1+w_{d})\left\{ \left[k^{2}+9H^{2}(c_{s}^{2}-c_{a}^{2})\right]\frac{\theta_d}{k^{2}}+\frac{\dot{h}}{2}\right\} -3H(c_{s}^{2}-w_{d})\delta_d\,, \nonumber \\
%
%
\frac{\dot{\theta}_d}{k^{2}} &=& -H(1-3c_{s}^{2})\frac{\theta_d}{k^{2}}+\frac{c_{s}^{2}}{1+w_{d}}\delta_d\,.
\label{scalarDE}
\end{eqnarray}

\noindent Here $\delta_d$ and $\theta_d$ are used in their usual meaning, i.e.
$\delta_d$ is the density perturbation and $\theta_d$ is the velocity
perturbation of the dark energy. $c_{a}$ is known as the adiabatic sound speed and
given by 
\begin{equation}
c_{a}^{2}=w_{d}-\frac{\dot{w}_{d}}{3H(1+w_{d})}\,,
\end{equation}

\noindent where $w_{d}$ is the dark energy equation of state and 
$c_{s}$ is the speed of sound in the dark energy and is given by 
$c_{s}^{2}=\frac{\delta p_{d}}{\delta\rho_{d}}$. 
For a perfect fluid $c_s^2=c_a^2$, in which case the DE perturbation equation is the same as the matter perturbation equation given by Eq.~\ref{matter_prtb}.

In the case of dark energy perturbation, we evolve these
equations along with the other perturbation equations.

\subsection{Theory of tensor perturbations}

The tensor perturbations are gauge-invariant quantities. According to Eq.~\ref{metricmerturbation} and Eq.~\ref{stressperturbation}, Einstein's equation for the tensor perturbation can be written as 

\begin{equation}
    \ddot{D}_{ij} + 2H\dot{D}_{ij} + \nabla^2D_{ij} = \frac{16\pi G}{c^2}a^2 \pi^T_{ij}
\end{equation}

\noindent Thus, the tensor metric perturbation can be described by two gravity wave polarization modes, $h_{+}$ and 
$h_{\times}$.  Here we have chosen the perturbations in the $x-y$ plane, which corresponds
to an implicit choice of the $z$ axis to be in the direction of the wave propagation.  
 As the perturbation equations concerning both polarization modes are the same, the metric perturbation can be denoted by a single variable, $h_q$, where $q \in ({+,\times})$. 
Cold dark matter and baryon, being non-relativistic, do not
contribute to any anisotropic stress and as a consequence are not involved in tensor perturbation. (for details check Appendix~\ref{AppendixA})

In Fourier space, the tensor perturbation equations take the form

\begin{equation}
\ddot{h}_q=-2\frac{\dot{a}}{a}\dot{h}_q-k^{2}h_q+\frac{16\pi G}{c^2}a^2\Pi^t\,,
\label{tensorperturbation}
\end{equation}

\noindent where, $\Pi$ is the anisotropic stress term and is given by  
$\Pi^t=2\left(\rho_{\gamma}\mathcal{S}^t_{\gamma}+\rho_{\nu}\mathcal{S}^t_{\nu}\right)$\, 
with ~\citep{Weinberg2006, Baskaran2006, Weinberg2008cosmology}

\begin{eqnarray}
\mathcal{S}^t_{\gamma}&=&\frac{\delta^t_{\gamma}}{15}+\frac{\Delta^t_{T2}}{21}+\frac{\Delta^t_{T4}}{35}\,, \\
\mathcal{S}^t_{\nu}&=&\frac{\delta^t_{\nu}}{15}+\frac{F^t_{\nu2}}{21}+\frac{F^t_{\nu4}}{35}\,.
\end{eqnarray}

\subsubsection{Photons}

The photons behave differently before and after recombination. 
In the tightly coupled era,
the photons and baryons evolve as a single fluid. 
However, after the epoch of decoupling, the photons and baryons evolve
independently. During the time when the photons and
baryons are not tightly coupled, the photon perturbations
are governed by the equations~\citep{Weinberg2006, Baskaran2006, Weinberg2008cosmology}

\begin{eqnarray}
\dot{\delta}^t_{\gamma}=\dot{\Delta}^t_{T\,0}=-k^{2}\theta^t_{\gamma}-an_{e}\sigma_{T}\delta^t_{\gamma}+an_{e}\sigma_{T}\Psi^t_{e}-\dot{h}_q \,, \\
\dot{\Delta^t_P}_{0}=-k^{2}\Delta^t_{P1}-an_{e}\sigma_{T}\Delta^t_{P0}-an_{e}\sigma_{T}\Psi^t_{e} \,,
\end{eqnarray}

\noindent For $l\geq1$ the equations are the same as those of the scalar case and
are given by 

\begin{eqnarray}
\dot{\Delta_T^t}_{l}\, & = & \frac{k}{2l+1}\left[l{\Delta_T^t}_{(l-1)}-(l+1){\Delta_T^t}_{(l+1)}\right]-an_{e}\sigma_{T}{\Delta_T^t}_{l}\,, \\
\dot{\Delta_P^t}_{l} & = & \frac{k}{2l+1}\left[l{\Delta_P^t}_{(l-1)}-(l+1){\Delta_P^t}_{(l+1)}\right]-an_{e}\sigma_{T}{\Delta_P^t}_{l} \,,
\end{eqnarray}

\noindent where, 

\begin{equation}
\Psi^t_{e}=\frac{\delta^t_{\gamma}}{10}+\frac{{\Delta_T^t}_{2}}{7}+\frac{3{\Delta_T^t}_{4}}{70}-\frac{3{\Delta_P^t}_{0}}{5}+\frac{6{\Delta_P^t}_{2}}{7}-\frac{3{\Delta_P^t}_{u4}}{70}\,.
\end{equation}

\noindent The truncation condition is the same as Eq.~\ref{eq:Photontrancation} used for scalar cases.

\subsubsection{Tight coupling approximation }

In the tight coupling limit, we take $\Delta^t_T = \Delta^t_P = 0$
for the $l\geq1$. Thus the equations for $l=0$ mode take the form 

\begin{equation}
\delta^t_{\gamma}=-\frac{4}{3}\frac{\dot{h}_q}{an_{e}\sigma_{T}}\,,
\quad\quad\quad\quad \Delta^t_{P0}=\frac{1}{3}\frac{\dot{h}_q}{an_{e}\sigma_{T}}\,
\end{equation}



\subsubsection{Massless neutrino}

The perturbation equations for the massless neutrinos are given by 

\begin{equation}
\dot{\delta}^t_{\nu}=\dot{F}^t_{\nu\,0} = - k^{2}\theta^t_{\nu}-\dot{h}_q\,,
\end{equation}

\noindent and for $l\ge 1$

\begin{equation}
\dot{F}^t_{\nu\, l}=\frac{k}{2l+1}\left[lF^t_{\nu\,(l-1)}-(l+1)F^t_{\nu\,(l+1)}\right] \,.
\label{neutrinoTesor}
\end{equation}

\noindent The truncation condition is the same as Eq.~\ref{eq:Fnu9}. The contribution from the massive neutrinos will be very small. Therefore, we ignore their contribution to massive neutrinos in the tensor perturbations.

\section{\label{sub:initialcond}Initial conditions}
 In the previous section, we discuss the perturbation equations for the scalar and the tensor perturbation. However, for solving the set of differential equations, we need the initial conditions.

\subsection{\label{sub:initialcond_scalar}Scalar perturbation}

The initial conditions in the universe are assumed to be set by inflation. In single field inflation models, the inflationary field later decays to produce all the constituents of the universe, i.e. baryons (including leptons), photons, neutrinos, CDM, etc. It can be shown that the perturbations produced during inflation stay only in the modes that existed at the end of inflation, for as long as the perturbations remain outside the horizon. The wavelengths that interest us are far outside the horizon during the era of reheating after inflation. Therefore, the perturbations in those modes will remain the same irrespective of whatever the constituents of the universe may become. If the scalar perturbations are adiabatic at the end of inflation, then reheating cannot generate entropic perturbations~\citep{Weinberg2008cosmology, Weinberg2003}. 

However, there are different multifield inflationary models such as double inflation, where one field decays to CDM and the other fields produce other constituents like baryons, photons, and neutrinos. Such a scenario can produce both the adiabatic and the isocurvature modes. In this particular case, it will produce the CDM isocurvature modes. 

In the present work, we want to calculate the CMB power spectrum for different initial conditions and are not interested in the theoretical details of the production of different isocurvature modes. There can be different types of isocurvature initial conditions, such as baryon isocurvature modes, CDM isocurvature modes, neutrino density isocurvature modes, and neutrino velocity isocurvature modes \citep{bucher2000general}. 

In scalar perturbation, the metric is perturbed by the perturbations in the primordial plasma. Therefore, if there is no plasma fluid perturbation, then metric perturbations should be $0$. We also assume that in the very early universe, even before the decoupling of neutrinos, everything was tightly coupled and hence the anisotropic stress terms were zero. Hence, we just need to set the initial values for $\delta$ and $\theta$ for all the $5$ components of the universe. In total, we will have $10$ different modes.  

As we are working with a set of linear differential equations, we can set all these $10$ parameters one by one, keeping the rest of them as $0$ and evolve the equations independently. Finally, we can add all these solutions to get the final solution. Some authors studied dark energy perturbations~\citep{Liu2010, Gordon2004}, but in this work we do not consider any dark energy isocurvature modes, i.e. we set $\delta_{DE} = 0$ and $\theta_{DE} = 0$ at $\tau = 0$.  Thus, we are left with a total of $8$ degrees of freedom. 

We are using the synchronous gauge, and it can be shown that the metric perturbation cannot be eliminated even when there is no density or velocity contrast. Therefore, an extra initial condition is required, making the total number of degrees of freedom to $9$. This extra mode is not a physical mode, but a gauge mode and is often eliminated (as we have done in this work) by setting the velocity of the cold dark matter to $0$. Thus, we will finally have a total of $8$ degrees of freedom corresponding to $8$ different modes. It can also be shown that three of these modes will be decaying modes, and hence will not contribute to the final power. As a result, we have a total of 5 degrees of freedom to choose. One of the decaying modes decays, as baryons and photons were tightly coupled and so $\theta_\gamma$ and $\theta_b$ cannot be independently chosen. The other two modes decay because of total nonzero velocity perturbation and density perturbation~\citep{bucher2000general}.

For defining these 5 modes we can define $5$ variables, which are $\delta_c$, $\delta_b$, $\delta_\nu$, $\theta_\nu$ and $\eta$. We are choosing $\eta$ instead of the photon variables $\delta_\gamma$ because this is standard in the literature. In such a scenario, we can choose 

\begin{equation}
    \delta_b = \delta_c = \frac{3}{4}\delta_\nu = \frac{3}{4}\delta_\gamma
\end{equation}

\noindent which is known as the adiabatic initial condition. 

We must set the initial conditions deep inside the radiation dominated era, after neutrino decoupling where the physics is known. In the radiation dominated era, we can get the initial conditions for $5$ different modes as: adiabatic initial conditions, baryon isocurvature initial conditions, CDM isocurvature initial conditions, neutrino density isocurvature and neutrino velocity isocurvature model. 

 \begin{figure}
\centering
\includegraphics[width=0.32\textwidth,trim = 70 100 90 130, clip]{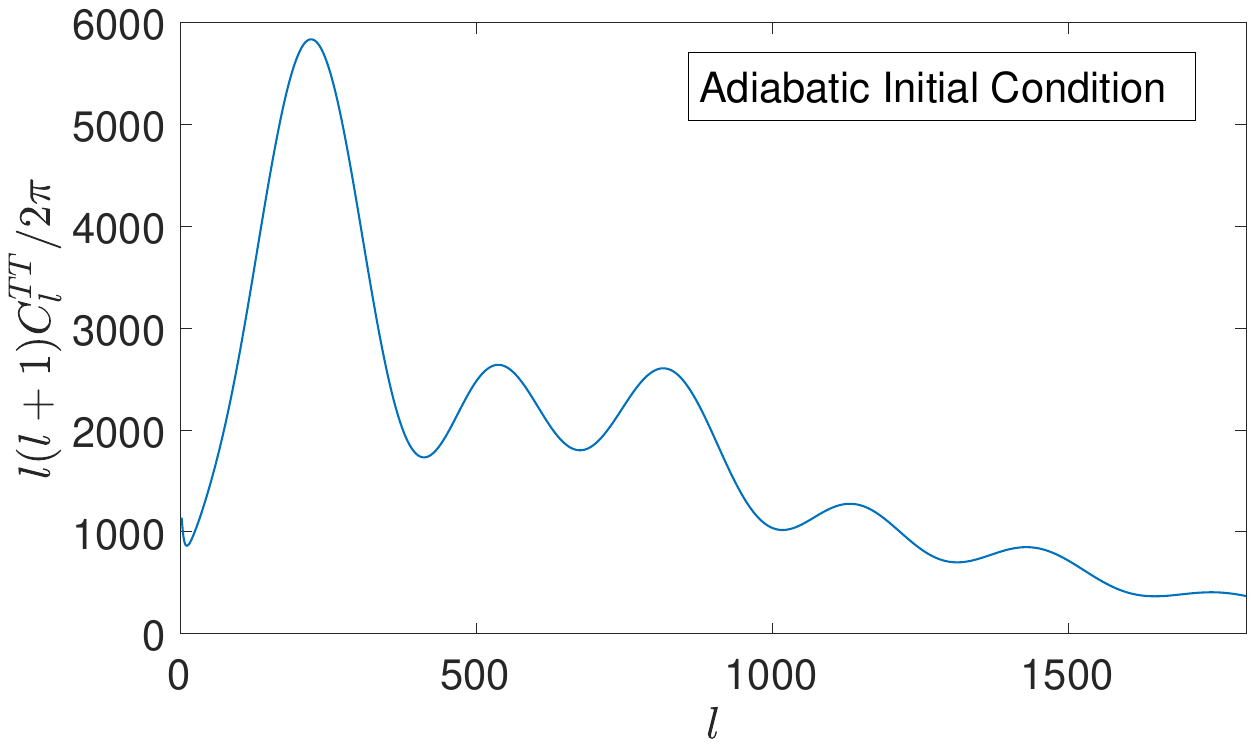}
\includegraphics[width=0.32\textwidth,trim = 70 100 90 130, clip]{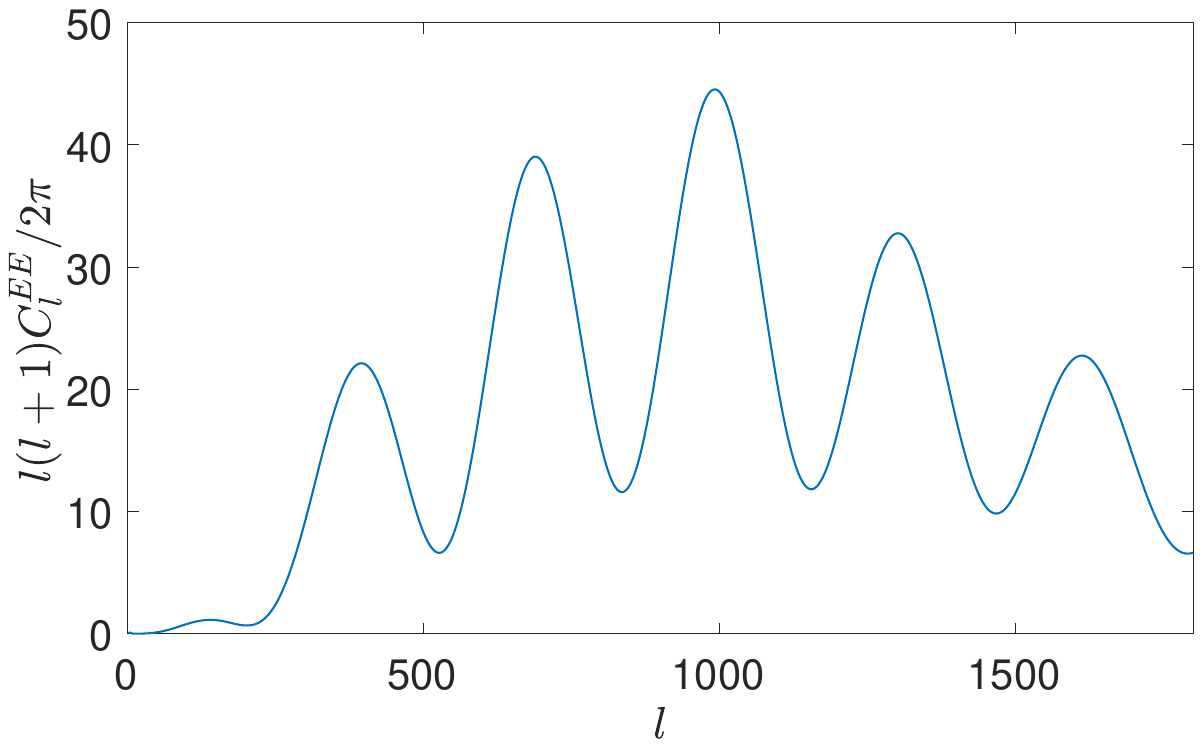}
\includegraphics[width=0.32\textwidth,trim = 70 100 90 130, clip]{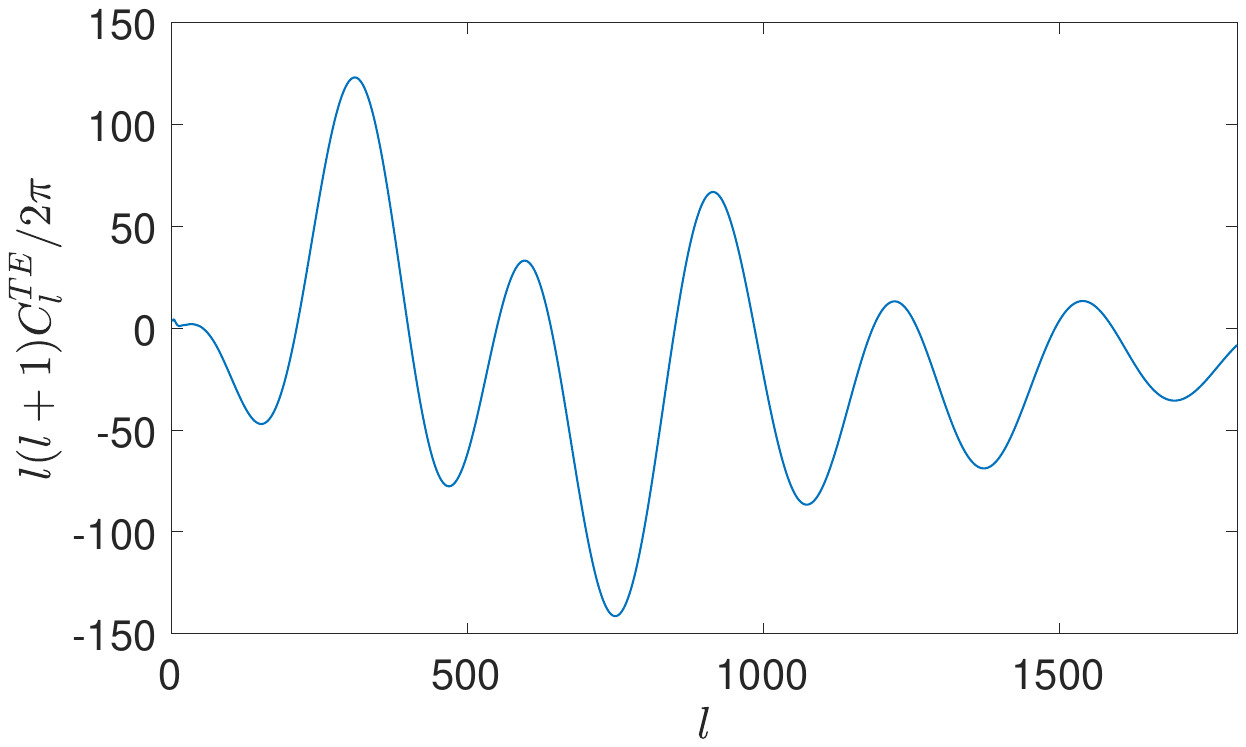}
\includegraphics[width=0.32\textwidth,trim = 70 100 90 130, clip]{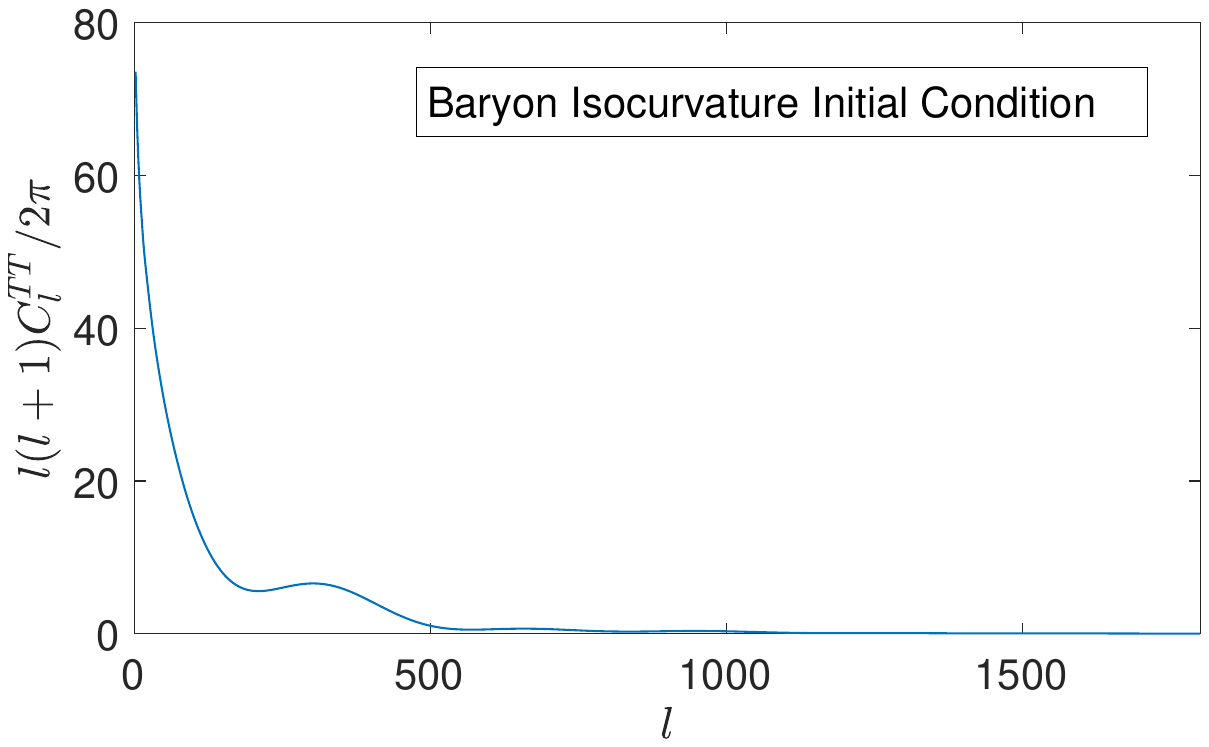}
\includegraphics[width=0.32\textwidth,trim = 70 100 90 130, clip]{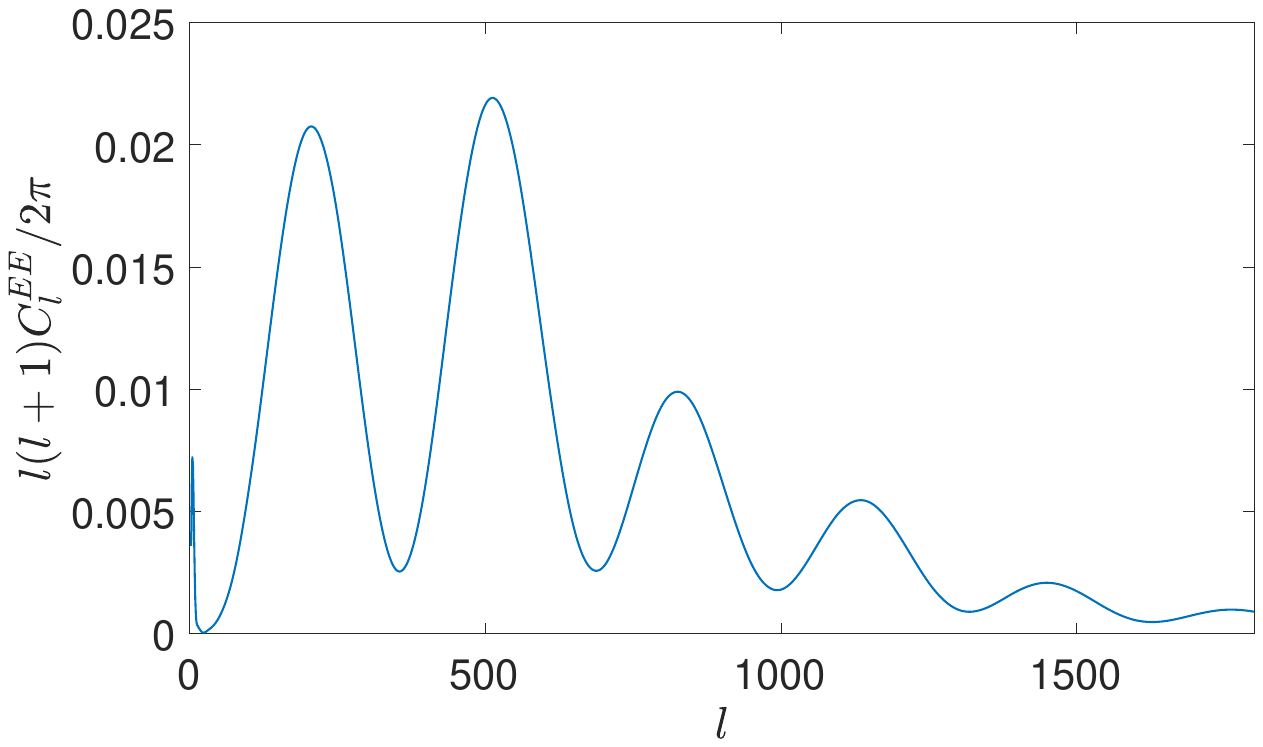}
\includegraphics[width=0.32\textwidth,trim = 70 100 90 130, clip]{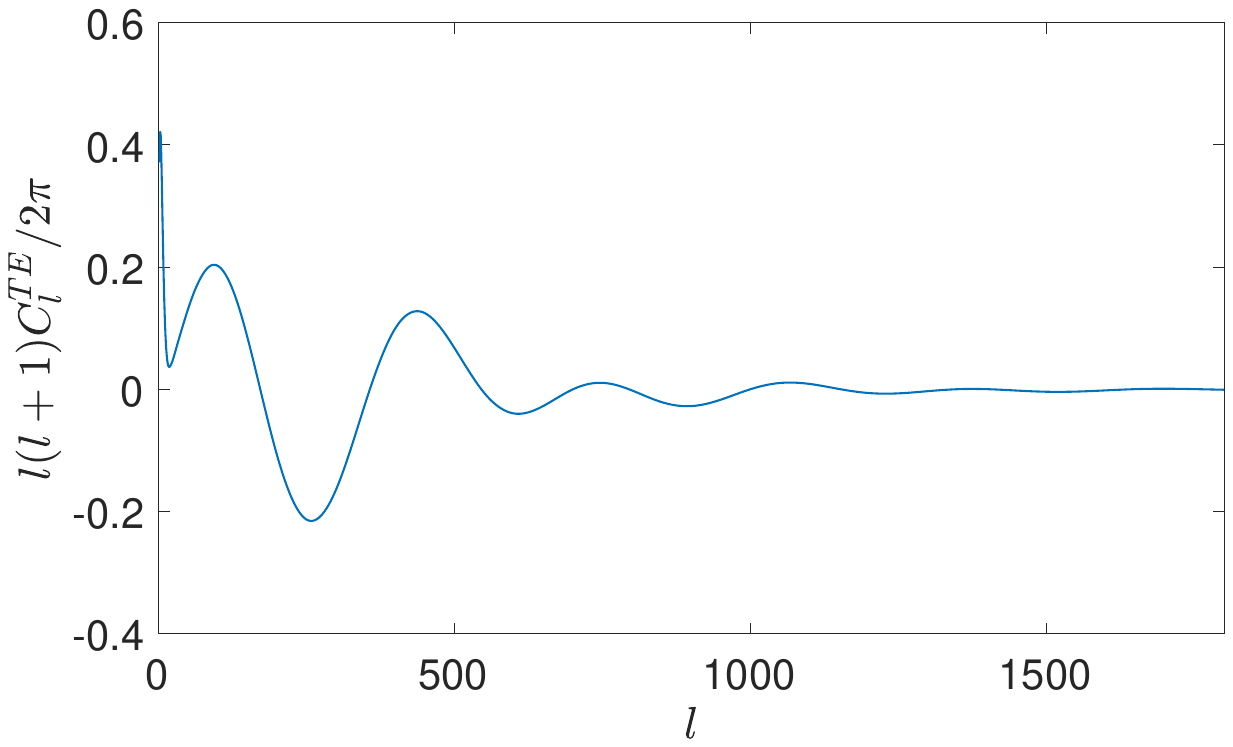}
\includegraphics[width=0.32\textwidth,trim = 70 100 90 130, clip]{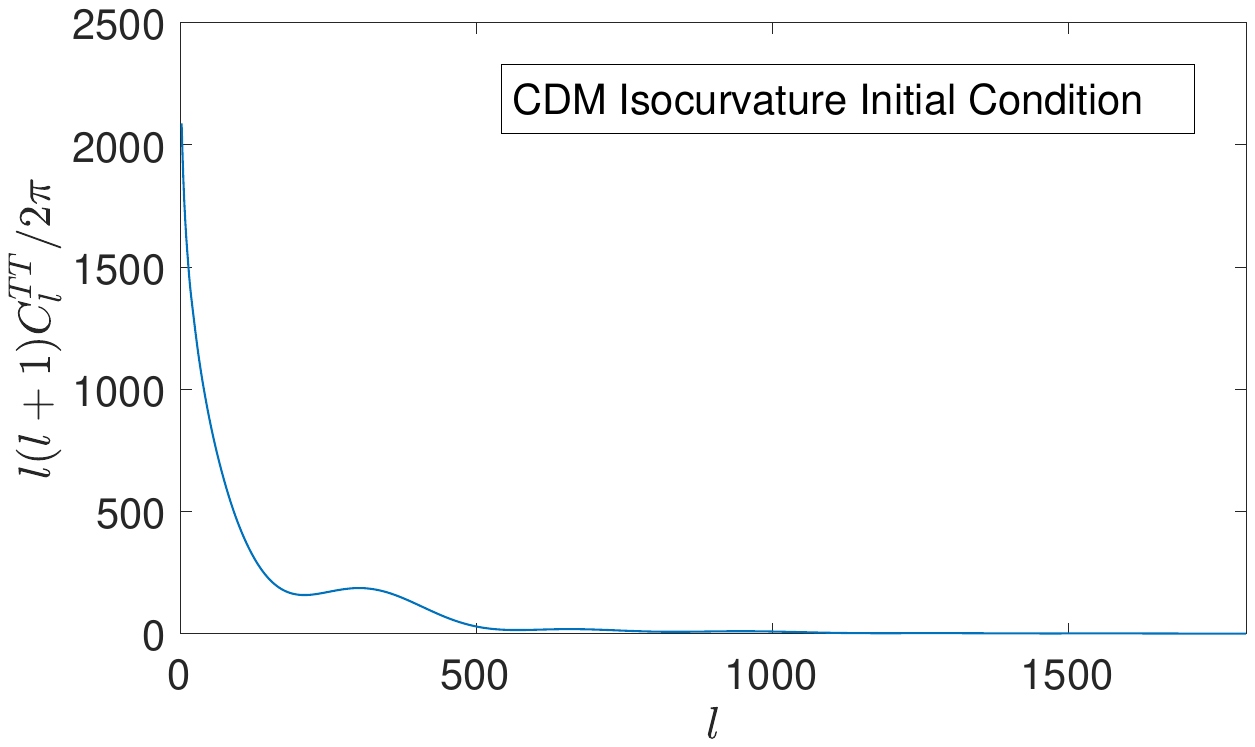}
\includegraphics[width=0.32\textwidth,trim = 70 100 90 130, clip]{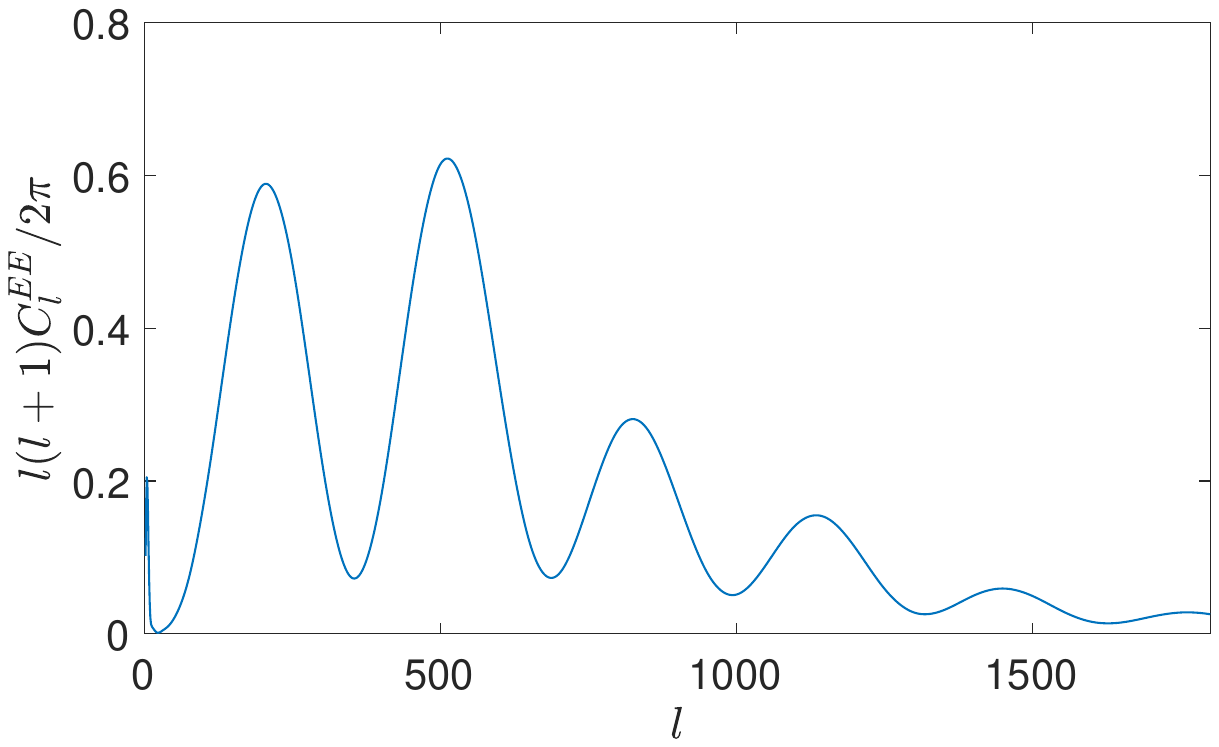}
\includegraphics[width=0.32\textwidth,trim = 70 100 90 130, clip]{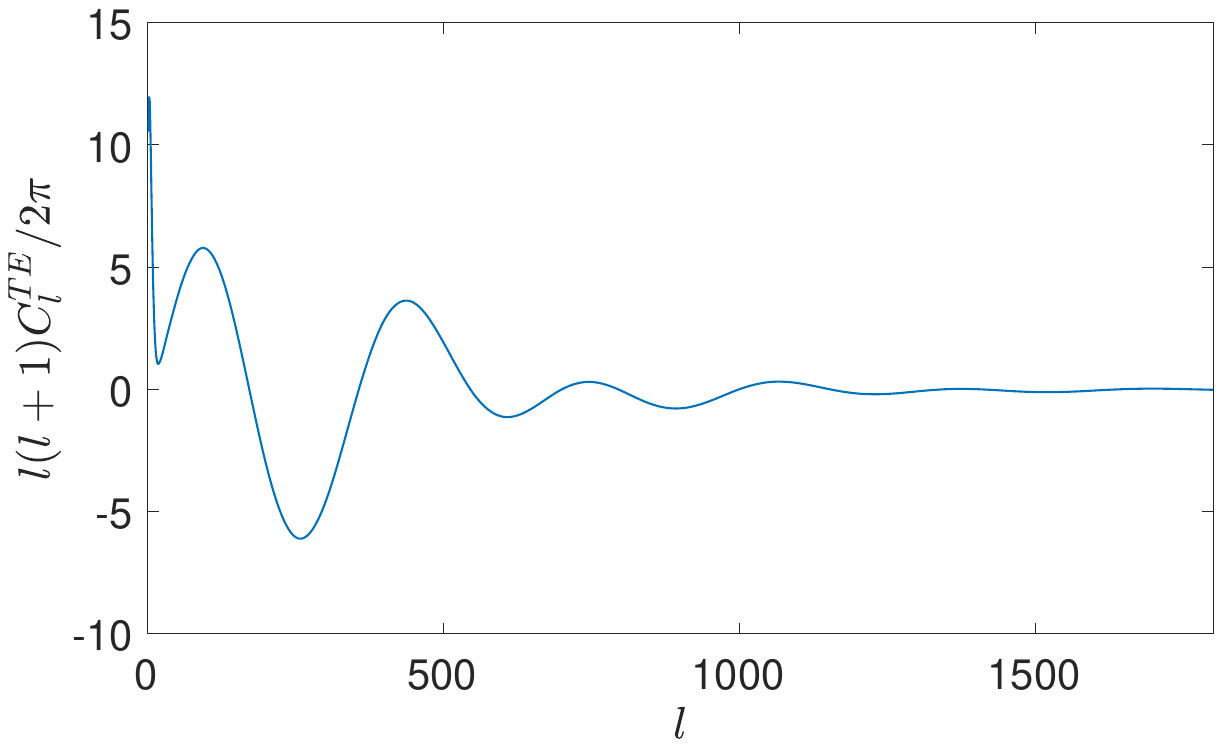}
\caption{\label{fig:ClAdiabatic}The plot shows unlensed CMB scalar power spectrum ($C_l$) for 
adiabatic, baryon Iso-curvature and CDM Iso-curvature initial conditions. We use $\Omega_b h^2 = 0.0223$, $\Omega_b h^2 = 0.1188$, $h = 67.74$ $\texttt{km/sec/Mpc}$, $n_s = 0.9667$, $\kappa = 0.08$. The plots show that the isocurvature CMB power spectrum decays at high $l$.}
\end{figure}

\subsubsection{Adiabatic initial conditions}
In this case $ \delta_b = \delta_c = \frac{3}{4}\delta_\nu = \frac{3}{4}\delta_\gamma$. At $\tau = 0$, we can set the density perturbations to zero and set $\eta$ to a nonzero quantity, i.e. 
\begin{eqnarray}
    &&\delta_b\vert_{\tau=0} = \delta_c\vert_{\tau=0} = \delta_\nu\vert_{\tau=0}=\theta_\nu\vert_{\tau=0} = 0 \nonumber \\
    &&\rm{Definable\;\; variable:} \;\;\;\;\; \eta
\end{eqnarray}

\noindent A few straight forward calculations show that deep inside the radiation dominated era at time $\tau$, the values of the perturbation variables are

\begin{eqnarray}
h & = & C(k\tau)^{2}\,,\qquad\eta=2C-\frac{5+4R_{\nu}}{6(15+4R_{\nu})}C(k\tau)^{2}\,,\;\;\;\;\; \delta_{\gamma}  =  -\frac{2}{3}h,\;\;\;
\;\;\;\delta_{c}=\delta_{b}=\frac{3}{4}\delta_{\nu}=\frac{3}{4}\delta_{\gamma}\,,\nonumber \\
\theta_{c} & = & 0\,,\;\;\quad\theta_{\gamma}=\theta_{b}=-\frac{1}{18}C(k^{4}\tau^{3})\,,\;\;\quad\theta_{\nu}=\frac{23+4R_{\nu}}{15+4R_{\nu}}\theta_{\gamma}\,,\;\;\;\;\;\;\;\;
\sigma_{\nu}  =  \frac{4C}{3(15+4R_{\nu})}(k\tau)^{2}\,.
\label{adiabaticinitialcond}
\end{eqnarray}

\noindent Here $R_\gamma$ and $R_\nu$ are the fractional contribution of the photon and neutrinos respectively in the early radiation dominated universe. For $N_\nu$ number of neutrino species we can define $R = \frac{7}{8} N_\nu\left(\frac{4}{11}\right)^\frac{4}{3}$, $R_\gamma = (1+R)^{-1}$ and $R_\nu = R(1+R)^{-1}$. 

\subsubsection{Baryon isocurvature initial conditions} 
The baryon isocurvature model was first proposed by Peebles \citep{Peebles1987,Peebles1987a} for explaining the galaxy peculiar velocity field
\citep{bucher2000general,langlois2003isocurvature,carrilho2018isocurvature}. For baryon isocurvature model

\begin{eqnarray}
    &&\delta_c\vert_{\tau=0} = \delta_\nu\vert_{\tau=0}=\theta\vert_{\tau=0} = \eta\vert_{\tau=0} = 0 \nonumber\\
    &&\rm{Definable\;\; variable:} \;\;\;\;\; \delta_b
\end{eqnarray}
\noindent The initial perturbations can be written as 
\begin{eqnarray}
h & = & {\mathcal Y}_r\times\left(\frac{1}{1+\Omega_{0,c}/\Omega_{0,b}}-\frac{{\mathcal Y}_r}{2}\right)\qquad
\delta_{b}  =  1-\frac{1}{2}h,\qquad \delta_{\gamma}=-\frac{2}{3}h\,,\delta_{c}=-\frac{1}{2}h\,,\delta_{\nu}=\delta_{\gamma}\nonumber\\
\theta_{c} & = & 0\,,\quad\theta_{\gamma}=\theta_{b}=\theta_{\nu}=-\frac{h}{12}k^{2}\tau\,,\qquad
\eta  =  -\frac{1}{6}h
\label{baryonisocurvature}
\end{eqnarray}
\noindent where ${\mathcal Y}_r = \frac{\rho_m}{\rho_r}$, the ratio of matter density to radiation density at that particular epoch.  In the second row of Fig.~\ref{fig:ClAdiabatic}, we show the temperature and the polarization power spectrum from the baryon isocurvature model.             

\subsubsection{CDM isocurvature initial conditions} 

As we have discussed before, the CDM velocity mode is set to $0$ to remove the extra gauge mode. CDM density isocurvature modes can arise in different inflationary models such as two field inflation or double inflation etc. In CDM isocurvature mode \citep{bucher2000general,langlois2003isocurvature, carrilho2018isocurvature}

\begin{eqnarray}
    &&\delta_b\vert_{\tau=0} = \delta_\nu\vert_{\tau=0}=\theta\vert_{\tau=0} = \eta\vert_{\tau=0} = 0 \nonumber \\
    &&\rm{Definable \;variable:} \;\;\;\;\; \delta_c
\end{eqnarray}
\noindent Therefore, substituting these values in the perturbation equations we can get
\begin{eqnarray}
h & = & {\mathcal Y}_r\times\left(\frac{1}{1+\Omega_{0,b}/\Omega_{0,c}}-\frac{{\mathcal Y}_r}{2}\right)\, \qquad
\delta_{c}  =  1-\frac{1}{2}h\,,\qquad\delta_{\gamma}=-\frac{2}{3}h\,,\qquad\delta_{b}=-\frac{1}{2}h\nonumber\\
\delta_{\nu} &=& \delta_{\gamma}\,,\qquad\theta_{c}  =  0\,,\quad\theta_{\gamma}=\theta_{b}=\theta_{\nu}=-\frac{h}{12}k^{2}\tau\,,\qquad
\eta  =  -\frac{1}{6}h
\label{CDMisocurvature}
\end{eqnarray}

 \begin{figure}
\centering
\includegraphics[width=0.32\textwidth,trim = 70 100 90 130, clip]{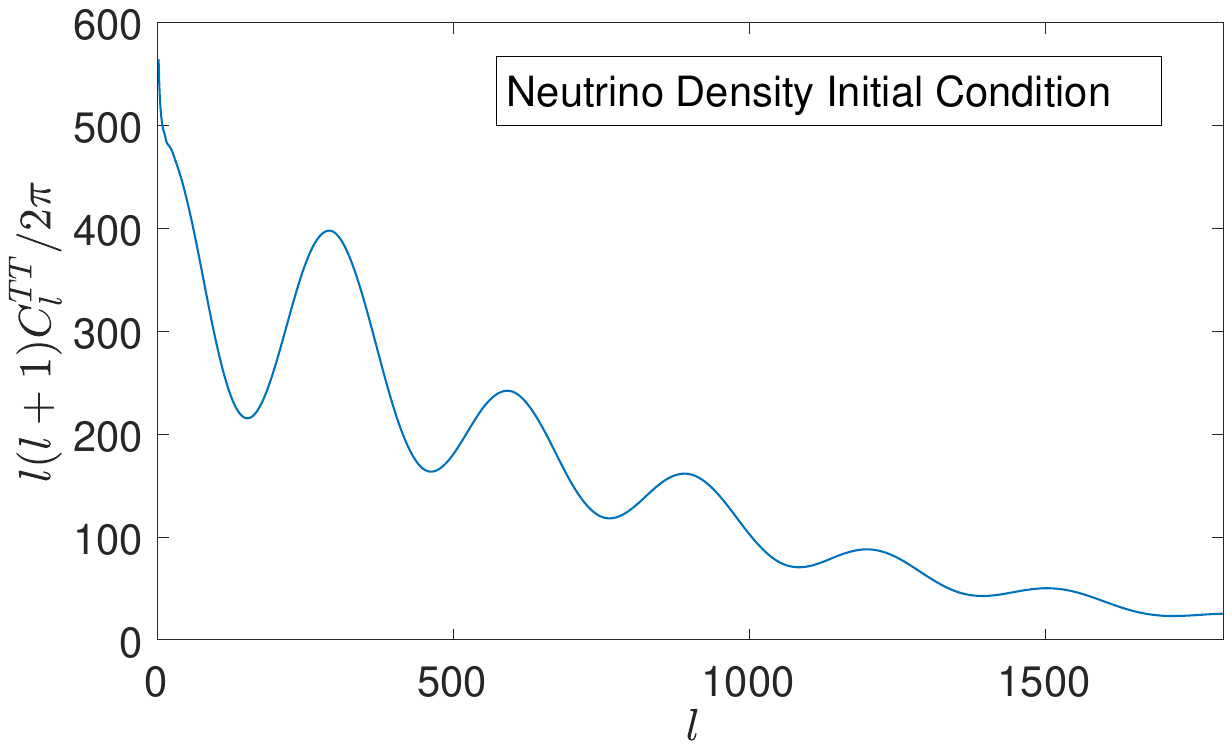}
\includegraphics[width=0.32\textwidth,trim = 70 100 90 130, clip]{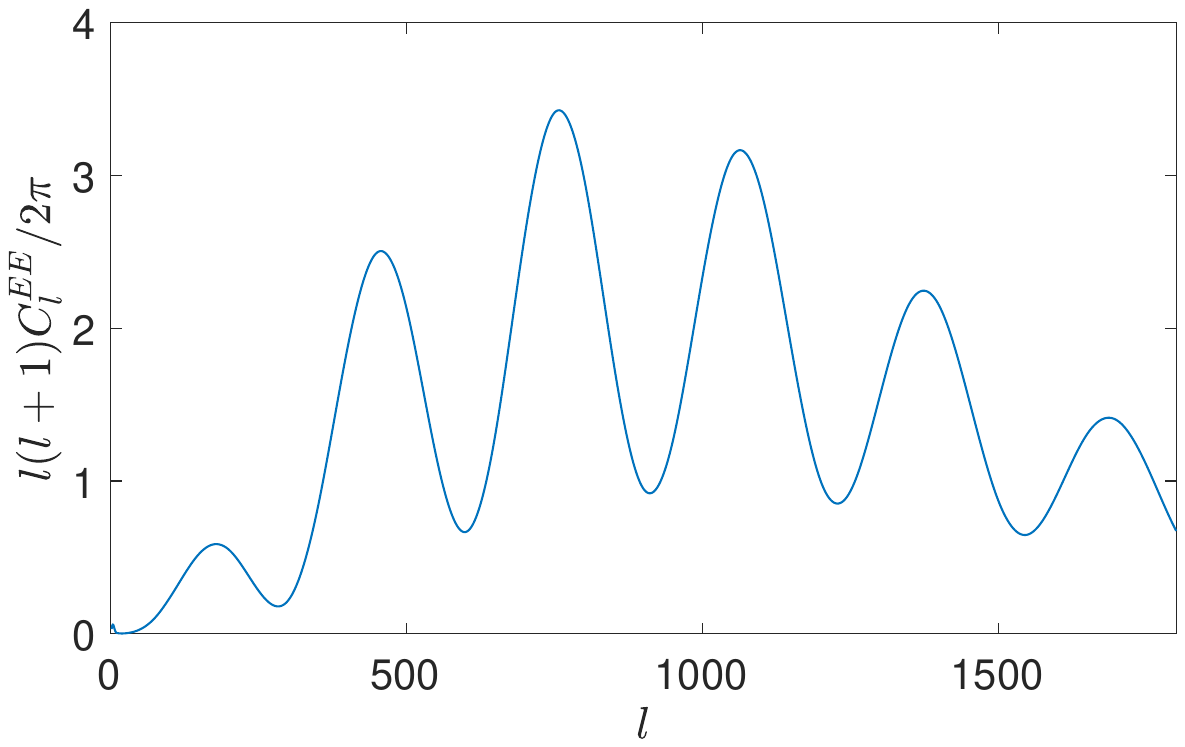}
\includegraphics[width=0.32\textwidth,trim = 70 100 90 130, clip]{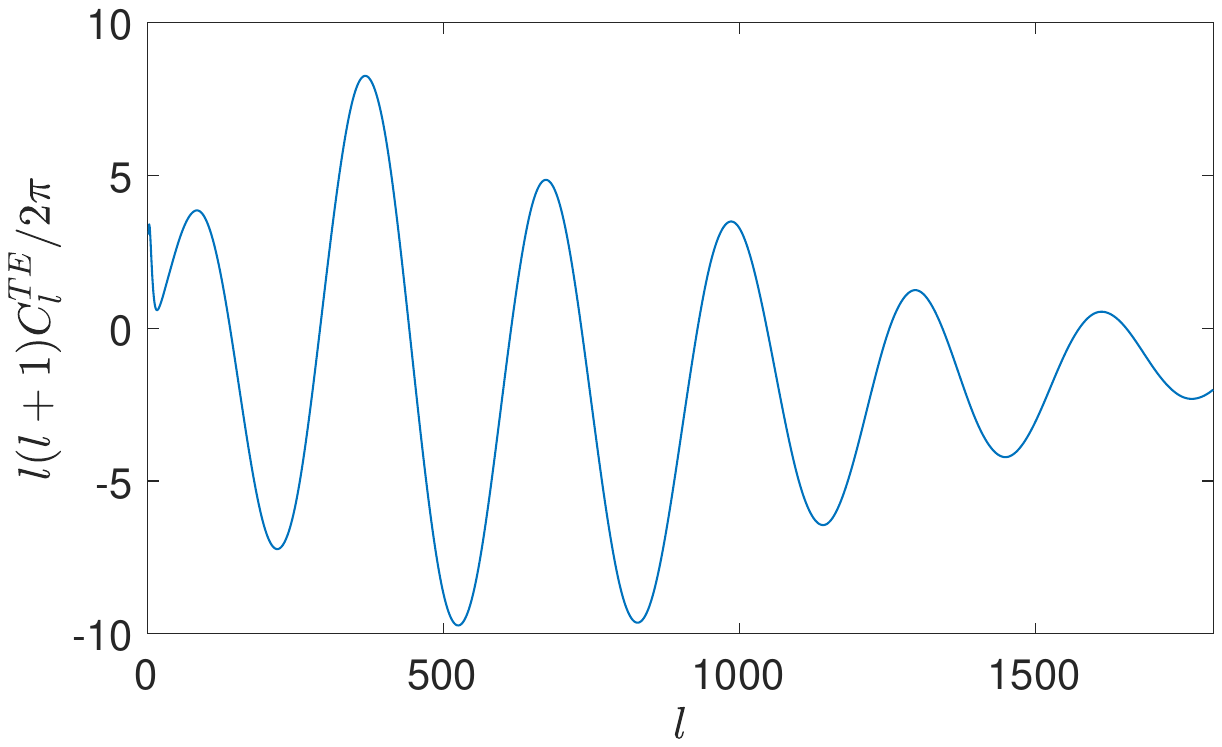}
\includegraphics[width=0.32\textwidth,trim = 70 100 90 130, clip]{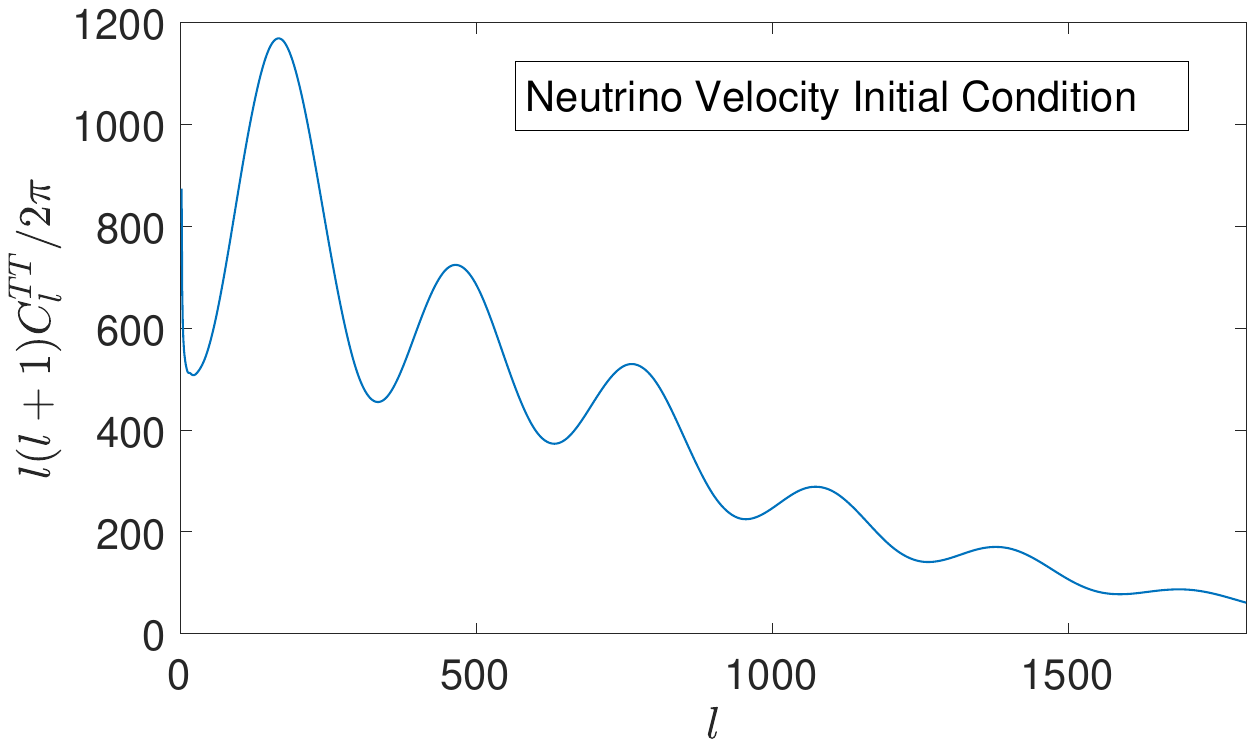}
\includegraphics[width=0.32\textwidth,trim = 70 100 90 130, clip]{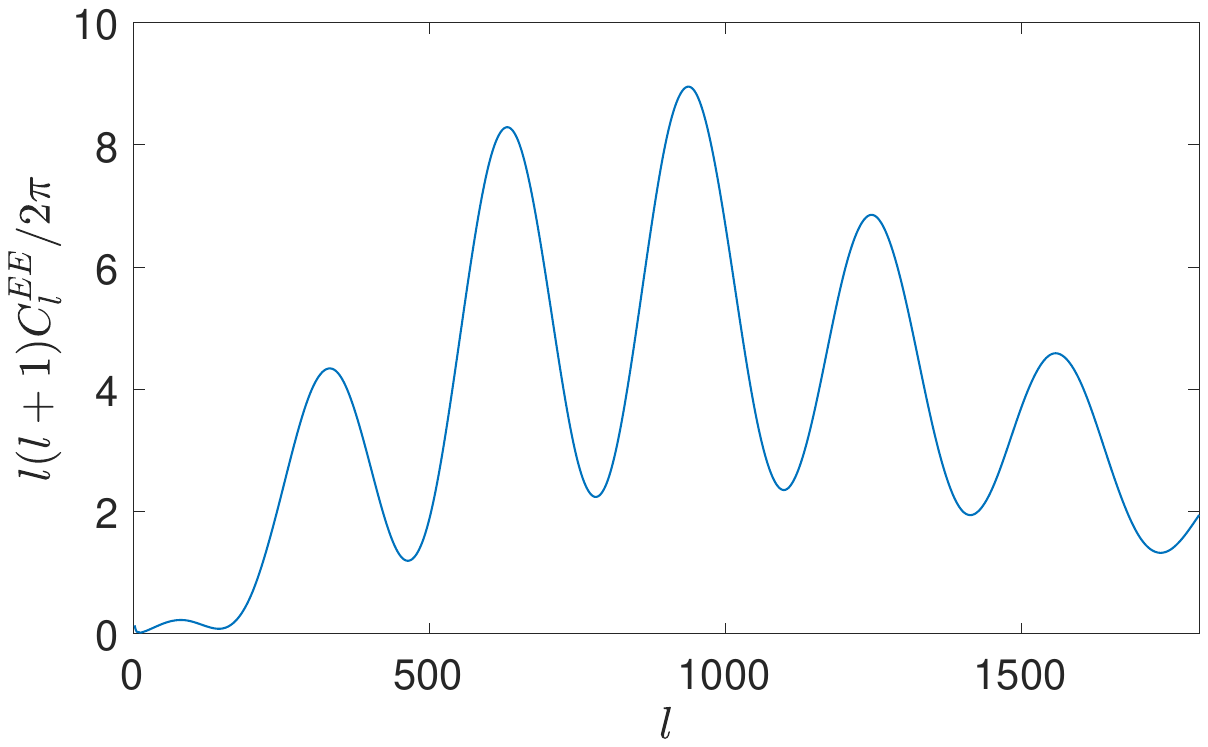}
\includegraphics[width=0.32\textwidth,trim = 70 100 90 130, clip]{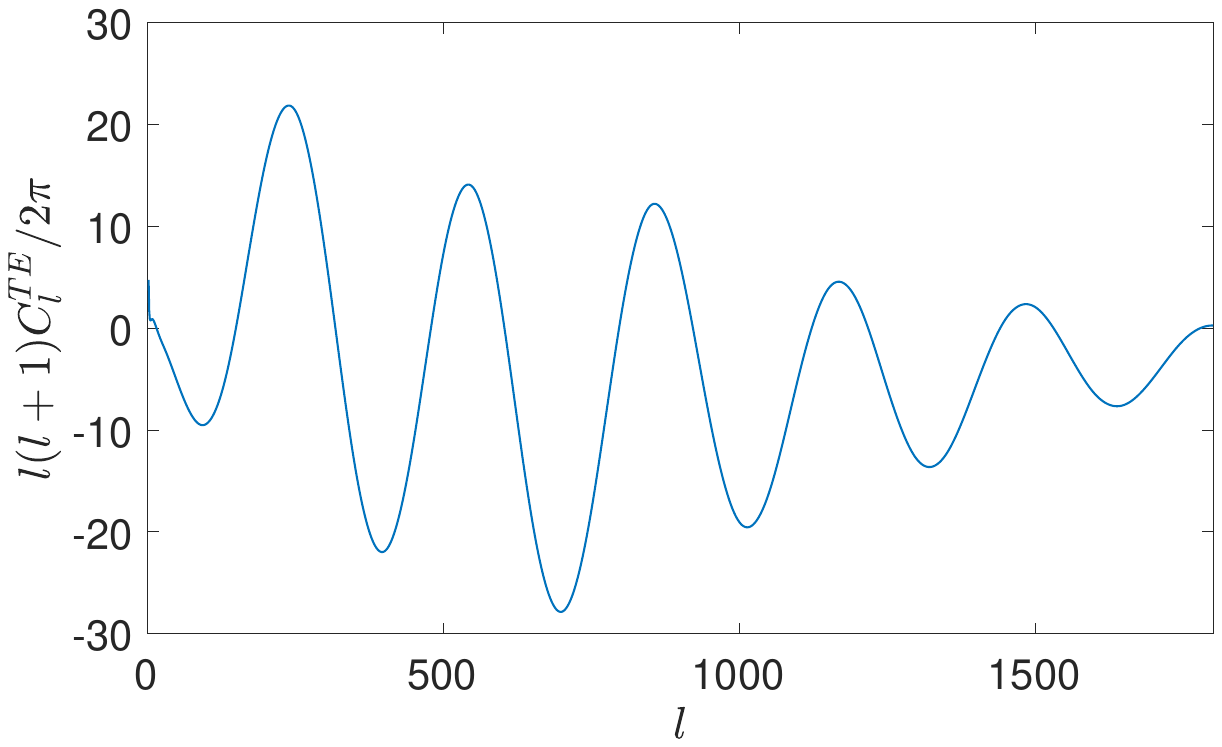}
\caption{\label{fig:ClIso}The plot shows the unlensed CMB scalar power spectrum ($C_l$) for 
neutrino density and neutrino velocity isocurvature initial conditions. We use $\Omega_b h^2 = 0.0223$, $\Omega_b h^2 = 0.1188$, $h = 67.74$ $\texttt{km/sec/Mpc}$, $n_s = 0.9667$, $\kappa = 0.08$. The plots show that the neutrino isocurvature TT power spectrum decays at high $l$.}
\end{figure}

\subsubsection{Neutrino density isocurvature} 
For neutrinos we will have both the neutrino velocity and density isocurvature modes. 
\begin{eqnarray}
    &&\delta_c\vert_{\tau=0} = \delta_b\vert_{\tau=0}=\theta\vert_{\tau=0} = \eta\vert_{\tau=0} = 0 \nonumber\\
    &&\rm{Definable\;\; variable:} \;\;\;\;\; \delta_\nu
\end{eqnarray}
\noindent For density modes we can start with an uniform energy density, with the total photon and neutrino density unperturbed. When the modes enter the horizon, the photon behaves as a perfect fluid while the neutrinos  freestream. Solving the set of perturbation equations with the above initial conditions we get
\begin{eqnarray}
h &=& \frac{\Omega_{0,b}R}{10}k^2\tau^3\qquad
\delta_c = -\frac{1}{2}h\qquad
\delta_b=\frac{1}{8}Rk^2\tau^2\qquad
\delta_\gamma = -R + \frac{4}{3}\delta_b\qquad
\delta_\nu = -\frac{1}{R}\delta_\gamma \nonumber \\
\theta_c &=& 0 \qquad
\theta_\gamma = \theta_b = -\frac{1}{4}Rk^2\tau+\frac{3\Omega_{0,b}R_\nu}{4R_\gamma^2}k^2\tau^2\qquad
\theta_\nu=\frac{1}{4}k^2\tau\qquad 
\eta=-\frac{R_{\nu}}{6(15+4R_{\nu})}k^{2}\tau^{2}\,\nonumber \\
\sigma_\nu &=& \frac{1}{2(15+4R_{\nu})}k^{2}\tau^{2}
\label{neutrinodelsity}
\end{eqnarray}

\subsubsection{Neutrino velocity isocurvature model} 
Unlike other terms, neutrinos can have velocity isocurvature modes. However, we need to start with a total of $0$ momentum, or otherwise the mode will decay. This can be done by carefully choosing to match the momentum of the neutrinos and photons in the early universe. For this particular case the definable variables are 
\begin{eqnarray}
    &&\delta_c\vert_{\tau=0} = \delta_\nu\vert_{\tau=0}=\delta_b\vert_{\tau=0} = \eta\vert_{\tau=0} = 0 \nonumber\\
    &&\rm{Definable\;\; variable:} \;\;\;\;\; \theta_\nu
\end{eqnarray}
Solving the perturbation equations with the above initial conditions gives us 
\begin{eqnarray}
h &=& \frac{3\Omega_{0,b}R}{2}k\tau^2\qquad
\delta_c = -\frac{h}{2}\qquad
\delta_b = Rk\tau - \frac{(3+2R)}{2}h\qquad
\delta_\gamma = \frac{4}{3}\delta_b\qquad
\delta_\nu = -\frac{4}{3}k\tau - \frac{2}{3}h \nonumber\\
\theta_c &=& 0 \qquad
\theta_\gamma = \theta_b = -Rk + 3\Omega_b R (1+R) k\tau + 2(1+R)\left(1-3\Omega_{0,b}(1+R)\right)h + \frac{R}{6}k^3\tau^2 \nonumber\\
\theta_\nu &=& k-\frac{(9+rR_\nu)}{3(5+4R_\nu)}k^3\tau^2\qquad
\sigma_\nu = \frac{4}{3(5+4R_\nu)}k\tau+\frac{16R_\nu}{(5+4R_\nu)(15+4R_\nu)}k\tau^2 \nonumber\\
F_\nu3 &=& \frac{4}{7(5+4R_\nu)}k^2\tau^2\qquad 
\eta = -\frac{4R_\nu}{3(5+4R_\nu)}k\tau+\left(\frac{-\Omega_{0,b}R}{4} + \frac{20R_\nu}{(5+4R_\nu)(15+4R_\nu)}\right)k\tau^2
\label{neutrinovelocity}
\end{eqnarray}

The neutrino velocity and density modes are shown in Fig.~\ref{fig:ClIso}. Both these modes decay at high $l$. However, the decay rate is much slower from the CDM or baryon isocurvature modes. 
 
\begin{figure}
\centering
\includegraphics[width=0.48\textwidth,trim = 300 300 360 300, clip]{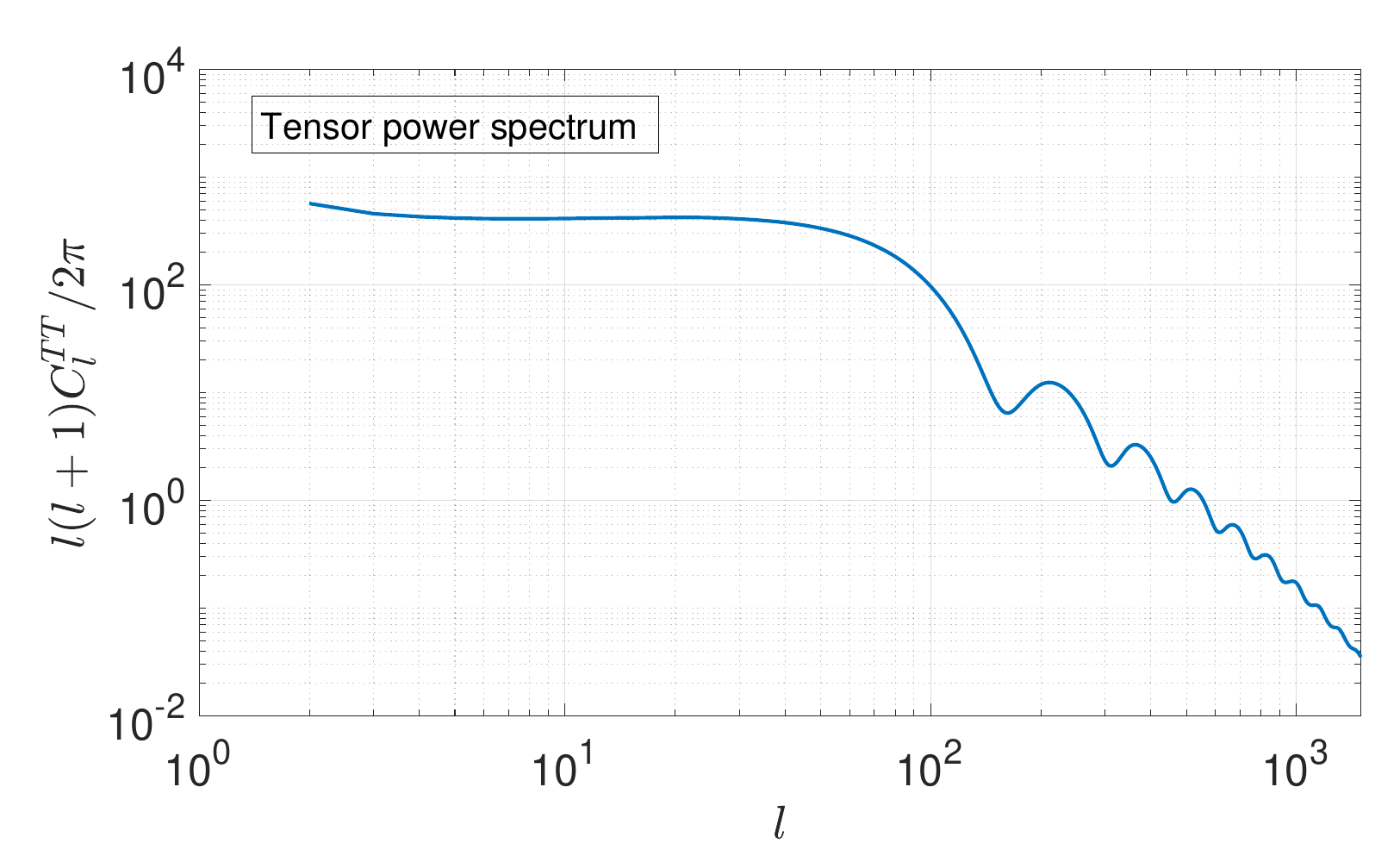}
\includegraphics[width=0.48\textwidth,trim = 300 300 360 300, clip]{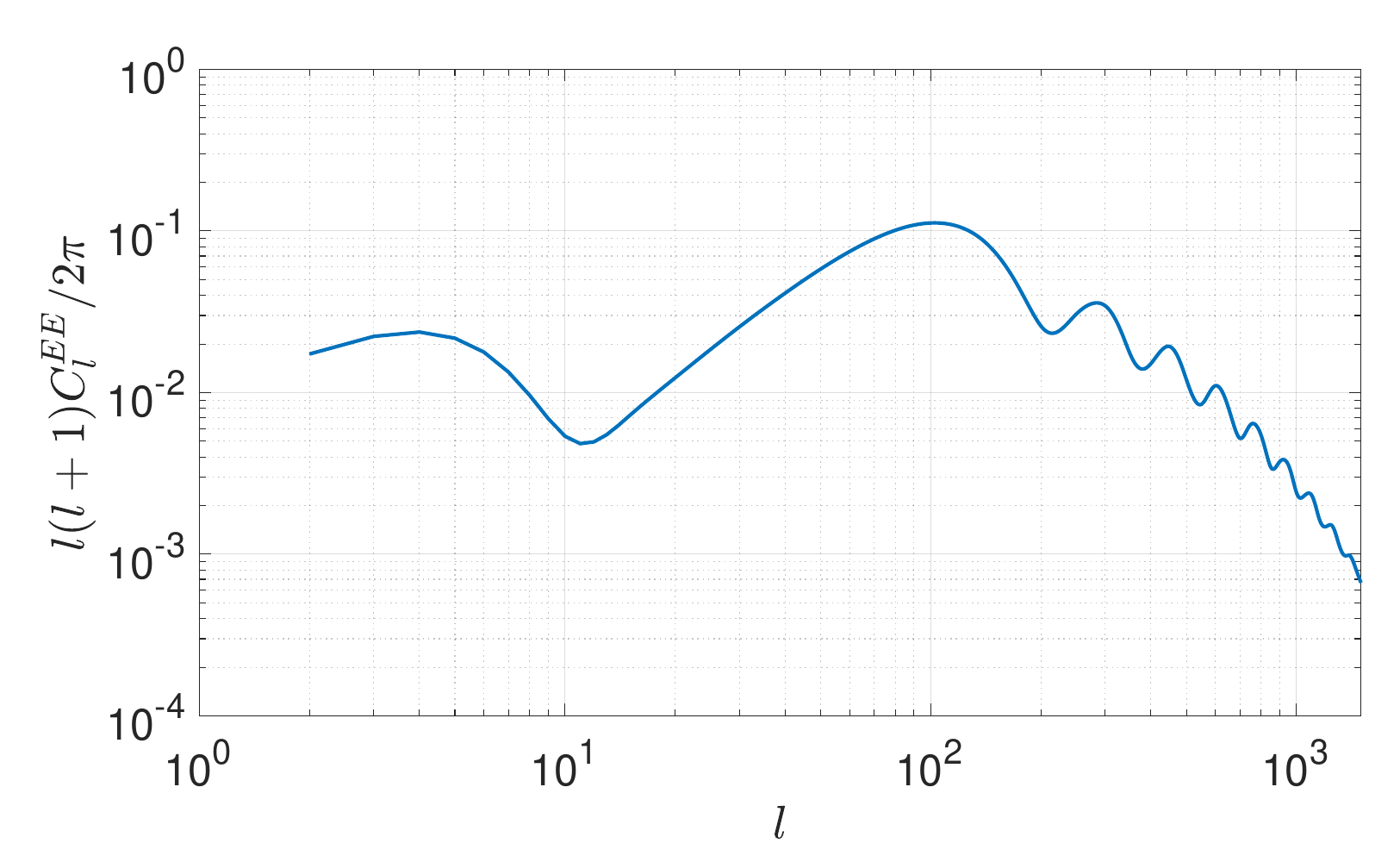}
\includegraphics[width=0.48\textwidth,trim = 300 300 360 300, clip]{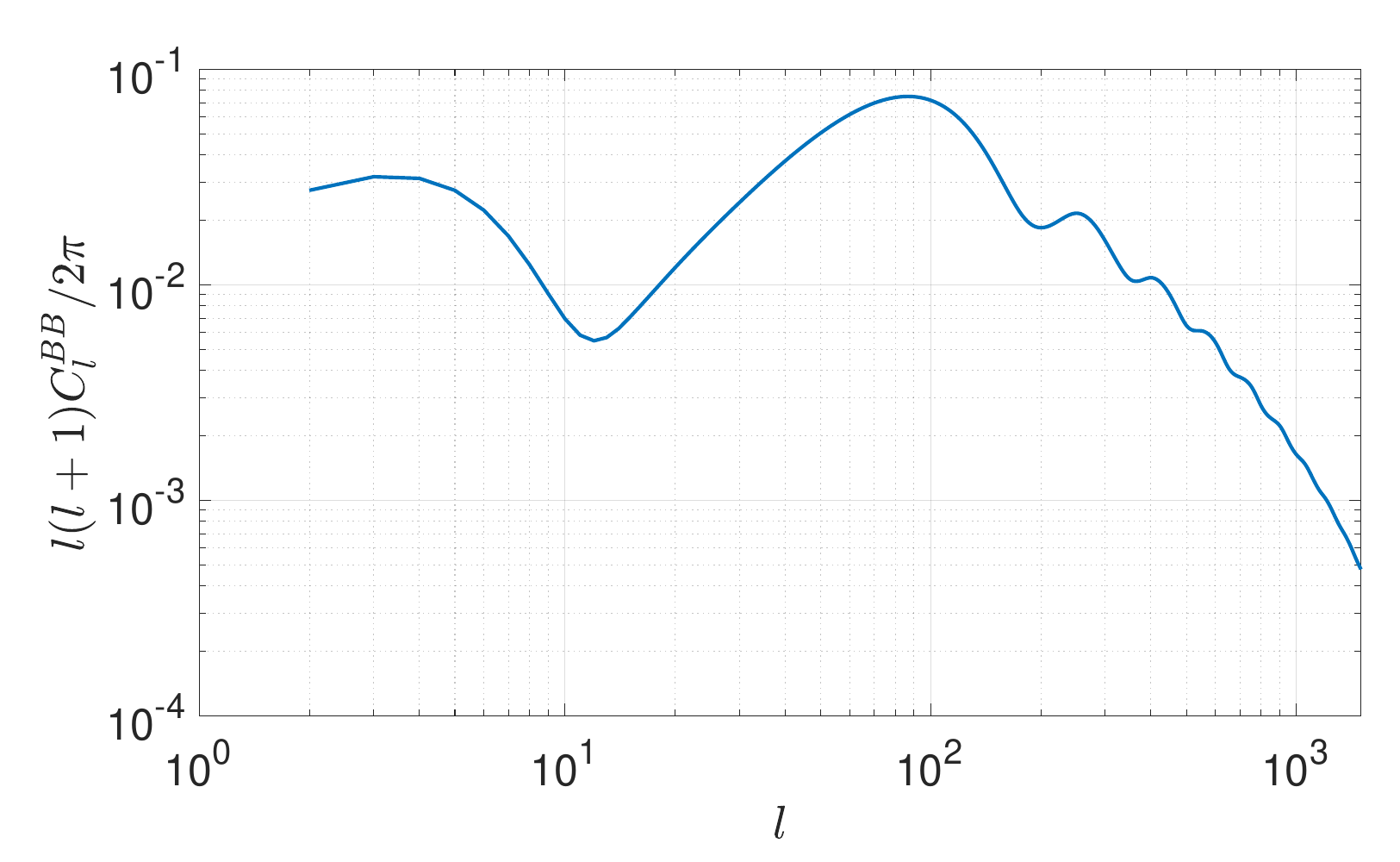}
\includegraphics[width=0.48\textwidth,trim = 300 300 360 300, clip]{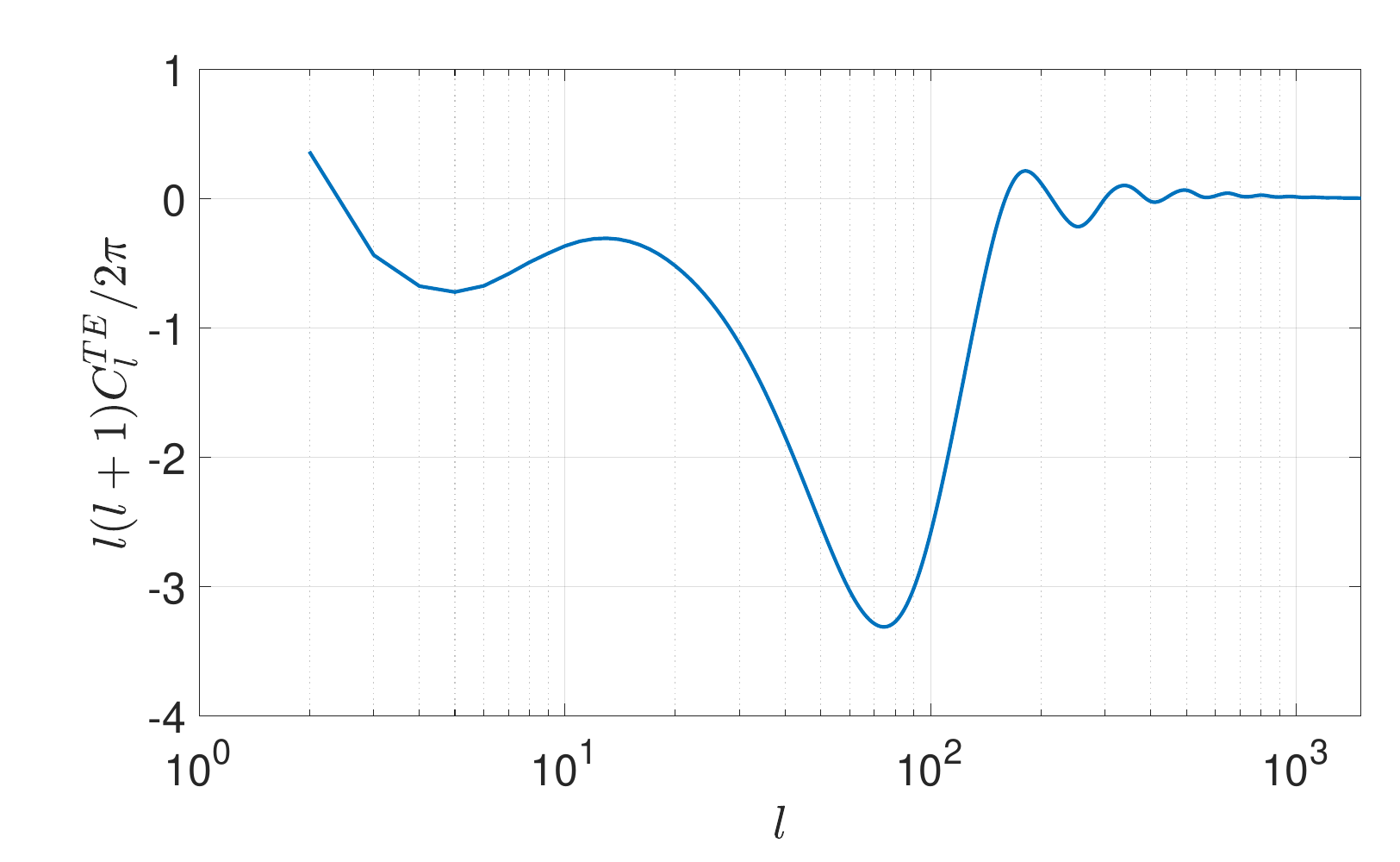}
\caption{\label{fig:Tensor}The plot shows the unlensed CMB tensor power spectrum ($C_l$). We use $\Omega_b h^2 = 0.0223$, $\Omega_b h^2 = 0.1188$, $h = 67.74$ $\texttt{km/sec/Mpc}$, $n_t = 0.04$, $\kappa = 0.08$. As $C_l^{TE}$ has negative values we plot the $y$-axis in linear scale.}
\end{figure}

\subsection{Tensor perturbation}
For the tensor perturbation, the anisotropic stress term in Eq.~\ref{tensorperturbation} has a very small contribution, and the modes from the anisotropic stress terms decays~\citep{Weinberg2008cosmology}. The only initial condition for tensor power spectra can be written as

\begin{equation}
h = 1, \;\;\;\;\;\;\;\;  \dot{h} = 0, \;\;\;\;\;\; \delta^t_\gamma = \theta^t_\gamma = \delta^t_\nu = \theta^t_\nu = 0 
\label{tensorinitialcondition}
\end{equation}
We show the tensor power spectrum in Fig.~\ref{fig:Tensor}. All the power spectra are plotted in the log-log scale except the $C_l^{TE}$ which is plotted in the log-linear scale, as it contains the negative values.

\subsection{\label{initialcondset}Setting the initial conditions}
We set the initial conditions when the wave is far outside the
horizon by taking $\tau_{i}^{hor}(k)=0.001/k$,
where $\tau_{i}^{hor}(k)$ is the point where the initial condition is
set for the wave number $k$ from the horizon crossing cut off. Secondly,
we set the initial conditions well within the tight coupling era and
the radiation dominated universe. This is confirmed by considering
$\tau_{i}(k)=\min(\tau_{i}^{hor}(k),0.1)$. Fig.~\ref{fig:initialcond} shows the positions where
the initial conditions are set for different $k$ modes (brown
line). It can be seen that for smaller $k$ modes, $\tau_{i}(k)$ is $0.1$,
and for higher $k$, which crosses the horizon earlier, the initial
conditions are set at $\tau_{i}^{hor}(k)$. If massive neutrinos are
present, then we set the initial conditions in an era where the massive
neutrinos are highly relativistic, which is given by $\tau_{h}=\left(1-3.0/m_{\nu}\right)/\dot{a}_{rad}$.
In that case, we choose $\tau_{i}(k)=\min\left(\tau_{h},\;\;\tau_{i}^{hor}(k),\;\;0.1\right)$ as the initial conditions of mode $k$. 

\section{Calculating the CMB power spectrum and BipoSH coefficients\label{Section-6}}
In Sec.~\ref{photonbrightnessfunction},  we calculate the photon multipole functions, $\Delta_{Tl}(k,\tau_0)$ and $\Delta_{Pl}(k,\tau_0)$. We can obtain the CMB scalar power spectrum just by convolving the multipole brightness functions with the primordial inflationary power spectrum, i.e.
\begin{equation}
C^{TT}_{l} = (4\pi)^{2}\int k^{2}\, \mathrm{d}k\, P^{s}(k)\left[\Delta_{Tl}(k,\tau_0)\right]^2\,, 
\end{equation}

\noindent Analogous expressions can also be obtained for the polarization and the tensor perturbations\citep{Ma1995}.

In the above expression the  $P^s(k)$ is the power spectrum for the primordial density fluctuations $\zeta(\vec{k})$. It is given by 
\begin{equation}
    \langle \zeta(\vec{k})\zeta(\vec{k'}) \rangle =P^s(k)\delta^3(\vec{k} - \vec{k}')
\end{equation}

Here we consider that $P^s(k)$ depends on the magnitude of $k$ and is completely rotationally invariant. However, in different anisotropic inflation models, the rotational in-variance is not preserved~\citep{ackerman2007imprints,groeneboom2010bayesian}. This leads to the statistical isotropy (SI) violation in the CMB sky. In presence of SI violation, the CMB angular power spectra is not sufficient and we need the BipoSH coefficients~\citep{hajian2003measuring,Das2014a,Joshi2012} to represent the full sky statistics. Along with the CMB power spectrum, \texttt{CMBAns} can also calculate the BipoSH coefficients for the anisotropic inflation model. 

The anisotropic inflation term is also important in the context of different topological models of the universe, where the promordial power spectrum will be discrete and direction dependent function of $k$. However, provided the size of the manifold is much larger than the distance to the last scattering surface,  we can approximate it as a contineous function. The detail calculation of the topological model is a future project and beyond the scope of the present paper. 


In accordance with~\cite{ackerman2007imprints}, the primordial power spectrum for the anisotropic inflation model is taken as

\begin{equation}
    P^s(\vec{k}) = P^s(k)(1+g(\hat{k}.\hat{n})^2)
\end{equation}

In any given coordinate system we can write this expression as $P^s(\vec{k}) = P^s(k)[\sum_{lm}g_{lm}(k)Y_{lm}(\hat{k})]$, where $g_{lm}$'s are the spherical harmonic coefficients of $g$ and $g_{00} = 1$. 

\begin{figure}
\centering
\includegraphics[width=0.48\textwidth,trim = 50 80 50 110, clip]{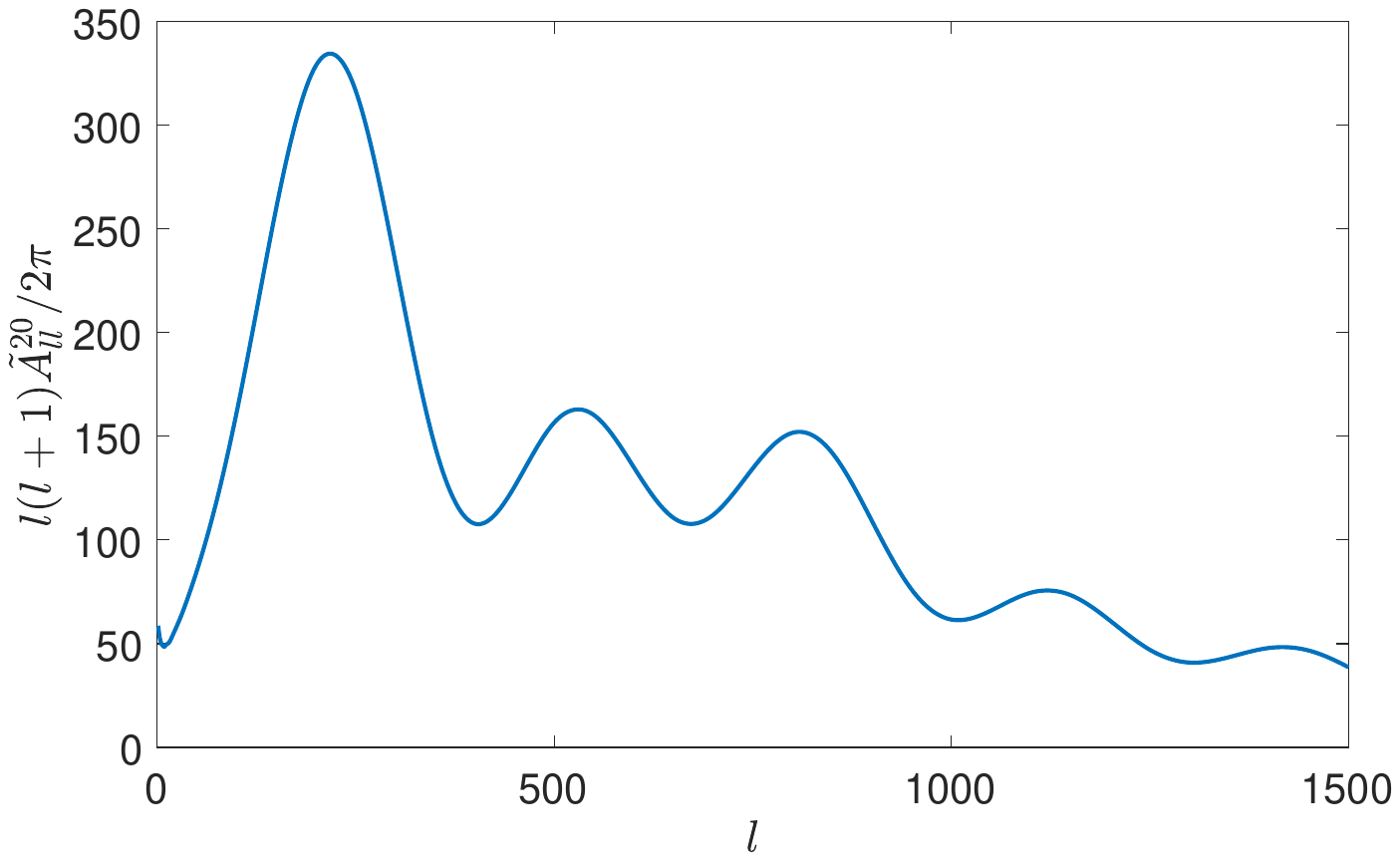}
\includegraphics[width=0.48\textwidth,trim = 50 80 50 110, clip]{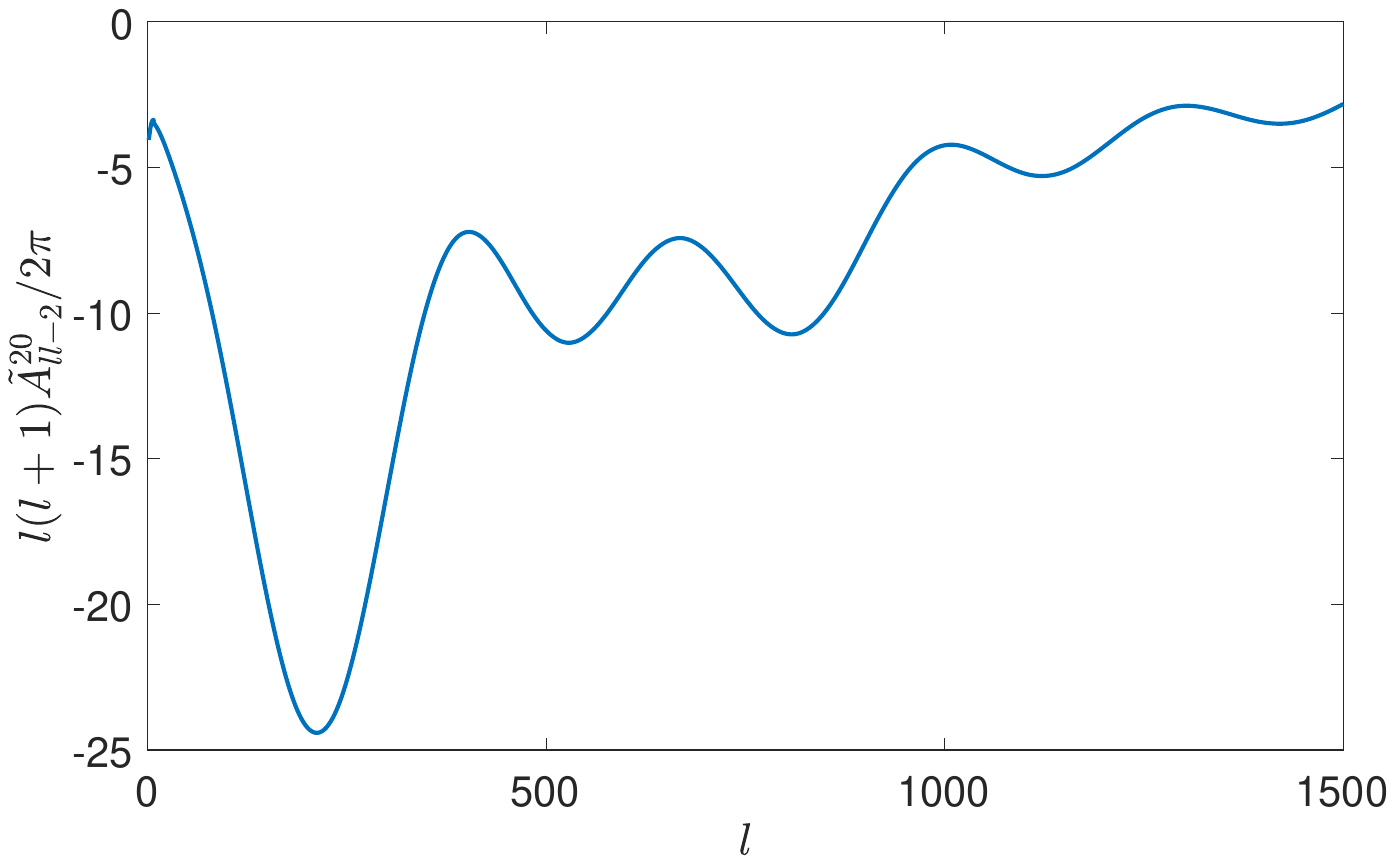}
\caption{\label{fig:BipoSH} BipoSH coefficients $\tilde{A}^{LM}_{ll}$ and $\tilde{A}^{LM}_{ll-2}$ are calculated for standard $\Lambda$CDM parameter and $g^{20} = 1.5$. The BipoSH coefficients are normalized in WMAP format.}
\end{figure} 

Few algebraic manipulation gives us

\begin{eqnarray}
    \langle a_{lm} a^*_{l'm'}\rangle &=& (4\pi)^{2}\int k^{2}\, \mathrm{d}k\, P^{s}(k)\Delta_{Tl}(k,\tau_0)\Delta_{Tl_1}(k,\tau_0)\sum_{l_2m_2}g_{l_2m_2}(k)\int Y_{lm}(\hat{k})Y^*_{l_1m_1}(\hat{k})Y_{l_2m_2}(\hat{k})\mathrm{d}\Omega_{\hat{k}} \nonumber \\
    &=&  (4\pi)^{2}\int k^{2}\, \mathrm{d}k\, P^{s}(k)\Delta_{Tl}(k,\tau_0)\Delta_{Tl_1}(k,\tau_0) \nonumber  \\
    && \times\sum_{l_2m_2}g_{l_2m_2}(k) (-1)^{m_1} \Bigg[\frac{1}{\sqrt{4\pi}}\sqrt{\frac{(2l+1)(2l_1+1)}{(2l_2+1)}}C^{l_2m_2}_{lml_1-m_1}C^{l_2 0}_{l0l_10}\Bigg]
\end{eqnarray}

\noindent Here $a_{lm}$'s are the coefficients of the spherical harmonics expansion of the CMB temperature fluctuations. In case of standard inflation all the $g_{lm}$s except $g_{00}$ are zero. This leads to the CMB angular power spectrum. However, in presence of the isotropy violation, $\langle a_{lm} a^*_{l'm'}\rangle \neq C_l\delta_{ll'}\delta_{mm'} $, i.e. we will have signal in the off-digonal terms of the covariance matrix. In such a case, we can expand $\langle a_{lm} a^*_{l'm'}\rangle = (-1)^{m'}\sum_{LM}C^{LM}_{lml'-m'}A^{LM}_{ll'}$. Here $C^{LM}_{lml'm'}$ are the Clebsch Gordan coefficients and $A^{LM}_{ll'}$s are the BipoSH coefficients. After some algebraic manipulation, we can calculate the Biposh coefficients as 

\begin{eqnarray}
A^{LM}_{ll'} &=& \int k^2 \mathrm{d}k P^{s}(k)\Delta_{Tl}(k,\tau_0)\Delta_{Tl'}(k,\tau_0)\epsilon^{LM}_{ll'} \\
\epsilon^{LM}_{ll'} &=& \sum_{mm_1}\sum_{l_2m_2}g_{l_2m_2}(k)  \Bigg[\frac{1}{\sqrt{4\pi}}\sqrt{\frac{(2l+1)(2l_1+1)}{(2l_2+1)}}C^{l_2m_2}_{lml_1-m_1}C^{l_2 0}_{l0l_10}\Bigg]C^{LM}_{lml_1-m_1} \nonumber\\
&=& g_{LM}(k) \Bigg[\frac{1}{\sqrt{4\pi}}\sqrt{\frac{(2l+1)(2l_1+1)}{(2L+1)}}C^{L 0}_{l0l_10}\Bigg]
\end{eqnarray}

\noindent Here we use the properties of Clebsch gordon coefficients $\sum_{mm_1}C^{l_2m_2}_{lml_1m_1}C^{LM}_{lml_1m_1} = \delta^{L}_{l_2}\delta^{M}_{m_2}$ ~\citep{varshalovich1988quantum}. If we use the WMAP re-normalization, i.e. $\tilde{A}^{LM}_{lml_1m_1}={A}^{LM}_{lml_1m_1} \Bigg[\sqrt{\frac{(2l+1)(2l_1+1)}{(2L+1)}}C^{L 0}_{l0l_10}\Bigg]^{-1}$, we get~\citep{Joshi2012} 
\begin{equation}
\tilde{A}^{LM}_{ll_1} =  (4\pi)^2\int k^2 \mathrm{d}k \frac{g_{LM}(k)}{\sqrt{4\pi}}P^{s}(k)\Delta_{Tl}(k,\tau_0)\Delta_{Tl'}(k,\tau_0) 
\end{equation}

So we can see that the isotropy violation in the primordial power spectrum directly reflects in the present CMB sky. In \texttt{CMBAns} along with the CMB angular power spectrum we can also calculate the BipoSH coefficients. At present \texttt{CMBAns} only calculates the BipoSH coefficients for the scalar power spectrum.  

Fig.~\ref{fig:BipoSH} shows the BipoSH coefficients $\tilde{A}^{20}_{ll}$ and $\tilde{A}^{20}_{ll-2}$, calculated using CMBAns for $g^{20}=1.5$, $g^{00} = 1$ and all other $g^{lm}=0$.

We can see that we can calculate both the $C_l$ and the BipoSH coefficients using the  brightness fluctuation functions. 
 However, if we use the expression for $\Delta_{Tl}$ from Sec.~\ref{photonbrightnessfunction}, then for calculating $C_l$ up to $l_\text{max}$, we need to solve $2\times l_\text{max}$ coupled differential equations for photons up to the present era, which will be highly time consuming. 
Therefore, Seljak and Zaldarriaga proposed a method in~\cite{Seljak1996, Zaldarriaga1998}, which analytically integrates the CMB perturbation terms. This is known as the line-of-sight integration method. We discuss the line of sight Integration in the next section. 

\subsection{Scalar power spectrum}\label{ScalerSourceTerm}
The Boltzmann equations for the perturbation in photon intensity and polarization are given by (for details check Appendix~\ref{AppendixA})
\begin{eqnarray}
\frac{\partial \Delta_{T}}{\partial\tau}+ik\mu \Delta_{T} + \frac{2}{3}\dot{h}+\frac{4}{3}(\dot{h}+6\dot{\eta})P_{2}(\mu) 
= \left(\frac{\partial \Delta_{T}}{\partial\tau}\right)_{C} \label{F1gamma}\\
\frac{\partial \Delta_{P}}{\partial\tau}+ik\mu \Delta_{P}  
= \left(\frac{\partial \Delta_{P}}{\partial\tau}\right)_{C} \label{Ggamma}
\end{eqnarray}

\noindent The terms on the right hand side are the collision terms due to the Compton scattering and are given by 
\begin{eqnarray}
\left(\frac{\partial \Delta_{T}}{\partial\tau}\right)_{C} & = & an_{e}\sigma_{T}\left[-\Delta_{T}+\Delta_{T0}-4\frac{i\theta_b}{k}P_1(\mu)-\frac{1}{2}(\Delta_{T2}+\Delta_{P0}+\Delta_{P2})P_{2}(\mu)\right]\\
\left(\frac{\partial \Delta_{P}}{\partial\tau}\right)_{C} & = & an_{e}\sigma_{T}\left[-\Delta_{P}+\frac{1}{2}(\Delta_{T2}+\Delta_{P0}+\Delta_{P2})\left(1-P_{2}(\mu)\right)\right]
\end{eqnarray}

\noindent where $n_{e}$ is the proper mean number density of free electrons and $\mu = \hat{v}_e\cdot\hat{k}$. Both the Eq.~\ref{F1gamma} and Eq.~\ref{Ggamma} are of the form 
\begin{eqnarray}
\frac{\partial \mathcal{Y}}{\partial\tau}+(an_e\sigma_T + ik\mu)\mathcal{Y} & = & \mathcal{Q}(\tau), \;\;\;\;\;\;\;\;\;\; \mathcal{Y}\in(\Delta_{T},\,\Delta_{P})\,
\end{eqnarray}
which can be solved as 
\begin{equation}
    \mathcal{Y} = e^{-\int\mathcal{P}(\tau)\mathrm{d}\tau}\int e^{\int\mathcal{P}(\tau)\mathrm{d}\tau}\mathcal{Q}(\tau)\,\mathrm{d}\tau
\end{equation}

\noindent where $\mathcal{P} = an_e\sigma_T + ik\mu$. By solving the temperature and the polarization perturbations, we get
\begin{eqnarray}
\Delta_T(\tau_0,k,\mu) &=& \int_0^{\tau_0}\mathrm{d}\tau e^{ik\mu(\tau - \tau_0)}e^{-\kappa}\Bigg[\dot{\kappa}\left(\Delta_{T0}-4\frac{i\theta_b}{k}P_1(\mu)-\frac{1}{2}\Pi P_2(\mu)  \right) - \frac{2}{3}\dot{h} - \frac{4}{3}(\dot{h}+6\dot{\eta})P_2(\mu)\Bigg] \nonumber\\ \Delta_P(\tau_0,k,\mu) &=& \int_0^{\tau_0}\mathrm{d}\tau e^{ik\mu(\tau - \tau_0)}e^{-\kappa}\frac{\dot{\kappa}}{2}\Pi (1-P_2(\mu))\,.
\label{lineofsightintegration}
\end{eqnarray}

\noindent Here $\kappa = \int_\tau^{\tau_0}an_e\sigma_T\mathrm{d}\tau$ is the optical depth at time $\tau$. $\Pi$ is the anisotropic stress term and is given by 
\begin{equation}
    \Pi = \Delta_{T2}+\Delta_{P2}+\Delta_{P0}\,.
\end{equation}

In the above expression, i.e. Eq.~\ref{lineofsightintegration}, the terms with $\mu$ can be eliminated by integration by parts and considering that the boundary terms can be dropped, because they will vanish
as $\tau \rightarrow 0$ and are unobservable at $\tau = \tau_0$. We can replace every occurrence of $\mu$ with $\frac{1}{ik}\frac{\mathrm{d}}{\mathrm{d}\tau}$. This gives us the scalar source terms as 
\begin{eqnarray}
S_{T}(k,\tau) & = & -g\left(\Delta_{T0}+2\ddot{\alpha}+\frac{\dot{\theta}_{b}}{k^2}+\frac{\Pi}{4}+\frac{3\ddot{\Pi}}{4k^{2}}\right)+e^{-\kappa}(\dot{\eta}+\ddot{\alpha})+\dot{g}\left(\frac{\theta_{b}}{k^2}+\frac{3\dot{\Pi}}{4k^{2}}\right)+\frac{3\ddot{g}\Pi}{4k^{2}}
\label{scalar_temp}\\
S_{P}(k,\tau) & = & \frac{3g\Pi(\tau,k)}{4k^2(\tau_0-\tau)^{2}}\,.\label{scalar_pol}
\end{eqnarray}

\noindent Here $\alpha = (\dot{h} + 6\dot{\eta})/2k^2$ \citep{zaldarriaga1997microwave,Zaldarriaga1998}. For obtaining the polarization term, we need to use the recursive property of the spherical Bessel function and $\lim_{x\rightarrow 0} j_l(x)/x^2 = \frac{1}{15}$.  

Expanding the $\Delta_T$ and $\Delta_P$ in Legendre polynomial and using the property  
\begin{equation}
    \int^{1}_{-1}\frac{\mathrm{d}\mu}{2}P_l(\mu)e^{ik\mu(\tau-\tau_0)} = \frac{1}{(-i)^l}j_l[k(\tau-\tau_0]
\end{equation}
we can express the brightness fluctuation functions for temperature and $E$ mode polarization as 
\begin{eqnarray}
\Delta_{Tl}(k)  =  \int_{0}^{\tau_{0}}\mathrm{d}\tau\,S_{T}(k,\tau)j_{l}(x)\,,\qquad
\Delta_{El}(k)  =  \sqrt{\frac{(l+2)!}{(l-2)!}}\int_{0}^{\tau_{0}}\mathrm{d}\tau\,S_{P}(k,\tau)j_{l}(x)
\label{brightnessscalar}
\end{eqnarray}

\noindent The temperature and the polarization power spectra are given by 

\begin{eqnarray}
C_{l}^{XX} = (4\pi)^{2}\int k^{2}\, \mathrm{d}k\, P^{s}(k)[\Delta_{Xl}^{s}(k)]^{2}\,, \qquad
C_{l}^{TE} = (4\pi)^{2}\int k^{2}\mathrm{d}k\, P^{s}(k)\,\Delta_{Tl}^{s}\,\Delta_{El}^{s} \,,
\end{eqnarray}

\noindent where, $X$ can be $T$ or $E$. 
$P^{s}(k)$ is the scalar primordial power spectrum
set by inflation. 

\subsection{Tensor power spectrum}
%
The power spectrum for the tensor perturbation can be calculated in a similar manner. However, due to the absence of the reflection symmetry, we have a nonzero $B$ mode polarization (check Appendix~\ref{AppendixA}).

The source functions for the tensor polarization, i.e.
$S_{T}^{t}(\tau,k)$, $S_{E}^{t}(\tau,k)$, $S_{B}^{t}(\tau,k)$ are
given by \citep{Yen2006}

\begin{eqnarray}
S_{T}^{t}(\tau,k)&=&\left(-\dot{h}_qe^{-an_{e}\sigma_{T}}+g\Psi^t_e\right)/x^{2}\,,
\label{tensor_temp}\\
S_{E}^{t}(\tau,k)&=&g(\tau)\left(-\Psi^t_e+\frac{\ddot{\Psi^t_e}}{k^{2}}+\frac{6\Psi^t_e}{x^{2}}+\frac{4\dot{\Psi^t_e}}{kx}\right)+\dot{g}(\tau)\left(\frac{2\dot{\Psi^t_e}}{k^{2}}+\frac{4\Psi^t_e}{kx}\right)+\ddot{g}(\tau)\frac{\Psi^t_e}{k^{2}}\,, \\
S_{B}^{t}(\tau,k)&=&g(\tau)\left(\frac{4\Psi^t_e}{x}+\frac{2\dot{\Psi^t_e}}{k}\right)+2\dot{g}(\tau)\frac{\Psi^t_e}{k}
\label{tensor_bpol}\,,
\end{eqnarray}

\begin{figure}
\centering
\includegraphics[width=0.46\textwidth,trim = 0 50 0 30, clip]{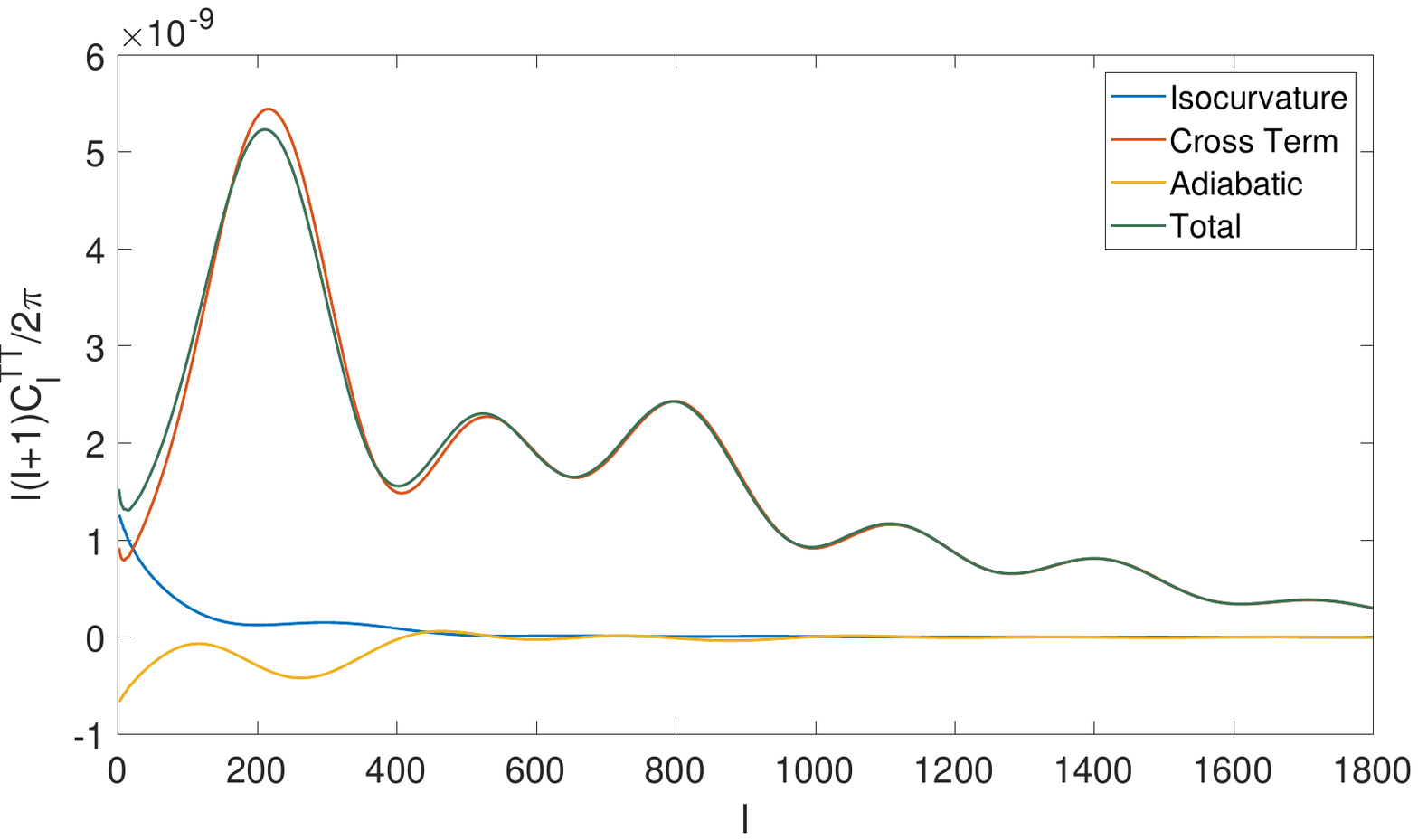}
\includegraphics[width=0.46\textwidth,trim = 0 50 0 30, clip]{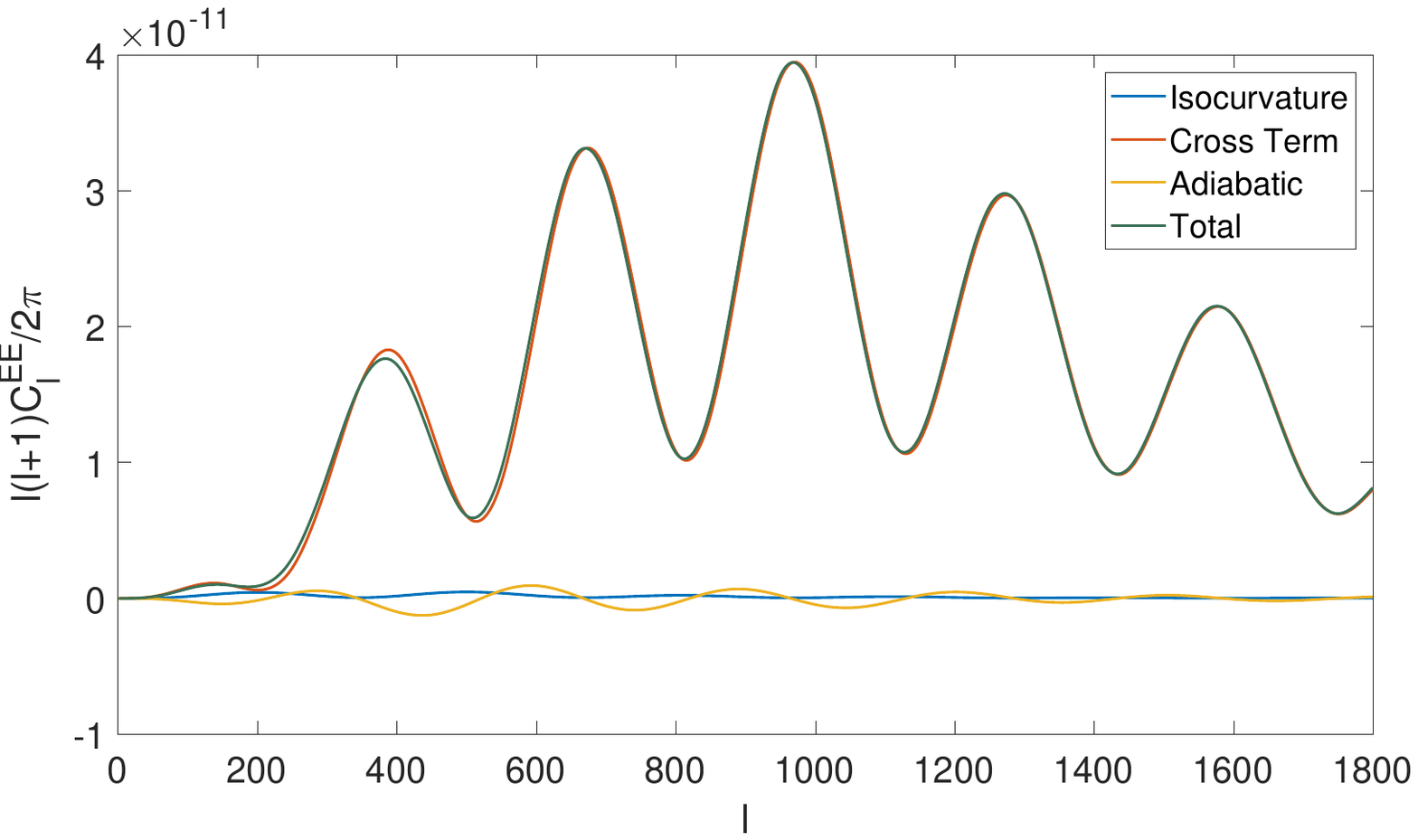}
\caption{\label{fig:twofield}The adiabatic, isocurvature and the cross term for $C_l^{TT}$ and $C_l^{EE}$ calculated for a two field inflationary model. We use $s_H=60$, $s_0=50$ and $R=5$ for this illustration }
\end{figure}

\begin{figure}
\centering
\includegraphics[width=0.99\textwidth,trim = 310 290 360 300, clip]{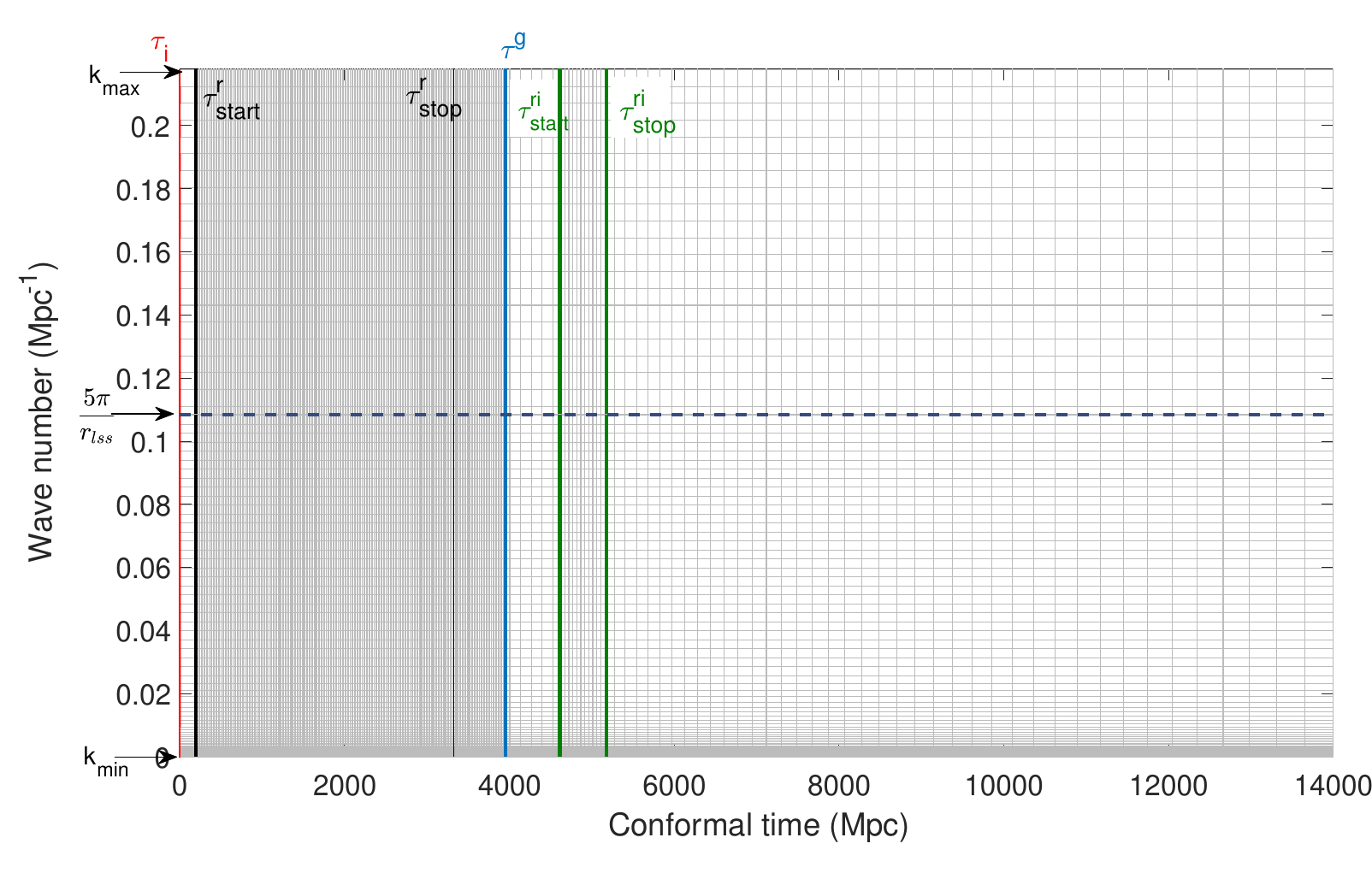}
\caption{\label{fig:Grid}A typical grid for calculating the source functions. For small weve-numbers ($k$) we use a logarithmic grid and for large wave numbers we use linear grid. The actual time grid is 10 times denser than the grid shown in the plot. $\tau_{i}$ is the point where we set the initial conditions. It is a function for $k$. However, as $\tau_{min}$ is very small, it gives an impression that the $\tau_{min}$ is constant. We use denser linear grid during the recombination and reionization. Elsewhere, we use a logarithmic grid. }
\end{figure}

\noindent where, $x=k(\tau_{0}-\tau)$. Once we get the source terms, we can calculate the brightness fluctuation functions. However, as the tensor fluctuations are spin 2 quantities, we get an extra $\sqrt{\frac{(l-2)!}{(l+2)!}}$ term in the brightness fluctuation functions. The brightness fluctuation functions for the tensor perturbation are 
\begin{eqnarray}
\Delta^t_{Tl}(k)  =  \sqrt{\frac{(l-2)!}{(l+2)!}}\int_{0}^{\tau_{0}}\mathrm{d}\tau\,S_T^{t}(k,\tau)j_{l}(x)\,,\qquad
\Delta^t_{E,Bl}(k)  =  \int_{0}^{\tau_{0}}\mathrm{d}\tau\,S_{E,B}^t(k,\tau)j_{l}(x)
\label{brightness_tensor}
\end{eqnarray}

\noindent The brightness fluctuation functions can be convolved with the primordial tensor power spectrum to get
\begin{eqnarray}
C_{l}^{tXX} = (4\pi)^{2}\int k^{2}\, \mathrm{d}k\, P^{t}(k)[\Delta_{Xl}^{t}(k)]^{2}\,, \qquad
C_{l}^{tXY} = (4\pi)^{2}\int k^{2}\mathrm{d}k\, P^{t}(k)\,\Delta_{Xl}^{t}\,\Delta_{Yl}^{t} \,,
\end{eqnarray}

\noindent where $(X, Y)\in (T, E, B)$.





\begin{figure}
\centering
\includegraphics[width=0.46\textwidth,trim = 230 290 280 300, clip]{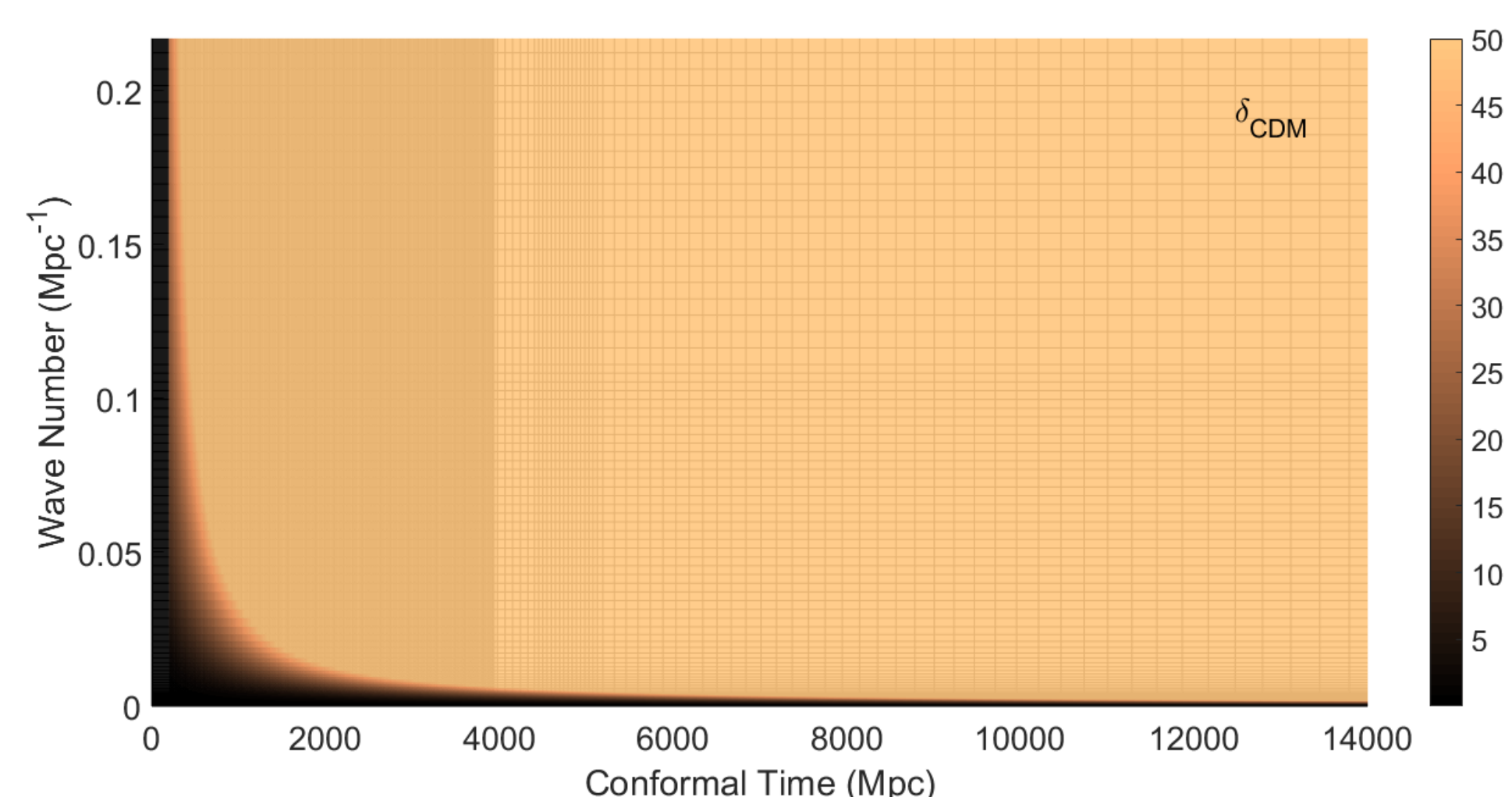}
\includegraphics[width=0.46\textwidth,trim = 230 290 280 300, clip]{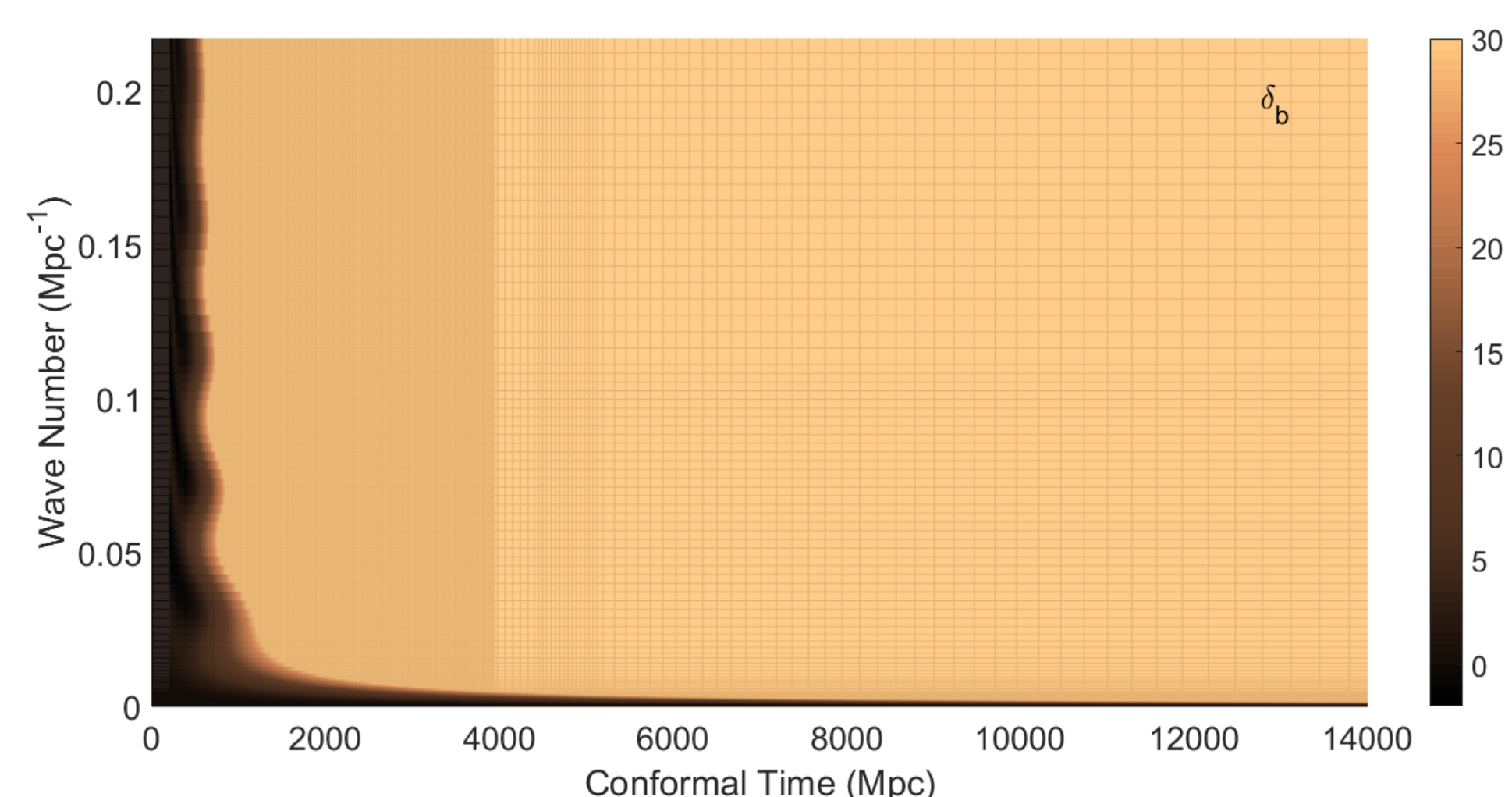}
\includegraphics[width=0.46\textwidth,trim = 230 290 280 300, clip]{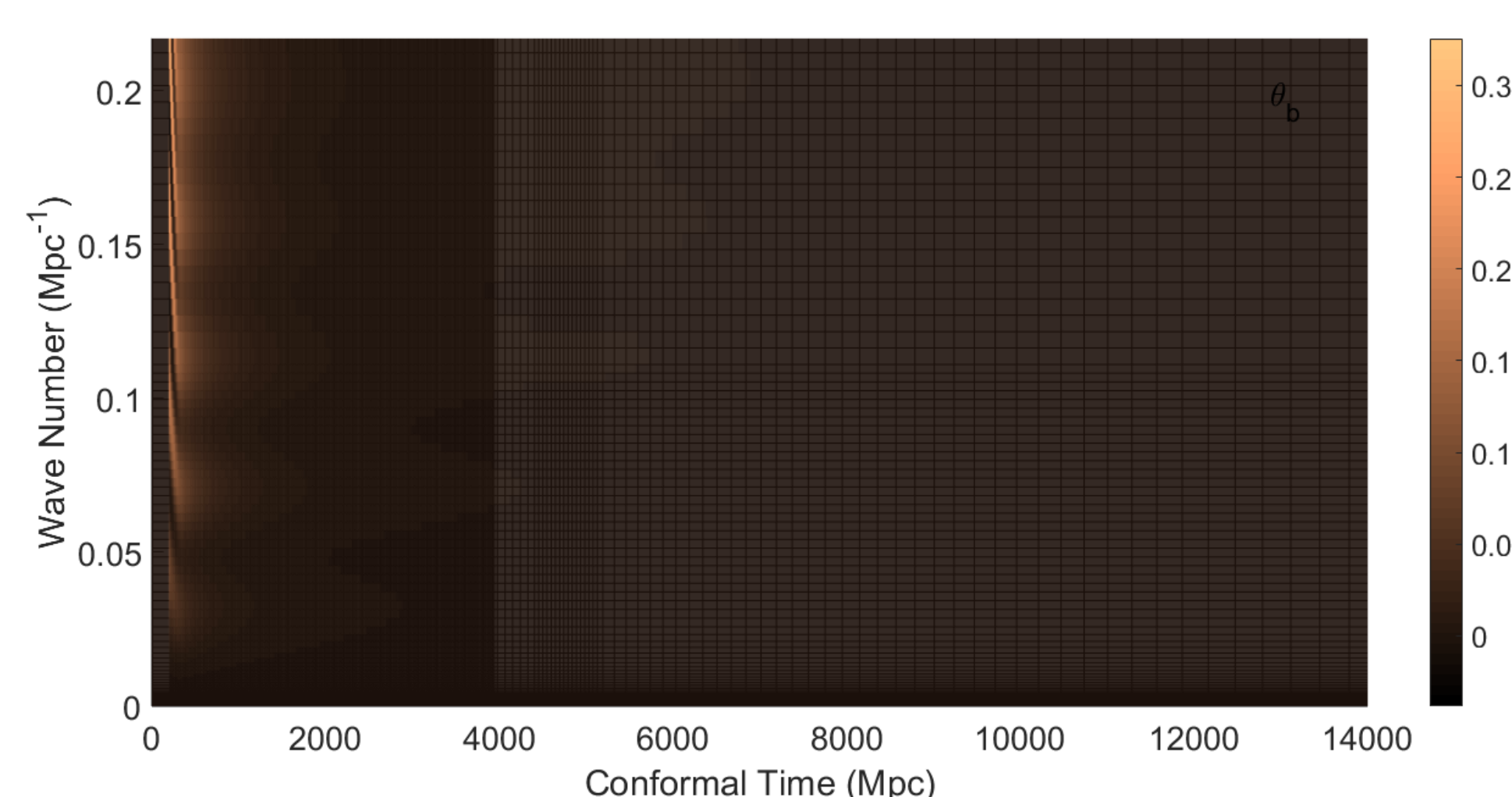}
\includegraphics[width=0.46\textwidth,trim = 230 290 280 300, clip]{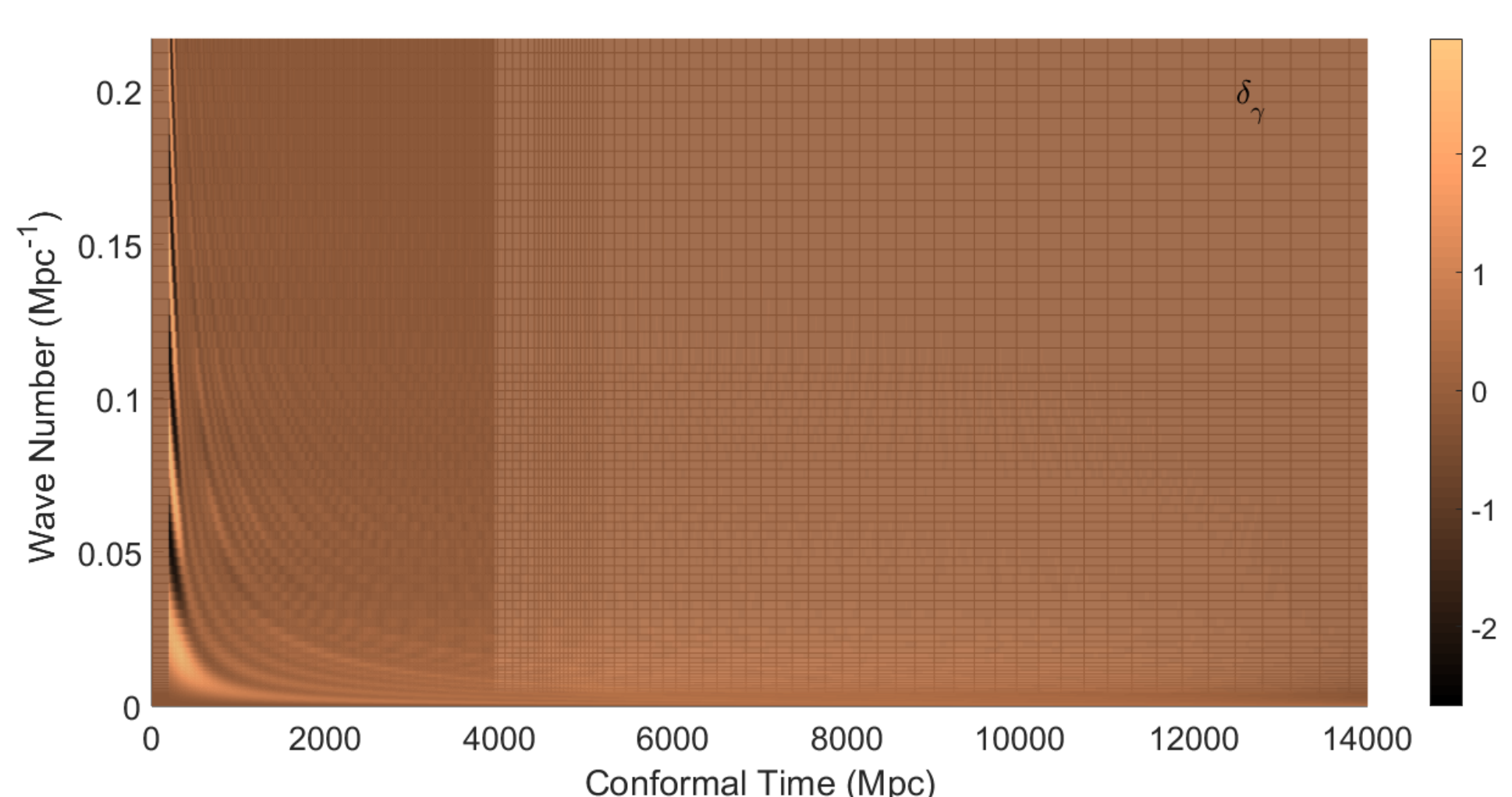}
\includegraphics[width=0.46\textwidth,trim = 230 290 280 300, clip]{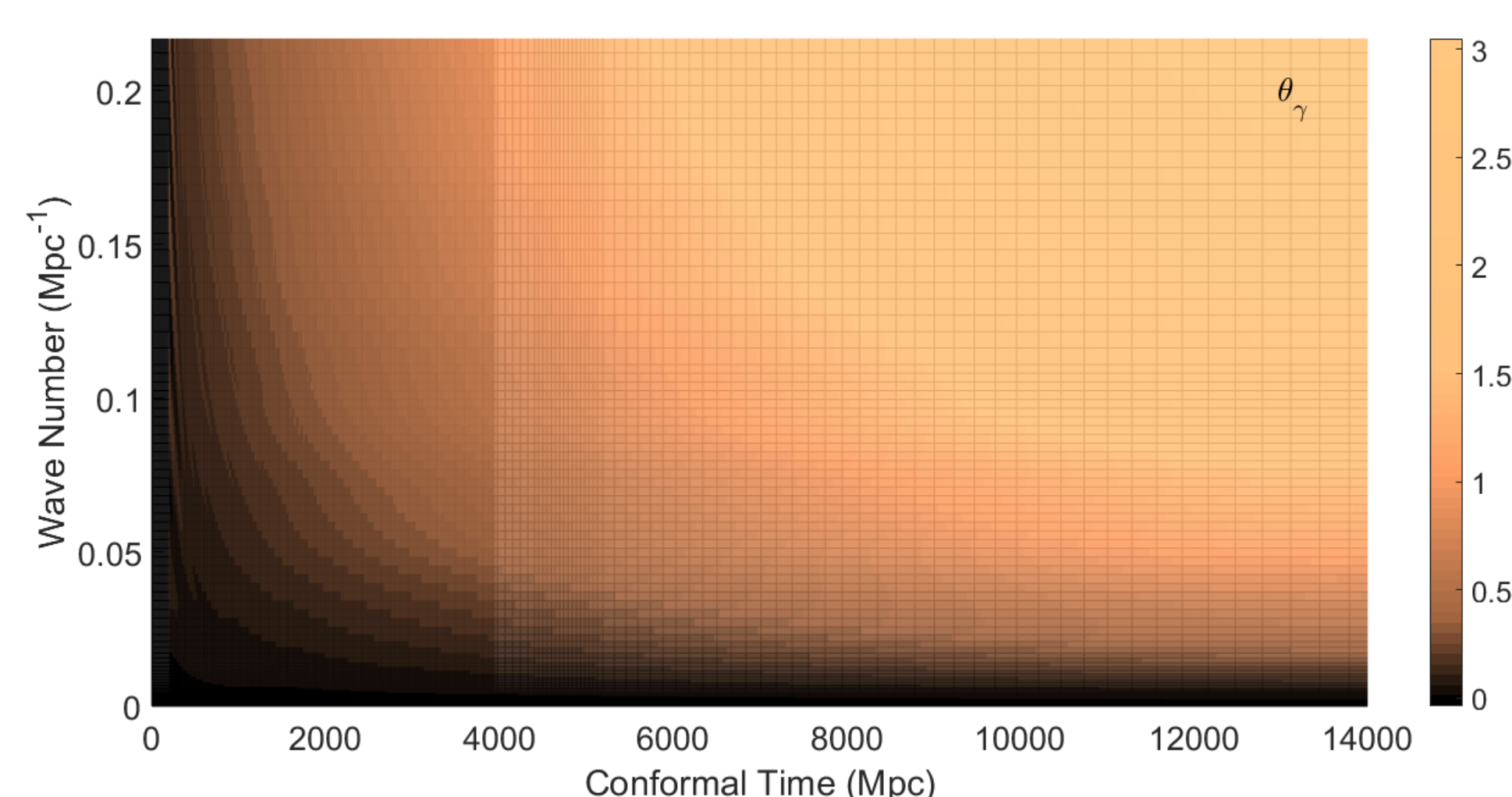}
\includegraphics[width=0.46\textwidth,trim = 230 290 280 300, clip]{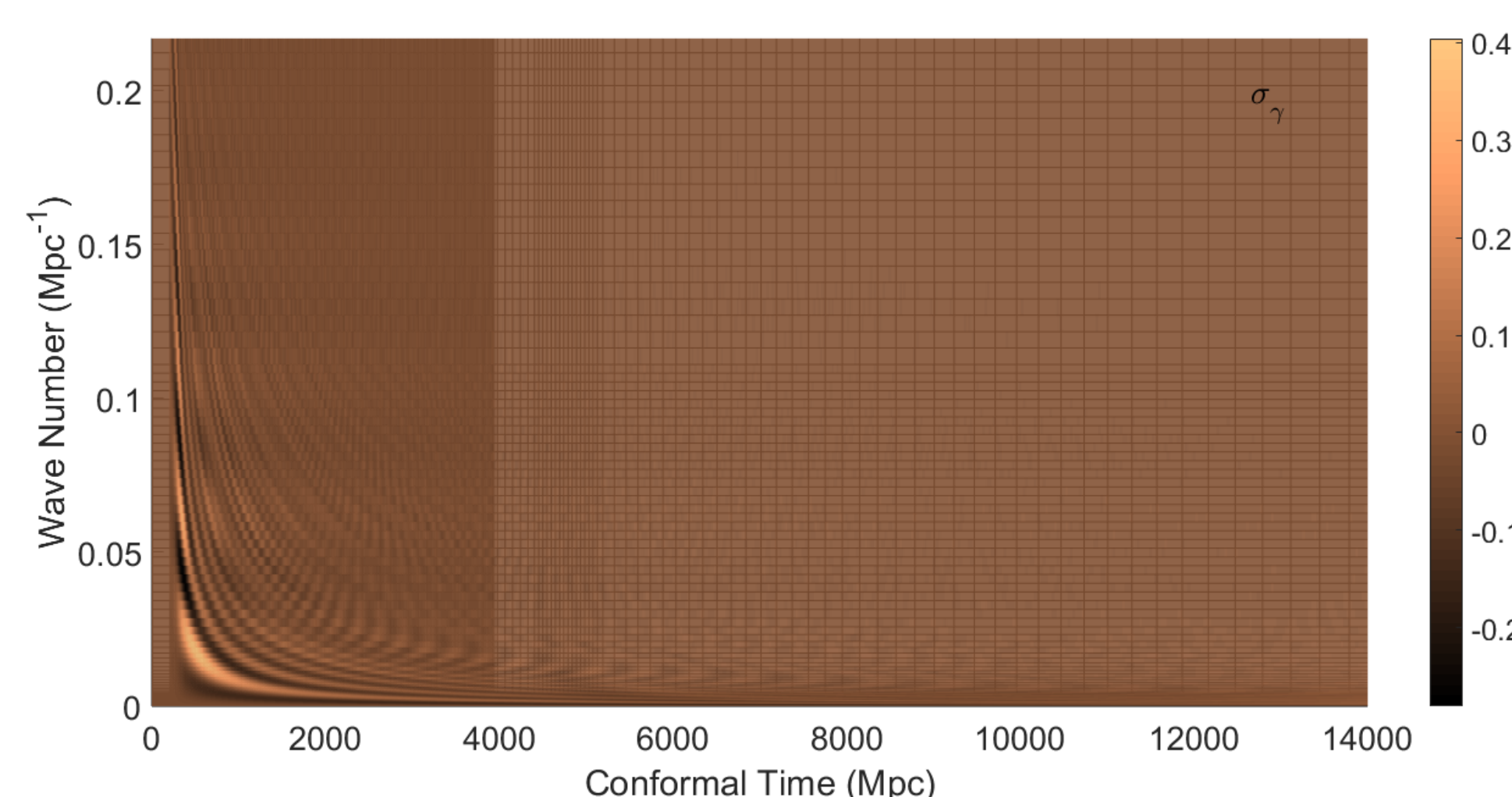}
\caption{\label{fig:perturbation}Plots of some of the scalar perturbation variables, $\delta_c$, $\delta_b$, $\theta_b$, $\delta_\gamma$, $\theta_\gamma$ and $\sigma_\gamma$. The $k-\tau$ grid, used for calculating the source terms, is plotted in the background. CDM is not coupled to any other constituents of the universe, and at late time $\delta_c$ grows exponentially. While ploting we truncate the values of $\delta_c>50$ for a better visualization. Baryons are coupled to photons during recombination, so we can see the acoustic oscillations in the early epoch. However, after decoupling, baryons follow the CDM. We truncate $\delta_b>30.0$. We can also see the similar oscillatory features in $\theta_b$, $\delta_\gamma$, $\theta_\gamma$ and $\sigma_\gamma$ during recombination. A smoother grid is used during recombination and reionization to capture these oscillating features in the source terms. For this analysis, we use adiabatic initial conditions with $\Omega_b h^2 = 0.0223$, $\Omega_b h^2 = 0.1188$, $h = 67.74$ $\texttt{km/sec/Mpc}$, $n_s = 0.9667$, $\kappa = 0.08$.}
\end{figure}

\section{Calculating $C_l$ for Two field inflation}
A single scalar field inflationary model can only generate a primordial adiabatic spectrum. Pure isocurvature perturbations are also ruled out as they produce high power in low multipoles. However if we consider more than one scalar inflationary fields, then it can produce both the modes in the power spectrum, i.e. using two field inflation we can produce a correlated mixture of adiabatic and isocurvature modes.

 The model was first proposed by \cite{Polarski:1992dq,langlois1999correlated}, in the context of the particle physics. Several authors consider this type of inflationary model. 
 \texttt{CMBAns} can also calculate the power spectrum from two field inflationary model. 
 
 In case of two field inflation the brightness fluctuation function have the components from both the adiabatic and the isocurvature model. Therefore, the brightness fluctuation function can be written as $\zeta(k)\Delta_l(k) = \zeta^{\rm{adb}}(k)\Delta_l^{\rm{adb}}(k)+ \zeta^{\rm{iso}}\Delta_l^{\rm{iso}}(k)$. This gives
 
 \begin{equation}
     C_l^{TT} = (4\pi)^2\int k^2 \mathrm{d}k \Bigg[P^S_{\rm{adb}}(k)(\Delta^{\rm{adb}}_{Tl}(k))^2 + P^S_{\rm{iso}}(k)(\Delta^{\rm{iso}}_{Tl}(k))^2 + P^S_{\rm{cross}}(k)\Delta^{\rm{adb}}_{Tl}(k)\Delta^{\rm{iso}}_{Tl}(k)\Bigg]
 \end{equation}
\noindent where $P^S_{\rm{adb}}=\langle\zeta^{\rm{adb}}(k)\zeta^{\rm{adb}}(k)\rangle$, $P^S_{\rm{iso}}=\langle\zeta^{\rm{iso}}(k)\zeta^{\rm{iso}}(k)\rangle$ and 
$P^S_{\rm{cross}}=\langle\zeta^{\rm{adb}}(k)\zeta^{\rm{iso}}(k)\rangle$. $\zeta(k)$ are the primordial density fluctuations. 

\texttt{CMBAns} can calculate the primordial power spectrum for a double inflation model, which is a class of two field inflationary models. 
The double inflation can be characterize by three parameters, namely $s_H$, $s_0$ and $R$. $s_H$ is the number of e-folds before the end of inflation when the scale corresponding to our Hubble radius today crossed the Hubble radius during inflation. $R$ is the ratio of the masses of the heavy and the light inflationary field. $s_0$ is a parameter that represents the phases of inflation where the heady field, or the light field dominate. $\texttt{CMBAns}$ can calculate the angular power spectra for a double inflation model~\citep{Langlois1999}.   

The inflationary potential for the two field inflation models are not specific. Therefore, for other types of two field inflationary models, the power spectra $P^S_\text{\rm{adb}}(k)$, $P^S_\text{\rm{iso}}(k)$, $P^S_\text{\rm{cross}}(k)$ can be calculated separately. It can then be  supplied to $\texttt{CMBAns}$ as a text file and calculate the CMB power spectrum. 

In Fig.~{\ref{fig:twofield}} we show the isocurvature, adiabatic and the cross term for $C^{TT}_l$ and $C^{EE}_l$ for $R=5$, $s_H=60$ and $s_0=50$ in a double inflationary scenario. All the other parameters are the standard $\Lambda$CDM parameters.

\section{Numerical calculations}

In previous sections, we discussed all the mathematical equations used in \texttt{CMBAns}. The calculation of the power spectrum is briefly done through four steps.
\begin{enumerate}
    \item  Calculate the perturbation variables for different wave numbers at different time. This is done by integrating the set of linear differential equations, given by Eq.~\ref{eq:cdmtheta} \-- Eq.~\ref{scalarDE} (for scalar) and Eq.~\ref{tensorperturbation} \-- Eq.~\ref{neutrinoTesor} (for tensor) using a numerical integrator. For integration, we require the initial conditions which are given by Eq.~\ref{adiabaticinitialcond}, Eq.~\ref{baryonisocurvature}, Eq.~\ref{CDMisocurvature}, Eq.~\ref{neutrinodelsity}, Eq.~\ref{neutrinovelocity} (for different types of scalar initial conditions) and Eq.~\ref{tensorinitialcondition} (for tensor initial conditions). 
    
    \item Calculate the temperature and polarization source functions given by Eq.~\ref{scalar_temp} \-- Eq.~\ref{scalar_pol} (for the scalar case) and Eq.~\ref{tensor_temp} \-- Eq.~\ref{tensor_bpol} (for the tensor case), using the perturbation variables. 
    
    \item Calculate the brightness fluctuation functions by convolving the source terms with the spherical Bessel functions (Eq.~\ref{brightnessscalar} for scalar and Eq.~\ref{brightness_tensor} for tensor). 
    
    \item Convolve the square of the brightness fluctuation function with the primordial power spectrum to get the $C_l$'s. 
\end{enumerate}

For scalar perturbations, there are total $8 + 2\times(1+l^\gamma_{max})+(1+l^{\nu}_{max}) + N_q^\nu\times(1+l^{\nu_m}_{max})$ perturbation variables (gravity, CDM, baryon, DE each has 2 equations), where $N^\nu_q$ is the number of discretizations of the momentum of massive neutrinos. For the tensor perturbations, we have $2 + 2\times(1+l^{\gamma t}_{max})+(1+l^{\nu t}_{max})$ perturbation variables ($2$ comes from gravity $h$ and $\dot{h}$). 
The initial conditions are set deep inside the radiation dominated era, as discussed in Sec.~\ref{initialcondset}.

For integrating the set of perturbation equations, we use a \texttt{C} version of the \texttt{dverk} integrator, originally used in \texttt{CMBFAST} and \texttt{CAMB}. It is a Runge-Kutta  (RK)
subroutine  based  on  Verner's fifth and sixth order pairs of formulae\footnote{Runge-Kutta pairs: For the solution of initial value problems, the step-size is allowed to vary by estimating the error produced in each step. To achieve this, it is standard practice to build method pairs, based on the same stages which produce an output answer of order $p$ and a second approximation of order $q$, where $q>p$. The difference of these two approximations will give an asymptotically correct estimate of the error in the output value. As for small $h$, the actual local error is approximately proportional to $h^{p+1}$, and the step size in the following step can be chosen to give a value close to that specified as a user tolerance. } for finding approximations to the solution of  a  system  of  first  order  ordinary  differential equations  with  initial  conditions. The integrator can solve non-stiff equations very efficiently. 

We calculate the perturbation equations and store the source functions in a $k-\tau$ grid, shown in Fig.~\ref{fig:Grid}. 
The choice of the grid is important for speed and the accuracy of the calculation.
For calculating the power spectrum, we need to integrate the brightness fluctuation functions for $k$ from $0$ to $\infty$. However, numerically we can't integrate up to $k\rightarrow \infty$, and thus we take $(k\tau)_{max}$ as an input to the program. For calculating $C_l$ up to $l_{max}$, the typical value of  $(k\tau)_{max}\approx 2l_{max}$. The reason is that if we assume that all the fluctuations occur at the last scattering surface, the angle corresponding to  $(k\tau)_{max}$ at last scattering surface will be approximately $\frac{2\pi}{(k\tau)_{max}}$, which is roughly equal to $\pi/l$. We choose $k_{max} = (k\tau)_{max}/\tau_0$.

 At low $k$, we take smaller logarithmic grid spacing. For calculating only the scalar perturbations we use the logarithmic grid spacing, $\delta (\ln k) = 0.2$. However, if tensor perturbations are requested, we use a smaller logarithmic grid size at low $k$, $\delta (\ln k) = 0.1$. This logarithmic grid is smoothly matched with a linear grid with grid spacing $\delta k = 0.8/\tau^r$, where $\tau^r$ is the conformal time difference between the present era and the last scattering surface (the point where the visibility function, $g$, is maximum). The smooth matching can be done by taking a logarithmic grid for $k < \frac{\delta k}{\delta (\ln k)}$, and a linear grid otherwise. For $k>\frac{5\pi}{r_{lss}}$, we use an even bigger grid spacing $\delta k = 1.5/\tau^r$. $r_{lss} = \int_0^{\tau_{lss}}c_s\mathrm{d}\tau$ is the sound horizon at the last scattering surface.  $\tau_{lss}$ is the conformal time at the last scattering surface. 
 
The equations for the source terms, i.e. Eq.~\ref{scalar_temp} \-- Eq.~\ref{tensor_bpol}, show that apart from the ISW term, all the other terms are either multiplied with the visibility function $g$ or its temporal derivatives. As the visibility function is nonzero only during recombination and reionization, the source terms are also nonzero only in those eras. The ISW term is non oscillatory and is important throughout the expansion history of the universe. 

For specifying the temporal grid, we need a fine grid during the recombination and reionization era. However, in the rest of the universe, we can use a larger grid. The grid is shown in Fig.~\ref{fig:Grid}. In Sec.~\ref{initialcondset}, we have specified the points, $\tau_{i} =  \min\left(\tau_{h},\;\;\tau_{i}^{hor}(k),\;\;0.1\right)$, where the initial conditions are set. For different $k$ the initial conditions are set at different $\tau_i$. Therefore, for specifying the grid we take $\tau_{min} = \min\left(\tau_{h},\;\;\tau_{i}^{hor}(k_{min}),\;\;0.1\right)$. We mark $\tau_i$ with a red-line in Fig.~\ref{fig:Grid}.

\begin{figure}
\centering
\includegraphics[width=0.48\textwidth,trim = 230 290 280 300, clip]{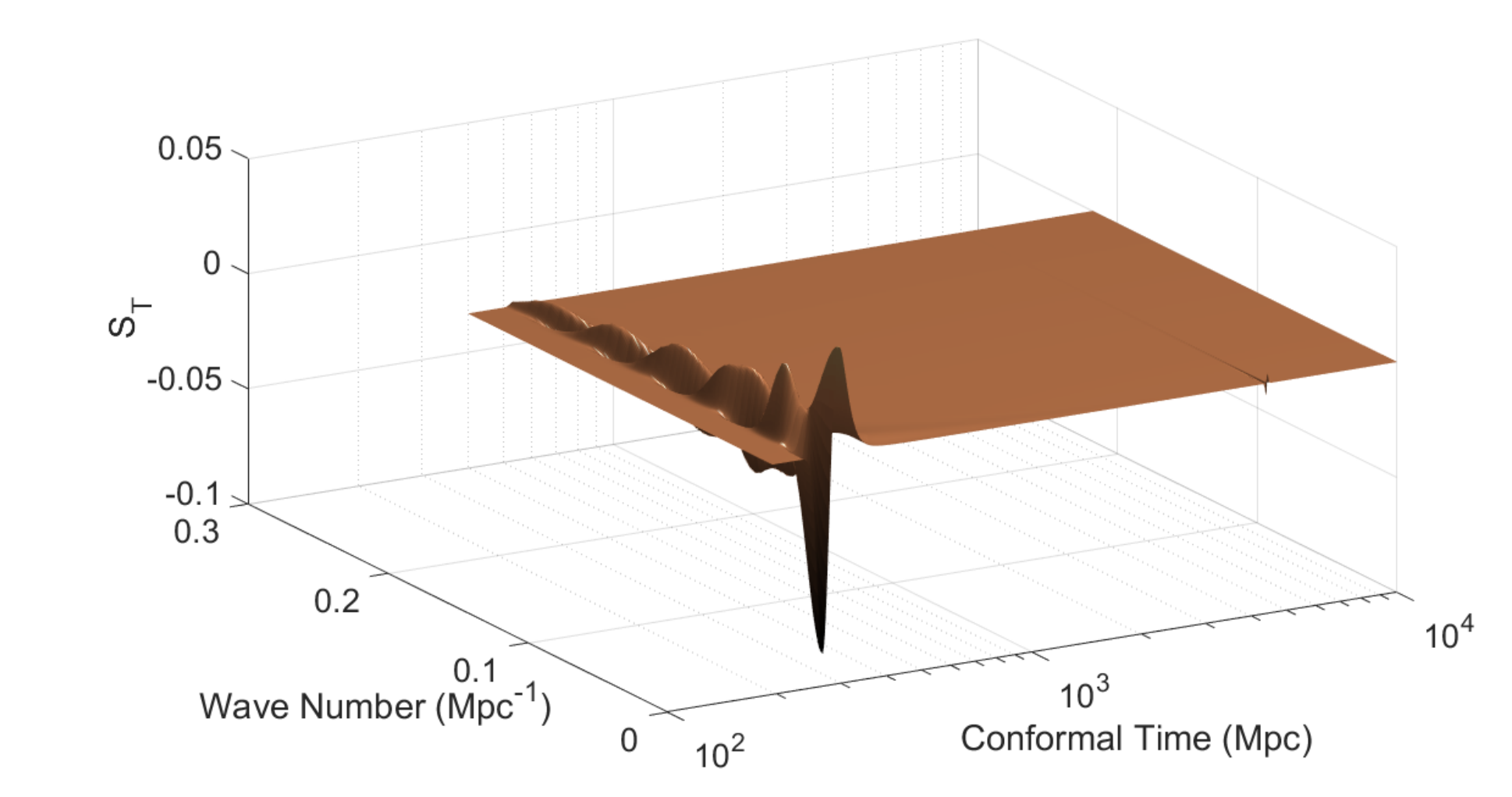}
\includegraphics[width=0.48\textwidth,trim = 230 290 280 300, clip]{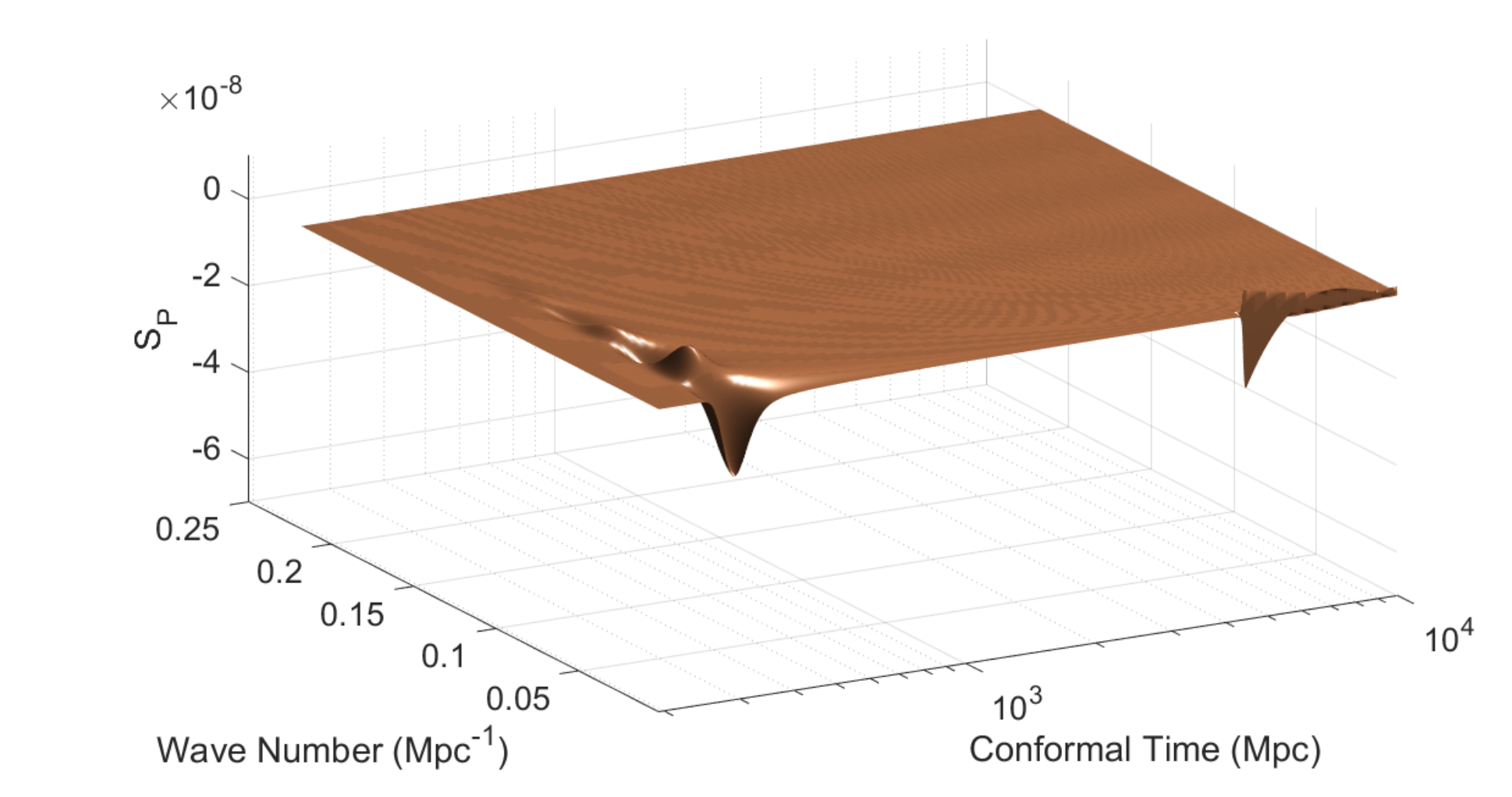}
\caption{\label{fig:Source}The temperature and polarization scalar source terms. The ISW term is not visible here as it is much smaller than the SW and velocity terms. We plot the time direction using a log scale. We see that the source terms have nonzero values only during recombination and reionization, because the visibility function $g \rightarrow 0$ in all the other places. This justifies the choice of grid. We use adiabatic initial conditions. We use $\Omega_b h^2 = 0.0223$, $\Omega_b h^2 = 0.1188$, $h = 67.74$ $\texttt{km/sec/Mpc}$, $n_s = 0.9667$, $\kappa = 0.08$.}
\end{figure}

The recombination start time is calculated by checking when $\int g'(\tau)\mathrm{d}\tau > 10^{-12}$.  To avoid any error, we consider $\frac{9}{10}$ of that time as the starting time for the recombination grid, i.e. $\tau^{r}_{start} = \frac{9}{10}\tau_{\int g'(\tau)\mathrm{d}\tau>10^{-12}}$. $\tau^{r}_{start}$ is shown in the figure using a thick black line.  We define the visibility function $g'(\tau)$ as $g'(\tau) = \dot{\kappa}\exp(\int_{\tau_{min}}^{\tau^{ri}_{start}} \dot{\kappa}\mathrm{d}\tau)$. The end of the recombination epoch is defined as the point when $\int g'(\tau)\mathrm{d}\tau > 0.9999$, i.e. $\tau^{r}_{stop} = \tau_{\int g'(\tau)\mathrm{d}\tau>0.9999}$. $\tau^{r}_{stop}$ is shown using a thin black line. After recombination, we want to smoothly change the grid to a logarithmic grid. This is done by redefining the end point of the recombination grid as $\tau^{g} = \max(\tau^{r}_{stop}, \delta\tau/\delta{\ln\tau})$. $\delta \tau$ is the linear grid spacing during the recombination era and $\delta\ln\tau$ is the logarithmic grid spacing after recombination. There may be cases where the reionization starts before $\tau^{g}$. To account for those models, we have taken $\tau^{g} = \min(\tau^{g},\frac{9}{10}\tau^{ri}_{start})$. 
 
For obtaining a smooth reionization, we join the ionization fraction before and after reionization using a hyperbolic tangent function. Given the optical depth to the last scattering surface, we find the reionization start redshift by calculating $\kappa = \int^{\tau_0}_{\tau^{ri}_{start}} a n_e \sigma_T \mathrm{d}\tau$. The reionization end redshift is taken as $0.8z^{ri}_{start} - 1$. At the end of reionization, we again smoothly match the grid with the logarithmic grid and make proper adjustment at the reionization end redshift to smoothly transform the grid to the logarithmic grid. The reionization grid start and end points are shown using the green line in Fig.~\ref{fig:Grid}.
 
 In Fig.~\ref{fig:perturbation}, we show some of the scalar perturbation variables, plotted over the grid. All the oscillatory features are located near the low $k\tau$. At high $k$ or high $\tau$, fluctuations in the perturbation variables are very small. In Fig.~\ref{fig:Source}, we show the scalar source terms. Most of the structures in the source terms are concentrated near recombination and the reionization. In the rest of the places, the source terms are almost $0$, except that the ISW term will be present in the scalar temperature source term. As the ISW effect is non-oscillatory and  much smaller in comparison with the SW and the velocity term, it is not clearly visible in the plot. This justifies the choice of the smooth grid during recombination and reionization. 

After calculating the source functions, we convolve the source terms with the spherical Bessel functions for calculating the brightness fluctuation functions, $\Delta_{Tl}(k)$ and $\Delta_{Pl}(k)$. Instead of calculating the brightness fluctuation functions for each and every multipole $l$, we choose some specific multipoles, and calculate the brightness fluctuation functions and then the $C_l$'s and BipoSH coefficients in those specific $l$'s. Later we interpolate the $C_l$'s to get the power spectrum at every multipole. 
We pre-compute the spherical Bessel functions in these particular $l$'s in a suitable linear grid. For calculating the brightness fluctuation functions, we interpolate the source functions for each $k$ into the spherical Bessel function grid using spline interpolation. Then we integrate it over $\tau$ using the trapezoidal rule. 

The square of the brightness fluctuations are multiplied with the primordial power spectrum, $P(k)$ for calculating the $C_l$'s in the specific multipoles, which are then interpolated to all the $C_l$'s using spline interpolation. We use COBE normalization techniques for normalizing the $C_l$ ~\citep{bunn480four}, which will soon be modified to make the normalization more relevant to WMAP and Planck results.

\section{Conclusion}

We develop a cosmological Boltzmann package for fast and accurate calculation of the CMB power spectrum for a flat ($\Omega_k = 0$) background cosmology. In this paper, we discuss all the equations and the approximation schemes and different truncation conditions used in \texttt{CMBAns}. The lensing calculations and the comparison of the results with other Boltzmann packages like \texttt{CAMB}, \texttt{CLASS} etc. are not discussed in this paper and are left for future publications in this series. 

\texttt{CMBAns} was initially written in 2010 and used as an internal software package. However the program has been restructured and several new features have been added since then. We have tested \texttt{CMBAns} for a wide range of different initial conditions, Hubble's parameters, $\Omega_b$, $\Omega_c$ etc. to check for the robustness of the program.  


 \texttt{CMBAns} is tested for different dark energy models. We can model the Hubble parameter using a Matlab GUI. In section~\ref{darkenergy} we have shown a special model of dark energy which can decrease the ISW effect at low multipoles. Details of such dark energy models and how they can affect the ISW effect are discussed in detail in \cite{Das2013a,Das:2013sca}. Without a GUI interface it would have been immensely difficult to explore such models. \texttt{CMBAns} can calculate the CMB power spectrum for perturbed and unperturbed dark energy models etc. For perturbed dark energy models it uses the fluid approximation for dark energy.
 
 Recently lots of works are also going on in the isotropy violation in the CMB sky. Along with calculating the angular power spectrum, \texttt{CMBAns} can also calculate the BipoSH spectra for anisotropic inflation model, which is a whole new feature of $\texttt{CMBAns}$. No other publicly available Boltzmann package has this ability. 

In \texttt{CMBAns}, we use a power law for the primordial power spectra. Users can change the scalar spectral index, $n_s$, running of scalar spectral index,  $\alpha_s$ and its running, $\frac{\mathrm{d}\alpha_s}{\mathrm{d}\ln k}$. However, the nature of the primordial power spectra can be modified easily. It has been tested for  perturbed power law model~\citep{mukherjee2015estimation}. Default tensor spectral index, and scalar to tensor ratio are taken as $n_t=n_s-1$ and  $r=7*(1-n_s)$, which can be changed to any values or modify the model of power spectrum model.  \texttt{CMBAns} can also calculate the CMB power spectrum for two field inflation (or double inflation) model where the inflationary field produce both isocurvature and adiabatic modes simultaniously. 

We use the \texttt{C} programming language for \texttt{CMBAns}. However, to make the program object oriented, we use the concept of class from \texttt{C++}. A similar technique is also used in \texttt{CLASS} code. Several stand alone codes, such as calculating the recombination history, power spectra evolution with different cosmological parameters, Bessel function calculation etc. are provided with the package. However, users are not limited to what already come with the program. The influx of precision CMB data means that CMB modeling tools must quickly evolve. Modularity, an important feature of \texttt{CMBAns}, offers a way to solve this problem. The modularity of \texttt{CMBAns} offers a lot of flexibility and let users quickly expand the functionalities of the package to include new cosmological models, by simply writing a new module or classes using the functionality provided in \texttt{CMBAns}.

\begin{appendices}


\section{Baryon temperature calculation \label{baryontempappendix}}
 The baryons and the photons were coupled in the early universe, mostly due to Compton scattering. Therefore, in the very early universe, the temperature of baryon and photons were equal, i.e. $T_b = T_\gamma$. After decoupling, the baryon's temperatyre slowly fall. The baryon in the universe was subjected to various sources of heating and cooling~\citep{lecnote}. However, in the context of CMB, the important heating and cooling mechanisms are adiabatic cooling and Compton heating of baryons. 

\subsection{Adiabatic cooling}
Due to the expansion of the universe, the photons and the baryons both undergo the adiabatic cooling. For photons the wavelength will increase as the universe expands, i.e. $\lambda_\gamma \propto a$. As for photons $T_\gamma \propto E_\gamma \propto \frac{1}{\lambda_\gamma}$, we have $T_\gamma \propto \frac{1}{a}$. 

However, for baryons, $T_b \propto E^K_b = \frac{p^2_b}{2m_b}$, where $E^K_b$ is the kinetic energy of the baryons. The de Broglie wavelength of the baryons varies in proportion to $a$. Therefore, for baryons $T_b \propto \frac{1}{a^2}$ provided there is no external heating or cooling.

Therefore, in an adiabatic condition, we get 
\begin{equation}
    \dot{T}_b = -2\left(\frac{\dot{a}}{a}\right)T_b 
    \label{adiabaticheating}
\end{equation}

\subsection{Compton heating}
In the early universe, the main source of external heating of baryons is Compton scattering. 
\begin{equation}
    e^- + \gamma \rightarrow e^- + \gamma 
\end{equation}

\subsubsection{Heating the electrons:}
Let's assume that a photon with momentum $\omega \hat{i}$ hits an electron at rest and is deflected to $\omega \cos(\theta) \hat{i} + \omega \sin(\theta) \hat{j}$. Hence, the momentum transfer to the electron is $q = \omega\sqrt{(1-\cos\theta)^2 + \sin^2\theta}$ (Here we assume that the change in energy of the photons is very small and the absolute value of the momentum remains almost the same before and after the collision.). The energy delivered to the electron is 
\begin{equation}
    \Delta E = \frac{q^2}{2m_e} = \frac{\omega^2}{m_e}(1-\cos\theta)   
    \;\;\;\;\;\;\;\;\;\;
    \implies
    \;\;\;\;\;\;\;\;\;\;
    \left< \Delta E\right> =  \frac{\left<\omega^2\right>}{m_e}\,.
\end{equation}

 Here, we assume that the electron is stationary at the beginning and the photons are isotropic around it. Therefore, the collision of photons with the electrons will be equally likely from all directions and the $\cos\theta$ term will vanish. 

The heating rate of the electrons will be 

\begin{equation}
    \Gamma = n_e n_\gamma \sigma_T\left< \Delta E\right> = n_e n_\gamma \sigma_T \frac{\left<\omega^2\right>}{m_e}\,,
\end{equation}

\noindent where $\sigma_T$ is the Thomson scattering cross section and $n_e$ and $n_\gamma$ are the number density of the electrons and photons. For simplifying the expressions, we take $c = \hbar = k_B = 1$.

\subsubsection{Energy loss by the electron:}
In the above calculation, we assume that the electrons are at rest. However, as the electrons have a temperature they cannot be at rest. Due to their motion, they will give away some energy to the photons via Compton drag.  

If an electron is moving at a speed $v_e$ along the x-direction, then in its rest frame the photons will have some net momentum. 
In a comoving frame, the photons stress energy tensor is $T^{\mu\nu} = \texttt{diag}\left( \rho_\gamma, \frac{1}{3}\rho_\gamma, \frac{1}{3}\rho_\gamma, \frac{1}{3}\rho_\gamma \right)$

In the comoving frame the electron's 4-velocity is 
$u^\mu = \frac{1}{\sqrt{1 - v_e^2}}\left(1, v_e, 0, 0\right)$
and it carries three spatial vectors
\begin{equation}
(e_1)^\mu = \frac{1}{\sqrt{1 - v_e^2}}\left(v_e, 1, 0, 0\right),\quad
(e_2)^\mu = \frac{1}{\sqrt{1 - v_e^2}}\left(0, 0, 1, 0\right),\quad
(e_3)^\mu = \frac{1}{\sqrt{1 - v_e^2}}\left(0, 0, 0, 1\right)\,.
\end{equation}
Therefore, in the electron's frame, the momentum density of the photons is 

\begin{equation}
    j_\gamma = - T_{\mu\nu}u^{\mu}(e_1)^{\nu} = - \frac{4v_e}{3(1-v_e^2)} \approx - \frac{4}{3}v_e \,.
\end{equation}
Due to the overall velocity of the photons, the electrons will feel some force 
\begin{equation}
    F = n_\gamma \sigma_T \left<p_\gamma\right>\,,
\end{equation}

\noindent where $\left<p_\gamma\right>$ is the average momentum of the photons  with respect to electron. However, $n_\gamma\left<p_\gamma\right> = j_\gamma$, is the photon momentum density. This gives, 

\begin{equation}
    F = \sigma_T j_\gamma = -\frac{4}{3}\sigma_T \rho_\gamma v_e = - \frac{4}{3}\sigma_T n_\gamma\left<\omega\right>v_e. 
\end{equation}

As the energy loss by the electron is given by $- F \cdot v_e$, the net energy loss rate is 

\begin{equation}
    \Lambda = -n_e\left<F.v_e\right> = \frac{4}{3}\sigma_T n_e n_\gamma \left<\omega\right>\left<v_e^2\right>\,
\end{equation}

\noindent where $\left<v_e^2\right>=3T_b/m_e$, assuming that the electrons follow a Maxwell distribution.

\subsubsection{Heating from stimulated Compton effect}
 The third process that will heat the photons is the stimulated Compton effect.  In any radiative process, the ratio of the stimulated to the spontaneous transition rate is equal to the ambient phase density of photons in the final stage, $f(\omega)$. 
 A scattering process can be thought of as absorption and emission of a photon~\citep{Gould1972, Dreicer1964}. We have seen that the total amount of energy that the photons are emitting is given by 
 
 \begin{equation}
         \Gamma = n_e n_\gamma \sigma_T \frac{\left<\omega^2\right>}{m_e} = \frac{n_e  \sigma_T}{h} \int \omega f(\omega)\mathrm{d}\omega
 \end{equation}
 
 \noindent The total amount of stimulated radiation will be 

 \begin{equation}
         \Gamma_{stim} = \frac{n_e  \sigma_T}{h} \int \omega^2 f(\omega) f(\omega) \mathrm{d}\omega = n_e n_\gamma \sigma_T \left< \omega^2 f(\omega) \right>\,.
 \end{equation}
\noindent Here $f(\omega)$ is the phase space distribution of photons. For a blackbody it will follow a Plankian distribution. 

\subsection{Total heating of electrons}
We assume that the universe behaves as a perfect blackbody. Therefore,  
the expectation values of the previous expressions are be given by 

\begin{eqnarray}
n_\gamma = \frac{2\zeta(3)}{\pi^2}T^3_\gamma \;\;\;\;\;\;\;\;
\left<\omega\right> = \frac{\pi^4}{30\zeta(3)}T_\gamma \;\;\;\;\;\;\;\;
\left<\omega^2\right> = \frac{12\zeta(5)}{\zeta(3)}T_\gamma^2 \;\;\;\;\;\;\;\;
\left<\omega^2f(\omega)\right> = \frac{4T_\gamma^2}{30\zeta(3)}\left[\pi^4-90\zeta(5)\right]\,. \nonumber
\end{eqnarray}

\noindent Replacing all these values we can get the energy that the electrons will receive from photon 

\begin{equation}
    \Gamma + \Gamma_{stim} - \Lambda = \frac{4\pi^2}{15}n_eT_\gamma^4\sigma_T\frac{T_\gamma - T_b}{m_e}
    \label{totalheat}
\end{equation}

The specific heat for mono-atomic gas at constant volume is $C_v = \frac{3}{2}n$, where $n$ is the total number density of the particles containing free electron, H, H$^+$, He, He$^+$, He$^{++}$. Therefore, the rate of change of the temperature can be calculated by $\Gamma + \Gamma_{stim} - \Lambda = C_v\dot{T}_b$.

\section{CMB Polarization Calculation}
\label{AppendixA}
The equation for the photons perturbation are more complicated than the neutrinos due to their scattering. The scattering can change the polarization of photons, so we can't write separate equation for different helicity states of photons. Instead we need to consider the perturbation in the $2\times 2$ density matrix of the photons, i.e. $\Delta\rho_{ij}(\hat{n})$. Here $\Delta\rho_{ij}(\hat{n})$ is considered to be normalized over the mean intensity $\rho_0$. In Eq.~\ref{radiationPerturbation}, we show the first order perturbation of the Boltzmann equation for photons and neutrinos. However, instead of $\Phi$ we will now have $\Delta\rho_{ij}$. This complicates the equation for the photon perturbation ~\citep{kosowsky1995cosmic,Zaldarriaga1998, zaldarriaga1997all,kamionkowski1997probe,kamionkowski1997statistics,Bond1984}. 

The density perturbation matrix  is a function of  the direction on the sky ($\hat{n}$) and its two perpendicular direction $\hat{e}_x$ and $\hat{e}_y$. We can rewrite these perturbations in terms of the perturbation in the four Stokes parameters $\Delta_I(\hat{n})$, $\Delta_Q(\hat{n})$, $\Delta_U(\hat{n})$ and $\Delta_V(\hat{n})$. They can be related as

\begin{eqnarray}
    \Delta_I(\hat{n}) =\frac{1}{2}\left[ \Delta\rho_{11}(\hat{n})+\Delta\rho_{22}(\hat{n})\right] \qquad
    \Delta_Q(\hat{n}) =\frac{1}{2}\left[ \Delta\rho_{11}(\hat{n})-\Delta\rho_{22}(\hat{n})\right]\nonumber \\
    \Delta_U(\hat{n}) =\frac{1}{2}\left[ \Delta\rho_{12}(\hat{n})+\Delta\rho_{21}(\hat{n})\right] \qquad
    \Delta_V(\hat{n}) =\frac{1}{2i}\left[ \Delta\rho_{12}(\hat{n})-\Delta\rho_{21}(\hat{n})\right]
    \label{stoke_pol_conversion}
\end{eqnarray}

\noindent If we rotate the coordinate perpendicular to $\hat{n}$ by an angle $\phi$, then $\Delta_Q$ and $\Delta_U$ will transform as 

\begin{eqnarray}
\label{QUTransformation}
 \Delta'_Q(\hat{n}) &=& \Delta_Q(\hat{n}) \cos(2\phi) + \Delta_U(\hat{n}) \sin(2\phi) \nonumber\\
 \Delta'_U(\hat{n}) &=&-\Delta_Q(\hat{n}) \sin(2\phi) + \Delta_U(\hat{n}) \cos(2\phi)
\end{eqnarray}

\noindent However, $\Delta_I$ and $\Delta_V$ remains invariant under such rotation. The above equations also show that $\Delta^2_Q + \Delta^2_U$ remains invariant under such coordinate transformation. Also, for unpolarized light $\Delta\rho_{11} = \Delta\rho_{22}$ and $\Delta\rho_{12} = \Delta\rho_{21} = 0$.

From Eq.~\ref{QUTransformation}, we can see that the $Q$ and $U$ components are direction-dependent quantities, i.e. they depend on $\hat{e}_x$ and $\hat{e}_y$ direction of space. Instead of taking $Q$ and $U$ component independently, if we form two  complex quantities $(\Delta_Q \pm i\Delta_U)$, then they will transform as spin $\pm 2$ quantities, i.e. 

\begin{eqnarray}
(\Delta'_Q \pm i\Delta'_U)(\hat{n}) = \exp(\pm i2\phi) (\Delta_Q \pm i\Delta_U)(\hat{n})\;.
\end{eqnarray}

In general, the intensity (temperature) field being a scalar field can be expanded in terms of the spherical harmonics as
\begin{equation}
    \Delta_I(\hat{n}) = \sum_{lm}a^I_{lm}Y_{lm}(\hat{n})\,.
\end{equation}

However, as the $\Delta_Q \pm \Delta_U$ behave as spin $\pm 2$ quantities, we have to expand them in spin-weighted spherical harmonics. This can be written as 

\begin{eqnarray}
    (\Delta_Q + i\Delta_U)(\hat{n})&=& \sum_{lm}\,_2a_{lm}\,_2Y_{lm}(\hat{n}) \\
    (\Delta_Q - i\Delta_U)(\hat{n}) &=& \sum_{lm}\,_{-2}a_{lm}\,_{-2}Y_{lm}(\hat{n})\,.    
\end{eqnarray}
\noindent We can use the spin raising ($\slashed{\partial}$) and lowering ($\bar{\slashed{\partial}}$) operators to construct some spin $0$ quantities as
\begin{eqnarray}
\slashed{\partial}^2 (\Delta_Q + i\Delta_U)(\hat{n})&=& \sum_{lm}\,_2a_{lm}\slashed{\partial}^2\,_2Y_{lm}(\hat{n}) 
= \sum_{lm}\left(\frac{(l+2)!}{(l-2)!}\right)^\frac{1}{2}\,_2a_{lm}Y_{lm}(\hat{n}) \\ 
    \bar{\slashed{\partial}}^2(\Delta_Q - i\Delta_U)(\hat{n}) &=& \sum_{lm}\,_{-2}a_{lm}\bar{\slashed{\partial}}^2\,_{-2}Y_{lm}(\hat{n})
= \sum_{lm}\left(\frac{(l+2)!}{(l-2)!}\right)^\frac{1}{2}\,_{-2}a_{lm}Y_{lm}(\hat{n})    
\end{eqnarray}

\noindent In the CMB literature, conventionally people use the two scalar fields $E$ and $B$ to represent the polarization, given by 
\begin{eqnarray}
\Delta_E(\hat{n}) &=& -\frac{1}{2}\left[\slashed{\partial}^2 (\Delta_Q + i\Delta_U)(\hat{n}) + \bar{\slashed{\partial}}^2(\Delta_Q - i\Delta_U)(\hat{n}) \right]\nonumber \\
&=&\sum_{lm}a^E_{lm}Y_{lm}
(\hat{n}) = \sum_{lm}\left(-\frac{1}{2i}\right)\left(\,_2a_{lm} + \,_{-2}a_{lm}\right)Y_{lm}(\hat{n})\\
\Delta_B(\hat{n}) &=& -\frac{1}{2}\left[\slashed{\partial}^2 (\Delta_Q + i\Delta_U)(\hat{n}) - \bar{\slashed{\partial}}^2(\Delta_Q - i\Delta_U)(\hat{n}) \right]\nonumber \\
&=&\sum_{lm}a^B_{lm}Y_{lm}
(\hat{n}) = \sum_{lm}\left(-\frac{1}{2i}\right)\left(\,_2a_{lm} - \,_{-2}a_{lm}\right)Y_{lm}(\hat{n})\,.
\label{equationEB}
\end{eqnarray}

\subsubsection*{Scalar Components}
We can calculate the perturbation equations for $\Delta\rho_{ij}$ and then use Eq.~\ref{stoke_pol_conversion} to calculate the perturbation in the Stokes parameters.
If we convert the scalar part of these perturbations in Stokes parameters in Fourier space, then we can get~\citep{kosowsky1995cosmic}

\begin{eqnarray}\label{SourceEquations}
 \frac{\partial \Delta_I}{\partial\tau}+ik\mu \Delta_I + \frac{2}{3}\dot{h}+\frac{4}{3}(\dot{h}+6\dot{\eta})P_{2}(\mu) 
&=& - an_{e}\sigma_{T}\left[\Delta_I - \Delta_{I0}-4\frac{i\theta_b}{k}P_1(\mu)-\frac{1}{2}( \Delta_{I2}-\Delta_{Q0}+\Delta_{Q2})P_{2}(\mu)\right] \label{Fgamma} \nonumber \\
\frac{\partial \Delta_Q}{\partial\tau}+ik\mu \Delta_Q  
&=& -an_e\sigma_T \left[ \Delta_Q + \frac{1}{2}\left(1-P_2(\mu)\right)\left(\Delta_I2+\Delta_{Q2}-\Delta_{Q0}\right) \right] \nonumber \\
\frac{\partial \Delta_U}{\partial\tau}+ik\mu \Delta_U  
&=& -an_e\sigma_T \Delta_U \nonumber\\
\frac{\partial \Delta_V}{\partial\tau}+ik\mu \Delta_V  
&=& -an_e\sigma_T \left[\Delta_V - \frac{3}{2}\mu\Delta_{V1}\right]\,.
\label{ScalarPOlarization}
\end{eqnarray}

\noindent The $\Delta$'s are the functions of ($k,\tau$). In the early universe, the photons were tightly coupled, so we can consider that the photons were not polarized at that early time. The above equation also shows that the $\Delta_U$ and $\Delta_V$ term don't have any source terms. If they were zero in the early universe, then they remain $0$ afterwards. As the $\Delta_U$ term is related to $\Delta_Q$ by coordinate transformation, we get nonzero $\Delta_U$ in late universe. However, $\Delta_V$  still remains $0$. This is also valid for the tensor perturbation.

There is another important property of the scalar perturbations. Let us consider a particular $k$ mode in the scalar perturbation in $\hat{k}$ direction. Now the density field produced by the single mode will have two important symmetry \-- azimuthal symmetry and reflection symmetry. The azimuthal symmetry implies that neither the temperature nor the stokes parameters depend on rotation around $\hat{k}$. 

Let's consider $\hat{e}_\theta$ and $\hat{e}_\phi$ as two perpendicular directions along $\hat{k}$. Under reflection $\hat{e}_\theta\rightarrow-\hat{e}_\theta$ and $\hat{e}_\phi\rightarrow-\hat{e}_\phi$. Eq.~\ref{QUTransformation} implies that under reflection, $\Delta_Q\rightarrow\Delta_Q$, i.e. it remains unchanged while $\Delta_U$ changes sign. Therefore, in linear theory, the scalar polarization can't have the $\Delta_U$ component, and the only polarization component that we can have is $\Delta_Q$. In terms of the $E$ and $B$ field, Eq.~\ref{equationEB} shows that $B$ will be zero and we can have only the $E$ field. As $\Delta_P = \Delta_Q$ ($\Delta_U$ being $0$), $\Delta_P$ will be given by Eq.~\ref{ScalarPOlarization}, and $\Delta_{El} = \left(\frac{(l+2)!}{(l-2)!}\right)^\frac{1}{2}\Delta_{Pl}$

\subsubsection*{Tensor Components}
The tensor components will have two polarisation, $\Delta^{+}$ and $\Delta^{\times}$ As we have done for the scalar part, we can also calculate the perturbation equations for $\Delta\rho_{ij}$ and then if we convert the tensor part of these perturbations in Stokes parameters in Fourier space, we can get the components of the tensor perturbations. The evaluation equations for the tensor perturbations can be simplified if we apply the following variable transformation \citep{basko1980polarization,Polnarev1985, Crittenden:1993,Crittenden1993}

\begin{eqnarray}
\Delta_I^+ = (1-\mu^2)\cos(2\phi)\tilde{\Delta}_I^+
 &\qquad&
\Delta_I^\times = (1-\mu^2)\sin(2\phi)\tilde{\Delta}_I^\times
\nonumber \\  
\Delta_Q^+ = (1-\mu^2)\cos(2\phi)\tilde{\Delta}_Q^+
&\qquad&   
\Delta_Q^+ = (1-\mu^2)\sin(2\phi)\tilde{\Delta}_Q^+
\nonumber \\    
\Delta_U^+ = 2\mu\sin(2\phi)\tilde{\Delta}_U^+
&\qquad&     
\Delta_U^\times = 2\mu\cos(2\phi)\tilde{\Delta}_U^\times
\end{eqnarray}

\noindent In terms of these new variables, the evaluation equations takes the form

\begin{eqnarray}
\frac{\partial\tilde{\Delta}^+_I}{\partial\tau} + ik\mu\tilde{\Delta}^{+}_I - 2\frac{\partial h^+}{\partial \tau} &=& -an_e\sigma_T\left(\tilde{\Delta}^+_I+\tilde{\Lambda}^+\right) \\ 
\frac{\partial\tilde{\Delta}^+_Q}{\partial\tau} + ik\mu\tilde{\Delta}^{+}_Q - 2\frac{\partial h^+}{\partial \tau} &=& -an_e\sigma_T\left(\tilde{\Delta}^+_Q-\tilde{\Lambda}^+\right) \\ 
\tilde{\Delta}^+_U &=& \tilde{\Delta}^+_Q \\ 
\frac{\partial\tilde{\Delta}^+_V}{\partial\tau} + ik\mu\tilde{\Delta}^{+}_V &=& -an_e\sigma_T\tilde{\Delta}^+_V
\end{eqnarray}

\noindent where 
\begin{equation}
    \tilde{\Lambda}^+ = -\frac{3}{70}\tilde{\Delta}^+_{I4} + \frac{1}{7}\tilde{\Delta}^+_{I2} - \frac{1}{10}\tilde{\Delta}^+_{I0}+\frac{3}{70}\tilde{\Delta}^+_{Q4} + \frac{6}{7}\tilde{\Delta}^+_{Q2}+\frac{3}{5}\tilde{\Delta}^+_{Q0}
\end{equation}

The $\times$ polariation also give the similar equations. Here also we can see that the $V$ perturbations don't have any source terms. Therefore, they will remain $0$ through out the history of the universe. Also for tensor case, both the $\Delta^{+,\times}_Q$ and $\Delta^{+,\times}_U$ will be nonzero due to the lack of reflection symmetry. We can convert the $Q$ and $U$ components to $E$ and $B$ modes using same technique as discussed in Eq.~\ref{equationEB}. The contribution from both the $+$ and $\times$ mode will be same and the total contribution will be the sum of both the quantities. However, each of $\Delta^{+}$, $\Delta^{\times}$ quantities are spin $\pm 2$ quantities. Therefore, we have to multiply with the spin raising and lowering operators, which eventually multiply all the $\Delta_l$'s with a factor of $\left(\frac{(l+2)!}{(l-2)!}\right)^\frac{1}{2}$. This gives the source terms used in Sec.~\ref{ScalerSourceTerm}.
\end{appendices}    

\section*{Acknowledgement}    
AP is supported by NASA NESSF Award 80NSSC17K0481P00002. SD is supported by NSF Award AST-1616554.  Authors wish to thank Prof. Peter Timbie for revision of the manuscript. SD wishes to thank Prof. Tarun Souradeep for many useful discussions throughout the course of this project.

\bibliographystyle{plainnat}
\bibliography{References}

\end{document}